%% file: main.tex
\documentclass[a4paper,11pt]{article}
\pdfoutput=1

\usepackage{jheppub} 

\usepackage{amssymb}
\usepackage{booktabs}
\usepackage{enumitem}
\usepackage{slashed}
\usepackage[T1]{fontenc}

\DeclareUnicodeCharacter{2212}{\ensuremath{-}}

\usepackage{url}
\usepackage{bm}
\usepackage{xcolor}
\usepackage[export]{adjustbox}
\usepackage[normalem]{ulem}
\usepackage[caption=false]{subfig}
\usepackage{float}
\usepackage{soul}
\usepackage{diagbox}
\usepackage{comment}
\usepackage[section]{placeins}   
\usepackage{booktabs}
\usepackage{footmisc}
\usepackage{tikz}

\usepackage[capitalise]{cleveref}
\crefformat{equation}{Equation~(#2#1#3)}
\crefformat{figure}{Figure~(#2#1#3)}
\crefformat{section}{Section~(#2#1#3)}
\crefformat{table}{Table~(#2#1#3)}

\usepackage[compat=1.1.0]{tikz-feynman}
\usetikzlibrary{quantikz}
\tikzset{every node/.append style={font=\small}}
\tikzset{>=latex}
\tikzset{every picture/.append style={line cap=round, line join=round}}
\tikzset{every path/.append style={>=latex, line width=0.8pt}}
\tikzset{
  fermion/.style={
    postaction={decorate},
    decoration={
      markings,
      mark=at position 0.50 with {\arrow{Stealth}}
    }
  },
  wboson/.style={
    decorate,
    decoration={snake,amplitude=1.1pt,segment length=5pt}
  },
  massinsert/.style={
  	inner sep=1pt,
  	line width=0.6pt
  }
}

\newcommand{\CharmBNV}[4]{%
\begin{tikzpicture}[x=1cm,y=1cm,thick,>=Stealth]

  \draw[rounded corners=0.25cm] (-0.1,0.9) rectangle (1.45,-0.9);
  \node at (0.43, 0.58) {$u_R$};
  \node at (0.43, 0.00) {$c_R$};
  \node at (0.43,-0.58) {$#1$};

  \draw[fermion] (0.78, 0.58) -- (1.45, 0.58)
    .. controls (2.65,0.58) and (3.55,0.58) .. (4.50,-0.58);
  \draw[fermion] (0.78, 0.00) -- (4.50,0.00);
  \draw[fermion] (0.78,-0.58) -- (1.45,-0.58)
    .. controls (2.65,-0.58) and (3.55,-0.58) .. (4.50,0.58);

  \draw[fermion] (7.15, 0.58) -- (6.4,0.58)
    .. controls (5.2,0.58) and (4.8,0.58) .. (4.50,0.58);
  \draw[fermion] (7.15, 0.00) -- (4.50,0.00);
  \draw[fermion] (7.15,-0.58) -- (5.55,-0.58)
    .. controls (5.10,-0.58) and (4.85,-0.58) .. (4.50,-0.58);

  \node[right] at (7.15, 0.58) {$\bar{\ell}^{\,n}_{L}(k_3)$};
  \node[right] at (7.15, 0.00) {$\bar{\nu}^{\,m}_{L}(k_2)$};
  \node[right] at (7.15,-0.58) {$#3$};

  \node[draw,circle,fill=white,minimum size=0.34cm,inner sep=0pt]
    at (3.95,0) {$\times$};
  \node[above=5pt] at (3.95,0.20) {$#2$};

  \node[anchor=north west,align=left] at (-0.85,-1.20) {$#4$};

\end{tikzpicture}%
}

\newcommand{\chis}[1]{\chi_{_{#1}}}

\newcommand{\mylength}{1.3cm}

\renewcommand{\arraystretch}{1.2}

\preprint{PITT-PACC-2625}

\title{A proposal for an LHC search for baryon number violation in complementarity with trilepton nucleon decay searches.}

\author[a,b]{Amit Bhoonah,}
\author[b]{Francis M. Burk,}
\author[b,c]{Da Liu,}
\author[d,e]{Tong Ou,}
\author[f,g]{Deepak Sathyan}

\affiliation[a]{Laboratory for Symmetry and Structure of the Universe, Department of Physics,
Jeonbuk National University, Jeonju, Jeonbuk 54896, Korea}
\affiliation[b]{Pittsburgh Particle Physics, Astrophysics, and
  Cosmology Center, Department of Physics and Astronomy,
  University of Pittsburgh, Pittsburgh, PA 15260, USA}
\affiliation[c]{School of Physics, Shandong University,
  Jinan, Shandong 250100, China (main)}
\affiliation[d]{Enrico Fermi Institute, Department of Physics,
  University of Chicago, Chicago, IL 60637, USA}
  \affiliation[e]{Institut für Theoretische Physik, ETH Zürich, Wolfgang-Pauli-Str. 27, 8093 Zürich, Switzerland}
\affiliation[f]{Maryland Center for Fundamental Physics, Department of
  Physics, University of Maryland, College Park, MD 20742, USA}
\affiliation[g]{Mitchell Institute for Fundamental Physics and
  Astronomy, Department of Physics and Astronomy, Texas A\&M
  University, College Station, TX 77843, USA}

\emailAdd{amit.bhoonah@jbnu.ac.kr}
\emailAdd{frb25@pitt.edu}
\emailAdd{liuda@sdu.edu.cn}
\emailAdd{tongou@phys.ethz.ch}
\emailAdd{dsathyan@tamu.edu}

\abstract{We investigate the compatibility of existing limits on trilepton nucleon decay by the Super-Kamiokande (Super-K) experiment with Baryon Number Violation (BNV) at the TeV scale and motivate a dedicated search using existing data at the LHC. Working within an effective field theory framework, we derive model-independent constraints on higher-dimensional operators associated with $B - L/3$ conserving texture of BNV.   
We demonstrate that existing Super-K limits on loop-level trilepton nucleon decay provide significantly stronger constraints compared to tree-level charm-baryon decays induced by these operators.
These constraints translate into bounds of $\mathcal{O}(1-30)$ TeV on the BNV scale, leaving open the possibility of an LHC search for $B - L/3$ type BNV. Motivated by this complementarity, we investigate the collider phenomenology of a representative UV completion featuring an $s$-channel diquark resonance that decays into a leptoquark--dilepton pair, producing a distinctive final state containing an anti-top quark, two charged anti-leptons, and missing transverse energy. 
Our results indicate that, although same-sign dilepton searches by ATLAS and CMS already place  indirect constraints on this model, a dedicated analysis based on the selection cuts we develop significantly extends the reach for our chosen benchmarks. 
In particular, using only Run 2 data, we demonstrate complementary reach to limits on trilepton nucleon decay searches set by the Super-K experiment and to the projected reach of Hyper-Kamiokande. The selection cuts are applicable beyond our representative model.}

\keywords{Beyond Standard Model, Baryon Number Violation, Lepton Number Violation}

\begin{document}
\maketitle
\flushbottom

\section{\label{sec:Introduction}Introduction and motivation}
The conservation of baryon ($B$) and lepton ($L$) number in the observed interactions of elementary particles is an established experimental fact. From a theoretical point of view, however, this raises many questions as there are no fundamental principles at play behind the existence of either of these global symmetries. Moreover, there are many reasons to consider them to be accidental effects, valid only up to the roughly 10 TeV energy scale that has been experimentally accessed so far. For example, it is predicted that global symmetries are explicitly broken by quantum gravitational effects \cite{Abbott:1989jw,Kallosh:1995hi,Lee:1988ge}. The validity of this argument in the absence of an experimentally verified quantum theory of gravity remains an open question. However, there are compelling arguments, based on robust theoretical foundations and, more importantly, experimental observations, in favour of a more complete theory of nature with additional elementary particles whose interactions violate either \emph{or both} $B$ and $L$.
Namely, baryon number violation is one of three necessary conditions for realizing \textit{baryogenesis} \cite{Sakharov:1967dj}, and lepton number violation is common in mechanisms that generate neutrino masses, which are necessary to explain the observed flavour oscillations in the neutrino sector \cite{Davis:1968cp, Super-Kamiokande:1998kpq, SNO:2001kpb}. \\

The most distinctive prediction of most theories with Baryon Number Violation (BNV) is, of course, that the proton --- or the bound neutron --- should decay. This is the subject of a rich experimental program, with modes such as $p \rightarrow \pi_{0} \ e^{+}/\mu^{+}$ and $p \rightarrow K^{+} \overline{\nu}$ having been extensively searched for and, so far, not observed \cite{Super-Kamiokande:2020wjk,McGrew:1999nd,Kamiokande-II:1989avz,Super-Kamiokande:2014otb,Super-Kamiokande:2005lev}. A variety of more exotic nucleon decay modes have also been considered --- see \cite{Heeck:2019kgr} for a summary --- but, even in these channels, experiments have yet to identify conclusive evidence for BNV. These null results put stringent constraints on the masses of new particles whose interactions violate $B$ and $L$, often pushing them to values above $10^{15}$ GeV, a scale characteristic of Grand Unified Theories (GUTs) \cite{Georgi:1974sy,Pati:1974yy,Fritzsch:1974nn}. \\

In light of this, one might be tempted to associate any BNV interaction with very high energy scales, inaccessible to even the most optimistic of future collider experiments. There is, however, one caveat to these conclusions: GUT theories typically violate $B$ and $L$ by one unit each, keeping the difference, $B - L$, as a conserved quantity. This, in turn, induces low energy BNV operators at dimension six. Given a measured proton lifetime of $\tau_{P}$ years, one can argue by naive dimensional analysis that the scale associated with BNV is roughly
\begin{equation*}
 M_{\text{BNV}} \sim \left(5000\times \frac{\tau_{P}}{10^{32} \ \text{yrs}} \right)^{\frac{1}{4}}  \times 10^{15} \ \text{GeV},  
\end{equation*}
where we take the proton mass to be 1 GeV for simplicity and neglect phase space factors. \\

It has, however, been shown that if instead of $B - L$, the underlying theory of nature\footnote{Perhaps, more accurately, an intermediate version of it valid for energies above the electroweak scale but below that of Grand Unification.} conserves a more general linear combination $B + aL, \text{ where} ~a  \in \mathbb{Q}$, then the BNV effective operator is generated at a dimension higher than six \cite{Weinberg:1980bf,Wilczek:1979hc}. This causes a suppression of proton-decay-inducing operators by multiple powers of $M_{\text{BNV}}$.  
In fact, for such a low energy effective operator of dimension $d>6$ and a corresponding proton lifetime $\tau_{P}$, the BNV scale is roughly
\begin{equation*}\label{eq:ndagenerald}
    M_{\text{BNV}} \sim \left(5000\times \frac{\tau_{P}}{10^{32} \ \text{yrs}} \right)^{\frac{1}{4}}  \times \left(10^{15}\right)^{\frac{4}{2d-8}} \ \text{GeV}.  
\end{equation*}
From this estimate, it can be seen that for sufficiently high $d$, the mass scale of new particles involved in BNV interactions may lie significantly below the GUT scale. As an example, $d$ = 11 sets $M_{\text{BNV}} \approx 10$ TeV, within reach of current or near-term future collider experiments such as the Large Hadron Collider (LHC). An interesting consequence of such theories, where the low energy operator for BNV is of such high dimension that current experimental limits predict $M_{\text{BNV}}$ around the $\mathcal{O}(1-10$) TeV scale, is that it should be possible to physically produce the new resonances responsible for BNV interactions at the LHC or other near-term future colliders, raising the possibility of testing UV complete BNV theories. \\

But what kind of BNV theory can be more efficiently probed at the LHC than at proton decay experiments? Ideally, we would prefer if they were independent of arbitrarily hierarchical flavour structures that suppress BNV interactions involving light quarks in order to increase the lifetime of the proton. To identify such theories, it is helpful to note that $B$ and $L$ symmetries cannot be violated by arbitrary amounts. To see this, let us reconsider the aforementioned $B + aL$ symmetries. Choosing $a = \mp1$ results in $B \mp L$ conservation and dimension six or seven operators typical of GUT theories ruling out as such theories as TeV-scale BNV candidates. It is convenient to continue this exercise by labeling the theories according to the particular combination of $B$ and $L$ that they conserve. Since BNV by one unit requires three quark fields, Lorentz invariance requires an odd number of additional fermion fields to form the corresponding effective operator. This rules out the possibility of having theories violating $B$ by one unit and $L$ by $\pm 2$ units. There is therefore no $B \pm L/2$ theory, with the caveat that we are assuming that no new exotic fermions charged under another global symmetry -- neither $B$ nor $L$ -- have been introduced. \\ 

The next available options are therefore those that violate $B$ by two units, or those that violate $B$ by one and $L$ by $\pm$ 3 units, conserving $B \pm L/3$. The former has been well studied before and, most notably, predicts neutron to anti-neutron, $n\rightarrow \overline{n}$, oscillations. This is the subject of an extensive experimental effort already \cite{Phillips:2014fgb,Theroine:2016chp,Shima:2025aam}. Although current limits from these experiments are consistent with BNV at the TeV scale \cite{Super-Kamiokande:2020bov}, explicit UV completions, such as the ones considered in \cite{Babu:2006xc,Babu:2013yca}, reveal a challenge for probing these theories at the LHC: the final states can involve multiple jets whose charge reconstruction is challenging experimentally, making it difficult to unambiguously identify such BNV signatures at the LHC. \\

In this work, we instead focus on theories with $B - L/3$\footnote{The conserved $B - L/3$ global symmetry is anomalous, but since we make no attempt to gauge it, this poses no serious implications for the model itself.} symmetries and show, using a \emph{representative model}, that in addition to being consistent with BNV at the TeV scale, such theories offer a more unambigous BNV signature at the LHC. The lowest dimensional BNV operators consistent with $B - L/3$ conservation occur at dimension nine, 
\begin{subequations}\label{eq:BNVOpt}
    \begin{equation}\label{eq:BNVOptL}
        \mathcal{O}^{\text{L}}_{\text{BNV}} = \frac{c^{ijklmn}_{9,L}}{M^{5}_{\text{BNV}}} \left(\overline{u}^{c}_{ic_{2}}u_{jc_{3}} \right)\left( \overline{Q}^{c}_{k\beta c_{1}} L_{l\alpha}\right)  \left(\overline{L}^{c}_{m\gamma}L_{n\delta}\right)\epsilon^{\alpha\beta}\epsilon^{\gamma\delta}\epsilon^{c_{1}c_{2}c_{3}},
    \end{equation}
    \text{and}
    \begin{equation}\label{eq:BNVOptR}
        \mathcal{O}^{\text{R}}_{\text{BNV}} = \frac{c^{ijklmn}_{9,R}}{M^{5}_{\text{BNV}}} \left(\overline{u}^{c}_{ic_{2}}u_{jc_{3}} \right) \left(\overline{u}^{c}_{k c_{1}} l_{l}\right)\left(\overline{L}^{c}_{m\gamma}L_{n\delta}\right)\epsilon^{\gamma\delta}\epsilon^{c_{1}c_{2}c_{3}},
    \end{equation}
\end{subequations}    
where $i,j,k,l,m,n$ are flavour indices, $\alpha,\beta,\gamma,\delta$ are $SU(2)_L$ indices, and $c_{1},c_{2},c_{3}$ are colour indices. $Q_{i} = (u_{i},d_{i})$, $L_{j} = (\nu_{j}, l_{j})$ are left handed quark and lepton doublets respectively, while $u_{k,l}$ represents a right handed up-type quark and $l_{j}$ a right-handed charged lepton.\footnote{If the theory involves light right-handed neutrinos with masses well below the weak scale, other BNV operators are allowed but we do not consider them in this work.} \\  
%
%

A feature of these operators is that they do not generate proton decay at leading order. This is not a consequence of fine-tuning or an accidental cancellation, but follows directly from $SU(3)$ invariance: the anti-symmetry of the Levi-Civita tensor  forces the two right-handed up-type quarks in these operators to carry different flavors, meaning at least one of them must be a charm or top quark. Consequently, at tree level, they predict charm baryon decays but kinematically forbid proton decay, which occurs at one loop. The flavour anti-symmetry is also true for the two left handed lepton pairs and is due to $SU(2)$ invariance. Interestingly, this implies that any theory generating the BNV operators~\eqref{eq:BNVOptL} and~\eqref{eq:BNVOptR} \emph{must} involve interactions, potentially Baryon and Lepton Number Conserving (BNC and LNC respectively), that violate quark and lepton flavour. Since the Wilson coefficients $c^{ijklmn}_{9,L}$ and $c^{ijklmn}_{9,R}$ are products of various BNC and BNV interactions, if $M_{\text{BNV}}$ is at the TeV scale one must therefore carefully consider how strong constraints from experiments searching for flavour violation in the leptonic and quark sectors impact the predicted proton lifetime. \\

Theories with this particular texture of $B$ and $L$ violation predict trilepton nucleon decay, for example:
\begin{equation*}
p \rightarrow e^{+}/\mu^{+}\overline{\nu}\overline{\nu},   
\end{equation*}
which has been searched for by the Super-Kamiokande (Super-K) experiment \cite{Super-Kamiokande:2014pqx}. Since no evidence of it was found, a corresponding upper limit of $\tau_{_{P}} >$ 1.7 $\times 10^{32}$ years if the charged anti-lepton is a positron and $\tau_{_{P}} >$ 2.2 $\times 10^{32}$ years if it is an anti-muon was set on the proton lifetime $\tau_{_P}$. \\

In this work, we make two key points. Firstly, we demonstrate,  through an explicit computation, that although the limits from the Super-K search for trilepton nucleon decay are extremely stringent, they are compatible with BNV at the TeV scale.  Secondly, we show that an LHC search for $B - L/3$ theories can be complementary to trilepton proton decay searches at future experiments such as Hyper-Kamiokande. We start by computing \emph{model independent} constraints on the scale of $M_{\text{BNV}}$ in Section~\ref{sec:ModelIndpConstraints} using the lower limits on the proton lifetime under trilepton decay, finding a maximal bound\footnote{Here, by maximal we mean that all dimensionless parts of the Wilson coefficients are set to 1.} in the $10 - 20$ TeV range. For completeness, we also examine charm baryon decays into three anti-leptons and show that, despite the one-loop suppression, proton decay remains the dominant probe in setting the scale of $M_{\text{BNV}}$. In Section~\ref{sec:Model}, we consider a representative \emph{minimal} model for $B - L/3$ consisting of a diquark, a leptoquark, and a dilepton singlet whose trilinear interactions violate BNV and generate both $\mathcal{O}^{\text{L}}_{\text{BNV}}$ and $\mathcal{O}^{\text{R}}_{\text{BNV}}$ as low energy effective operators. We then incorporate all the BNC constraints that impact the proton lifetime and show that the corresponding limits of $M_{\text{BNV}}$ are in the $2 - 4$ TeV range. Since this is within reach of the LHC, we identify a particular benchmark consistent with Super-K limits that can be searched for at the LHC. In Section~\ref{sec:OptimalProcessSelection} we survey the various processes at the LHC that can probe our benchmark and identify the optimal one, which we study in Section~\ref{sec:collidermain}. There, after examining various kinematical quantities, we identify a set of selection cuts that makes it possible to have an LHC search for $B - L/3$ BNV that is in principle background free\footnote{We study only the dominant background and identify selection cuts that reduce it to less than one expected event at 139 fb$^{-1}$, leaving a more detailed examination of other backgrounds to a future dedicated search.}. In Section~\ref{sec:Generalizations}, we argue that our proposed search are also applicable to five of the seven models for $B - L/3$ identified in \cite{Heeck:2026dmh} and conclude in Section~\ref{sec:conclusion}.  
\section{Model Independent low energy constraints on $\bf{B-L/3}$}\label{sec:ModelIndpConstraints}
\begin{figure}[htbp]
    \centering
    \input{tikz/charmbaryon_1}\\
    \vspace{0.5cm}
    \input{tikz/charmbaryon_2}\\
    \vspace{0.5cm}
    \hspace{2cm}\input{tikz/proton_decay_q_fig1}\\
    \vspace{0.5cm}
    \hspace{2.5cm}\input{tikz/proton_decay_l_fig1}
    \caption{Summary of the various low energy probes of $B - L/3$ conserving BNV discussed in this work. The top diagram represents a singly charged charm baryon decaying into an anti-lepton and two anti-neutrinos, mediated only by $\mathcal{O}^{\text{L}}_{\text{BNV}}$. The second diagram shows a doubly charged charm baryon decaying into two anti-leptons and an anti-neutrino, mediated by both $\mathcal{O}^{\text{L}}_{\text{BNV}}$ and $\mathcal{O}^{\text{R}}_{\text{BNV}}$. Finally, the third and the fourth diagrams show the loop-induced proton decay process. The third diagram exists for $\mathcal{O}^{\text{L}}_{\text{BNV}}$ only, while the fourth can be mediated by either $\mathcal{O}^{\text{L}}_{\text{BNV}}$ or $\mathcal{O}^{\text{R}}_{\text{BNV}}$. The black crosses on the $u^k$ propagators denote mass insertions that enforce the right chirality in both $\mathcal{O}^{\text{L}}_{\text{BNV}}$ and $\mathcal{O}^{\text{R}}_{\text{BNV}}$ diagrams. The red cross on $\bar{\ell}^j$ denotes mass insertion that is only necessary for the $\mathcal{O}^{\text{R}}_{\text{BNV}}$ diagram.}
    \label{fig:BNV_processes}
\end{figure}

Proton decay experiments provide the most stringent constraints on BNV that determine whether the new particles responsible for these interactions can be probed directly at colliders. For $B - L$ theories, experimental limits on the proton lifetime imply a characteristic scale for BNV interactions close to the GUT scale, well beyond the reach of the LHC. For $B - L/3$, it has been suggested that the high dimensionality of the corresponding operators could permit BNV physics at the TeV scale, making direct searches at the LHC an attractive possibility \cite{Durieux:2012gj,Fonseca:2018ehk}. Our goal in this section is to establish this quantitatively with a detailed, \emph{model independent}, computation of the predicted proton lifetime in the presence of $\mathcal{O}^{\text{L}}_{\text{BNV}}$ and $\mathcal{O}^{\text{R}}_{\text{BNV}}$. \\

However, for $B - L/3$, there is also a strong motivation to expand the low-energy phenomenology beyond proton decay due to the flavour structure of these operators. As already mentioned, at least one of the two right handed up type quarks must be of charm or top flavour. The phenomenological consequence is that these operators predict tree level BNV decays of charm baryons. Proton decay is kinematically forbidden at the same order but arises radiatively through the flavour-changing diagrams shown in Figure~\ref{fig:BNV_processes}. Despite this loop suppression, however, it is not immediately obvious that charm baryon decays can provide better reach on the scale of BNV compared to proton decay, especially given the extraordinary statistics achievable at experiments searching for the latter. A summary of the main low energy BNV processes mediated by $\mathcal{O}^{\text{L}}_{\text{BNV}}$ and $\mathcal{O}^{\text{R}}_{\text{BNV}}$ is given in Figure~\ref{fig:BNV_processes}. \\

More importantly, these two processes probe complementary aspects of the underlying BNV interactions. Charm baryons can decay into final states containing a $\tau$ anti-lepton, which is kinematically forbidden in proton decay. Furthermore, at leading order, individual charm-baryon decay amplitudes probe Wilson coefficients with definite quark flavours, whereas proton decay is only sensitive to CKM-weighted combinations of coefficients summed over the internal up-type quarks. Charm-baryon decays therefore offer a valuable means of probing the flavour structure of $B - L/3$ independently of proton decay. Motivated by this complementarity, in addition to deriving the lower limits on $M_{\text{BNV}}$ from null results of the trilepton nucleon decay search by Super-K, we investigate the prospective sensitivity of dedicated charm baryon trilepton decays in Section~\ref{sec:charm_baryon_BNV} before proceeding to the proton decay case in Section~\ref{ssec:protondecayamp} and summarizing the limits on the BNV scale from such decays in Section~\ref{ssec:upperlimitslowenergy}. \\

\subsection{Charm baryon BNV decays}
\label{sec:charm_baryon_BNV}
We first examine singly charged charm baryon BNV decays, mediated by only the $\mathcal{O}^{\text{L}}_{\text{BNV}}$ operators as shown in the first panel of Figure~\ref{fig:BNV_processes}. Written out explicitly, the particular operator responsible for singly charged charm baryon decays reads
\begin{equation}\label{eq:CharmOpt1}
        \mathcal{O}^{\text{L}}_{\text{charm}} = -\frac{2c^{uc}_{uu}(V^{T}c_{LQ}U)^{il}(U^{T}c_{LL})^{mn}}{M^{5}_{\text{BNV}}}\epsilon^{IJK} \left(\overline{u}^{c}_{I}P_{R}c_{J} \right)\left(\overline{d}^{c}_{ iK} P_{L}\nu_{l}\right)  \left( \overline{\nu}^{c}_{m}P_{L}l_{n}\right),
\end{equation}
where the flavour indices for the down type quarks are $i = d$ for $\Lambda^{+}_{c}$ or $ \Sigma^{+}_{c}$  and $i = s$ for $\Xi^{+}_{c}$ or $\Xi^{\prime +}_{c}$ while $m,n, j = e, \mu, \tau$ with $m\neq n$. Here $V$ and $U$ are the CKM \cite{pdg_ckm_matrix,Cabibbo:1963yz,Kobayashi:1973fv} and PMNS \cite{pdg_pmns_matrix,Pontecorvo:1957qd,Maki:1962mu} matrices defined as
\begin{align}
    V_{\rm CKM}\equiv V_L^{u} V_L^{d\dagger},~~~~U_{\rm PMNS}\equiv V_{L}^{\ell} V_{L}^{\nu\dagger},
\end{align}
where $V_{L}^{f}$ is the transformation matrix from the gauge basis to the mass basis: $f_L = V_L^f f_L^{(g)}$. Without loss of generality, we diagonalise the up-type quark and charged-lepton Yukawa matrices, so that the $V$ and $U$ matrices appear entirely as rotations of the left-handed down quarks and neutrinos respectively. The factor of 2 above comes from using the anti-symmetry of the couplings of the $\overline{L}^{c}L$ term, which implies that $c_{LL}$ as defined in Equation~\eqref{eq:CharmOpt1} is a strictly upper triangular matrix. \\

Since we will always average over the spin states of the parent baryon, the differential decay rate for the process reads 
\begin{equation}
    d\Gamma = \frac{1}{\left(2\pi\right)^{3}}\frac{1}{32m^{3}_{\text{B}}} \vert \overline{\mathcal{M}} \vert^{2}_{\text{B} \rightarrow \overline{l}\overline{\nu}\overline{\nu}} \ dm^{2}_{12} \ dm^{2}_{23},
\end{equation}
where $m_{\text{B}}$ is the mass of the parent baryon, $\vert \overline{\mathcal{M}}\vert^{2}_{\text{B} \rightarrow \overline{l}\overline{\nu\nu}}$ is the spin averaged square of the amplitude for particular process in question and
\begin{equation}
\begin{split}
    m^{2}_{12} & = \left(k_{1} + k_{2}\right)^{2}, \\ 
    m^{2}_{23} & = \left(k_{2} + k_{3}\right)^{2},  
\end{split}
\end{equation}
are kinematic invariants computed in the rest frame of the parent baryon and under the assumption that neutrinos are massless. \\

The amplitude for the process
\begin{equation*}
B_{c}(P) \rightarrow \overline{\nu}_{l}(k_{1}) \ \overline{\nu}_{m}(k_{2}) \ l^{+}_{n}(k_{3}),   
\end{equation*}
where $B_{c} = \Lambda^{+}_{c}$ or $\Sigma^{+}_{c}$  if $i = d$ and $B_{c} = \Xi^{+}_{c}$ if $i=s$ \footnote{Since the decay width for  $\Xi^{\prime +}_{c}$ if $i=s$ is not available, we do not consider it here.}, reads
\begin{equation}\label{eq:amp_Charm_B}
\begin{split}
    \mathcal{M}^{\text{L}}_{\Lambda^{+}_{\text{c}} \rightarrow \overline{l\nu\nu}} ={}& -\frac{2c^{uc}_{uu}(V^{T}c_{LQ}U)^{il}(U^{T}c_{LL})^{mn}}{M^{5}_{\text{BNV}}} \\
    &\times \Bigl[ \overline{v}_{m}(k_{2}) P_{L} v_{n}(k_{3})\, \langle 0\vert \epsilon^{IJK} P_{L}\, \overline{d}^{c}_{iI}\, \overline{u}^{c}_{J} P_{R} c_{K} \vert \Lambda^{+}_{\text{c}} \rangle P_{L} v_{l}(k_{1}) \\
    &\qquad + \delta_{lm}\, \overline{v}_{m}(k_{1}) P_{L} v_{n}(k_{3})\, \langle 0\vert \epsilon^{IJK} P_{L}\, \overline{d}^{c}_{iI}\, \overline{u}^{c}_{J} P_{R} c_{K} \vert \Lambda^{+}_{\text{c}} \rangle P_{L} v_{l}(k_{2}) \Bigr] ,
\end{split}
\end{equation}
where the second line takes into account the possibility that the neutrino flavours $l$ and $m$ are the same. The interference between these two terms vanishes in the massless neutrino limit and will be neglected in the remainder of this computation.  Since we were unable to find an explicit lattice computation for the  hadronic matrix element above, we follow the methods discussed in \cite{QCDSF:2008qtn} and approximate it as
\begin{equation}
    \langle 0\vert \epsilon^{IJK} P_{L} 
   \overline{d}^{c}_{iI} \overline{u}^{c}_{J}P_{R}c_{K} \vert B_{\text{c}} \rangle  \simeq \begin{cases} - P_{L} \frac{m_{\Lambda_{\text{c}}}f_{\Sigma_{\text{c}}}}{4} \overline{v}_{B_{c}}(P) \ \text{if} \ i = d \ \text{and} \ B_{c} = \ \Lambda_{c}, \\
   - P_{L} \frac{m_{\Sigma_{\text{c}}}f_{\Lambda_{\text{c}}}}{4} \overline{v}_{B_{c}}(P) \ \text{if} \ i = d \ \text{and} \ B_{c} = \ \Sigma_{c}, \\ 
   - P_{L} \frac{m_{\Xi_{\text{c}}}f_{\Xi^{\prime}_{\text{c}}}}{4} \overline{v}_{B_{c}}(P) \ \text{if} \ i = s \ \text{and} \ B_{c} = \ \Xi_{c}.  
    \end{cases}
    \label{eq:CharmBaryonHadronicME}
\end{equation}
From the lattice computation reported in \cite{Li:2025zfw}, 
\begin{equation}\label{eq:CharmBaryonLatticeConstants}
\begin{gathered}
    f_{\Lambda_{c}} = 1.71 \times 10^{-2}~\text{GeV}^{2} , \qquad
    f_{\Sigma_{c}} = 2.33 \times 10^{-2}~\text{GeV}^{2} , \\
    f_{\Xi_{c}} = 1.90 \times 10^{-2}~\text{GeV}^{2} , \qquad
    f_{\Xi^{\prime}_{c}} = 2.58 \times 10^{-2}~\text{GeV}^{2},
\end{gathered}
\end{equation}
omitting statistical and systematic uncertainties. \\

It is important to caution the reader once again that these hadronic matrix elements in Equation~\eqref{eq:CharmBaryonHadronicME} are \emph{estimates} and that the derivation makes use of chiral perturbation theory, which has questionable validity for charm baryons. We only employ this estimate since a proper lattice simualtion to determine the corresponding hadronic matrix is beyond the scope of this work. To reflect these uncertainties, we will use two values for the decay constants $f_{B_{c}}$ in Equation~\eqref{eq:CharmBaryonHadronicME}: the values in Equation~\eqref{eq:CharmBaryonLatticeConstants} and 
\begin{equation}
f_{B_{c}} = m^{2}_{B_{c}},
\end{equation}
and show a corresponding range in our final results in Table~\ref{tab:BNV_bounds}. \\

Assuming the hadronic matrix elements for charm baryons are given by Equation~\eqref{eq:CharmBaryonHadronicME}, the total amplitude is
\begin{equation}
\mathcal{M}^{\text{L}}_{B_{c} \rightarrow \overline{l\nu\nu}} \simeq{} \frac{2c^{uc}_{uu}(V^{T}c_{LQ}U)^{il}(U^{T}c_{LL})^{mn}}{M^{5}_{\text{BNV}}} \, \frac{m_{B_{\text{c}}}f_{B^{\prime}_
{c}}}{4} \mathcal{S}_{lmn},
\end{equation}
where $B^{\prime}_{c} = \Sigma_{c}$ if $B_{c} = \Lambda_{c}$ or $B^{\prime}_{c} = \Xi^{\prime}_{c}$ if $B_{c} = \Xi_{c}$, and vice versa for both. Here, we have introduced the spin structure
\begin{equation}\label{eq:spinorstructureL}
    \mathcal{S}_{lmn} \equiv \overline{v}_{m}(k_{2}) P_{L} v_{n}(k_{3})\, \overline{v}_{P}(p) P_{L} v_{l}(k_{1}) 
    + \delta_{lm}\, \overline{v}_{m}(k_{1}) P_{L} v_{n}(k_{3})\, \overline{v}_{P}(p) P_{L} v_{l}(k_{2})
\end{equation} 
for notational clarity. The decay width for this particular set of final states is then  
\begin{equation}
\begin{split}
\Gamma^{lmn}_{B_{c}\rightarrow \overline{l\nu\nu}}  \simeq \ \frac{\vert c^{uc}_{uu}(V^{T}c_{LQ}U)^{il}(U^{T}c_{LL})^{mn}\vert^{2}}{3072\left(2\pi\right)^{3}} \frac{m^{7}_{B_{\text{c}}}f^{2}_{B^{\prime}_{\text{c}}}}{M^{10}_{\text{BNV}}}h\left(\frac{m_{l_{n}}}{m_{B}}\right),   
\end{split}
\end{equation}
where 

\begin{equation}
h(x)=
1-8x^{2}+8x^{6}-x^{8}
-12x^{4}\log\left(x^{2}\right),
\qquad
x\equiv\frac{m_{\ell_n}}{m_P}.
\end{equation}

The corresponding branching fractions are, approximately,

\begin{equation}
\label{eq:BrLimit_C_Baryon}
\begin{aligned}
\mathrm{Br}^{lmn}_{B_c\to\bar{\nu}_l\bar{\nu}_m\ell_n^+}
\simeq {}&
\frac{
\left|
(V^T c_{LQ} U)^{dl}
(U^T c_{LL})^{mn}
\right|^2
}{
(M_{\mathrm{BNV}}/\mathrm{TeV})^{10}
}
\,h\!\left(\frac{m_{\ell_n}}{m_{B_c}}\right)
\\[4pt]
&\times
\begin{cases}
[7.1\times10^{-26},\,3.6\times10^{-21}],
& B_c=\Lambda_c^+,\\[3pt]
[8.9\times10^{-35},\,1.1\times10^{-29}],
& B_c=\Sigma_c^+,\\[3pt] 
[1.4\times10^{-37},\,1.2\times10^{-29}], & B_c=\Xi_c^+,  \\
\end{cases}
\end{aligned}
\end{equation}
where $l, m, n = e, \mu, \tau$ and we have used the measured lifetimes reported in the Baryon Summary Table in Ref.~\cite{ParticleDataGroup:2022pth}. The ranges in Equation~\eqref{eq:BrLimit_C_Baryon} reflect the theoretical uncertainty associated with our approximate treatment of the non-perturbative charmed-baryon matrix elements. \\ 

While we will derive the precise corresponding limits from the non-observation of such decay modes at charm baryon experiments in Section~\ref{ssec:upperlimitslowenergy}, the reader should anticipate, given the numerical pre-factors in Equation~\eqref{eq:BrLimit_C_Baryon}, that they are likely to be weak, with limits on $M_{\text{BNV}}$ well below the TeV scale.  In light of this, we do not pursue our analysis of charm baryon decays further, especially given the much stronger constraints that we will derive next on $M_{\text{BNV}}$ from trilepton proton decay. However, before closing this discussion of charm baryon decays, we mention that the BNV decays of doubly charged charm baryons, such as   
\begin{equation}\label{eq:DoublyChargedBaryonBNVdecay}
        B^{++}_{c}(P) \rightarrow l^{+}_{l}(k_{1}) \ \overline{\nu}_{m}(k_{2}) \ l^{+}_{n} (k_{3}), 
    \end{equation}
where $B^{++}_{c} = \Sigma_{c}^{++} \ \text{or}\ \Xi_{cc}^{++}$ are the only low energy probe of the individual Wilson coefficients $c^{uc}_{uu}c^{il}_{lu}c^{mn}_{LL}$ ($i=u$ for $\Sigma_{c}^{++}$ and $i=c$ for $\Xi_{cc}^{++}$) for the $\mathcal{O}^{\text{R}}_{\text{BNV}}$ operator. Unfortunately,  the derived lower limits on $M_{\text{BNV}}$ are expected to be comparable to our results for singly charged charm baryon decays, and hence quite weak.

\subsection{Proton decay}\label{ssec:protondecayamp}
Next, we consider the loop-induced proton decay process shown in the third and fourth panels of Figure~\ref{fig:BNV_processes}. Unlike for the singly charged charm baryon decays considered above, it is mediated by both $\mathcal{O}^{\text{L}}_{\text{BNV}}$ and $\mathcal{O}^{\text{R}}_{\text{BNV}}$, with the third and fourth panels mediated by the former but only the fourth one mediated by the latter operator. This can be seen by noting that the two operators differ in that $\mathcal{O}^{\text{L}}_{\text{BNV}}$ contains a left-handed quark-lepton doublet while $\mathcal{O}^{\text{R}}_{\text{BNV}}$ only contains a coupling between a right handed up-type quark and a right handed charged lepton, forbidding the third panel. \\

We start by computing the relevant amplitudes for proton decay mediated by these operators, considering them separately for simplicity. The one from $\mathcal{O}^{\text{L}}_{\text{BNV}}$ can be split into two sub-amplitudes,
\begin{equation}
    \mathcal{M}^{\text{L}}_{p \rightarrow \overline{l}\overline{\nu}\overline{\nu}} = \mathcal{M}^{{\text{L}}q}_{p \rightarrow \overline{l}\overline{\nu}\overline{\nu}} + \mathcal{M}^{{\text{L}}l}_{p \rightarrow \overline{l}\overline{\nu}\overline{\nu}}, 
\end{equation}
where $\mathcal{M}^{^{\text{L}q}}_{p \rightarrow \overline{l}\overline{\nu}\overline{\nu}}$ corresponds to diagram shown on the third panel of Figure~\ref{fig:BNV_processes} and $\mathcal{M}^{^{\text{L}l}}_{p \rightarrow \overline{l}\overline{\nu}\overline{\nu}}$ to the fourth one. As with the charm baryon computation, if the outgoing antineutrino flavours $ l= m$, each of these diagrams will have a companion with $k_{1} \leftrightarrow k_{2}$, with the interference between them being proportional to neutrino masses and therefore negligible. \\

As detailed in Appendix~\ref{sec:ProtonDEccayAmplitudes} the sub-amplitudes for proton decay induced by the $\mathcal{O}^{\text{L}}_{\text{BNV}}$ operator are 
\begin{align}
    \mathcal{M}^{\text{L}q}_{p \rightarrow \overline{l\nu\nu}} &= -\,\mathcal{C}^{(q)}_{lmn}\, \mathcal{S}_{lmn} , \ \text{and}\label{eq:ampprotondecayquarkL} \\
    \mathcal{M}^{\text{L}\ell}_{p \rightarrow \overline{l\nu\nu}} &= \mathcal{C}^{(\ell)}_{lmn}\, \mathcal{S}_{lmn} , \label{eq:ampprotondecaylepton}
\end{align}
respectively, where the coefficients are
\begin{align}
    \mathcal{C}^{(q)}_{lmn} &= \mathcal{K} \left(c_{LL}U\right)^{mn} \frac{m_{P}}{3}\sum_{j,k} V^{T kd} c^{uk}_{uu}\, m_{u_{k}}\, f_{\text{loop}}(m_{d_{j}},m_{u_{k}})\, V^{*uj} \left(V^{T}c_{LQ}U\right)^{jl} , \\
    \mathcal{C}^{(\ell)}_{lmn} &= \mathcal{K} \left(c_{LL}U\right)^{mn} \frac{m_{P}}{3}\sum_{j,k} V^{T kd} c^{uk}_{uu}\, m_{u_{k}}\, f_{\text{loop}}(m_{l_{j}},m_{u_{k}})\, U^{\dagger jl} c^{uj}_{LQ} , \\
        \mathcal{K} &\equiv \frac{2\sqrt{2}}{16\pi^{2}}\, \frac{\alpha_{\text{QCD}}\, G_{F}}{M^{5}_{\text{BNV}}}.
\end{align}
Here, $G_{F}$ is the Fermi constant, $\alpha_{\text{QCD}} = -0.014~\text{GeV}^{3}$ is taken from the lattice computations reported in \cite{Aoki:2017puj}. The factor of $m_{P}/3$ is due to a simplifying assumption regarding consituent quark momentum distributions inside the proton, as detailed in Appendix~\ref{sec:ProtonDEccayAmplitudes}. \\

The amplitude from $\mathcal{O}^{\text{R}}_{\text{BNV}}$, discussed in greater detail in Appendix~\ref{sec:ProtonDEccayAmplitudes}, is given by
\begin{equation}\label{eq:ampprotondecayquarkR}
    \mathcal{M}^{\text{R}}_{p \rightarrow \overline{l\nu\nu}} = \mathcal{C}^{(\text{R})}_{lmn}\, \mathcal{S}_{lmn} ,
\end{equation}
with
\begin{equation}\label{eq:coeffR}
    \mathcal{C}^{(\text{R})}_{lmn} = \mathcal{K} \left(c_{LL}U\right)^{mn} \sum_{j,k} V^{T kd} c^{uk}_{uu}\, m_{u_{k}}\, m_{l_{j}}\, f_{\text{loop}}(m_{u_{k}},m_{l_{j}})\, U^{\dagger jl} c^{uj}_{ul} ,
\end{equation}
where we note a mass insertion $m_{l_{j}}$, which is due to the different chirality structure of the $\mathcal{O}^{\text{R}}_{\text{BNV}}$ compared to $\mathcal{O}^{\text{L}}_{\text{BNV}}$. \\

We simplify the evaluation of the loop function $f_{\text{loop}}$ by dropping all terms with momentum or mass scale comparable to the proton mass, which is justified since $m_{P} \ll m_{W}$. This is equivalent to evaluating the loop integral in the ``hard'' region\footnote{Importantly, since the loop integral is finite, we do not need to use the method of regions to isolate UV divergences.}, where the loop function is the same for both operators and reads

\begin{equation}\label{eq:fLLoop}
f_{\text{loop}}(m_k,m_j) =
\begin{cases}
\displaystyle
\frac{\log\left(X_{kW}^{2}\right)}
{1-X_{kW}^{2}},
& \text{if } j = 1 \ \text{or} \ j = 2
\\[3ex]
\displaystyle
\frac{1}{1-X_{kW}^{2}}
\left[
\frac{\log\left(X_{jW}^{2}\right)}
{1-X_{jW}^{2}}
-
\frac{X_{kW}^{2}
\log\left(X_{jW}^{2}/X_{kW}^{2}\right)}
{X_{kW}^{2}-X_{jW}^{2}}
\right],
& j = 3
\end{cases}
\end{equation}
with $X_{ij} \equiv m_{i}/m_{j}$. \\

The proton decay rate induced from the different operators can then be written as
\begin{equation}
    \Gamma_{\text{L}/\text{R}} = \frac{1}{6144\pi^{3}}\frac{G^{2}_{F}\alpha^{2}_{\text{QCD}}}{32\pi^{4}}\frac{m^{9}_{P}}{M^{10}_{\text{BNV}}}h\left(\frac{m_{l_{n}}}{m_{P}}\right) \vert \mathcal{C}^{lmn}_{\text{L}/\text{R}} \vert^{2}_{p \rightarrow \overline{l}\overline{\nu}\overline{\nu}},
\end{equation}
with L or R corresponding to $\mathcal{O}^{\text{L}}_{\text{BNV}}$ and $\mathcal{O}^{\text{R}}_{\text{BNV}}$ respectively and  
\begin{subequations}
\begin{equation}
\begin{split}
    \mathcal{C}^{lmn}_{\text{L}} ={}& \frac{\left(c_{LL}U\right)^{mn}}{m^{2}_{P}} \sum_{j}\sum_{k} V^{T kd}c^{uk}_{uu}m_{u_{k}}m_{P}/3 \\
    & \Bigl( f^{jk}_{\text{loop}}\left(m_{u_{k}},m_{l_{j}}\right)U^{\dagger jl}c^{uj}_{LQ}
    - f_{\text{loop}}(m_{d_{j}},m_{u_{k}})V^{*uj}\left(V^{T}c_{LQ}U\right)^{jl} \Bigr),
\end{split}
\end{equation}
\begin{equation}
    \mathcal{C}^{lmn}_{\text{R}} = \frac{1}{m^{2}_{P}}\sum_{j} \sum_{k}  V^{T kd}c^{uk}_{uu}m_{u_{k}}m_{l_{j}}\left(f^{jk}_{\text{loop}}\left(m_{u_{k}},m_{l_{j}}\right)U^{\dagger jl}\right)c^{uj}_{LQ}\left(c_{LL}U\right)^{mn}, 
\end{equation}
\end{subequations}
the predicted proton lifetime is then
\begin{equation}
    \tau^{\text{L}/\text{R}}_{p \rightarrow \overline{l}_{n}\overline{\nu}_{l}\overline{\nu}_{m}} \simeq 4.5 \times10^{32} \ \text{years} \times \left(\frac{M_{\text{BNV}}}{\text{15 TeV}}\right)^{10}\times \left\vert \mathcal{C}^{lmn}_{\text{L}/\text{R}}\right\vert^{-2},
\end{equation}
where $l,m = e, \mu, \tau$ and $n = e, \mu$. 

\subsection{Lower limits on the BNV 
scale from low energy experiments}\label{ssec:upperlimitslowenergy}

\begin{table}[htbp]
    \centering
    \small
    \renewcommand{\arraystretch}{1.2}

    \begin{tabular}{l c c c}
        \toprule
        Proton decay mode &
        Experimental bound &
        Operator &
        Lower limit on $M_{\mathrm{BNV}}$ [GeV] \\
        \midrule

        $p \rightarrow e^{+}\bar{\nu}\bar{\nu}$
        & $\tau \geq 1.7 \times 10^{32}\,\mathrm{yr}$
        & $\mathcal{O}^{\mathrm{L}}_{\mathrm{BNV}}$
        & 12 $\times 10^{3}$ \\

        $p \rightarrow e^{+}\bar{\nu}\bar{\nu}$
        & $\tau \geq 1.7 \times 10^{32}\,\mathrm{yr}$
        & $\mathcal{O}^{\mathrm{R}}_{\mathrm{BNV}}$
        & 21 $\times 10^{3}$ \\

        $p \rightarrow \mu^{+}\bar{\nu}\bar{\nu}$
        & $\tau \geq 2.2 \times 10^{32}\,\mathrm{yr}$
        & $\mathcal{O}^{\mathrm{L}}_{\mathrm{BNV}}$
        & 12 $\times 10^{3}$ \\

        $p \rightarrow \mu^{+}\bar{\nu}\bar{\nu}$
        & $\tau \geq 2.2 \times 10^{32}\,\mathrm{yr}$
        & $\mathcal{O}^{\mathrm{R}}_{\mathrm{BNV}}$
        & 22 $\times 10^{3}$ \\

        \midrule
        Charm baryon decay mode &
        Projection &
        Operator &
        Lower limit on $M_{\mathrm{BNV}}$ [GeV] \\
        \midrule

        $\Lambda_c^{+}\rightarrow e^{+}\bar{\nu}\bar{\nu}$
        & $\mathcal{B}\leq 10^{-6}$
        & $\mathcal{O}^{\mathrm{L}}_{\mathrm{BNV}}$
        & 6 -- 18 \\

        $\Sigma_c^{+}\rightarrow e^{+}\bar{\nu}\bar{\nu}$
        & $\mathcal{B}\leq 10^{-6}$
        & $\mathcal{O}^{\mathrm{L}}_{\mathrm{BNV}}$
        & 1 -- 4 \\

        $\Xi_c^{+}\rightarrow e^{+}\bar{\nu}\bar{\nu}$
        & $\mathcal{B}\leq 10^{-6}$
        & $\mathcal{O}^{\mathrm{L}}_{\mathrm{BNV}}$
        & 8 -- 23 \\

        $\Lambda_c^{+}\rightarrow \mu^{+}\bar{\nu}\bar{\nu}$
        & $\mathcal{B}\leq 10^{-6}$
        & $\mathcal{O}^{\mathrm{L}}_{\mathrm{BNV}}$
        & 7 --20 \\

        $\Sigma_c^{+}\rightarrow \mu^{+}\bar{\nu}\bar{\nu}$
        & $\mathcal{B}\leq 10^{-6}$
        & $\mathcal{O}^{\mathrm{L}}_{\mathrm{BNV}}$
        & 1 -- 5 \\

        $\Xi_c^{+}\rightarrow \mu^{+}\bar{\nu}\bar{\nu}$
        & $\mathcal{B}\leq 10^{-6}$
        & $\mathcal{O}^{\mathrm{L}}_{\mathrm{BNV}}$
        & 9 -- 26 \\

        $\Lambda_c^{+}\rightarrow \tau^{+}\bar{\nu}\bar{\nu}$
        & $\mathcal{B}\leq 10^{-6}$
        & $\mathcal{O}^{\mathrm{L}}_{\mathrm{BNV}}$
        & 5 -- 14 \\

        $\Sigma_c^{+}\rightarrow \tau^{+}\bar{\nu}\bar{\nu}$
        & $\mathcal{B}\leq 10^{-6}$
        & $\mathcal{O}^{\mathrm{L}}_{\mathrm{BNV}}$
        & 1 -- 4 \\

        $\Xi_c^{+}\rightarrow \tau^{+}\bar{\nu}\bar{\nu}$
        & $\mathcal{B}\leq 10^{-6}$
        & $\mathcal{O}^{\mathrm{L}}_{\mathrm{BNV}}$
        & 10 -- 30 \\

        \bottomrule
    \end{tabular}

    \caption{
        Lower limits on the effective BNV scale $M_{\mathrm{BNV}}$ from experimental proton-decay constraints rounded to the nearest TeV (upper panel) and projected sensitivities to BNV charm-baryon decays (lower panel) rounded to the nearest GeV, assuming a branching-fraction sensitivity of $10^{-6}$. For charm baryon decays, these limits are at scales where the EFT description breaks down. Given the much stronger limits from proton experiments, for which EFT validity is well established, it can be concluded that charm baryon decays are not an optimal probe of $B - L/3$ conserving BNV. The pessimistic reader will likely infer from these limits that a similar conclusion applies to the LHC. However, we would like to remind them that although proton decay yields substantially stronger constraints despite being loop suppressed, these ones shown here assume maximal dimensionless Wilson-coefficient factors. Relaxing this assumption can significantly weaken the inferred bounds on $M_{\mathrm{BNV}}$, leaving TeV-scale new physics accessible to the LHC. Consequently, collider searches remain a complementary probe of $B -L/3$ conserving theories, particularly in UV completions where production and decay processes depend on different coupling combinations, especially flavour hierarchical ones.
    }
    \label{tab:BNV_bounds}
\end{table}

Since none of the various baryon decay mode discussed above have been observed so far, these null results imply lower limits on the scale $M_{\text{BNV}}$. To derive the precise, \emph{model dependent}, constraints one should consider the different dimensionless pieces of the Wilson coefficients separately and incorporate any relevant constraints that, while not baryon number violating on their own, directly affect the proton decay rate. Often, these can be quite stringent and further enhance the proton lifetime, as we will illustrate in the next section using a representative model. However, for the analysis in the present section, we will simply set $c_{uu}$, $c_{LQ}$, and $c_{LL}$ to unity and, in this way, derive the most conservative lower bound on $M_{\text{BNV}}$ from heavy flavour and proton decay experiments. More precisely,

\begin{enumerate}
    \item For limits from charm baryon decays, in the absence of dedicated searches for these modes, we will require that the BNV branching fraction not exceed $10^{-6}$ as a representative target sensitivity, chosen to reflect what a dedicated analysis at a charm factory might plausibly achieve,  and use the corresponding measured lifetimes reported in \cite{LHCb:2019ldj}. 
    \item For proton decay, we saturate the lower limits on the proton lifetime reported by Super-K.
\end{enumerate}

The results, obtained by computing the neutrino-flavor-inclusive decay widths,
\[
\Gamma_B^n \equiv \frac{1}{2}\sum_{l,m}\Gamma_B^{lmn},
\]
are summarized in Table~\ref{tab:BNV_bounds}. Our central conclusion is that, despite the high dimensionality of these effective operators, null results from proton decay experiments still provide surprisingly stringent constraints on $B - L/3$ conserving BNV interactions, with lower limits in the $10 - 30$ TeV range. Additionally, despite being derived from a loop level process, these are three orders of magnitude stronger than the limits derived from tree level charm baryon decays, reflecting the substantial statistical advantage  experiments designed to search for proton decay. In fact, our projected sensitivity from charm baryon trilepton decays are at a scale where the validity of an EFT limit is questionable, as can be seen in Table~\ref{tab:BNV_bounds}. Therefore, we also make the unfortunate conclusion that despite the unique complementarity that charm baryon trilepton decay searches represent, the statistical limitations inherent to heavy flavour experiments make them sub-optimal as a probe of $B - L/3$. \\

The same, however, cannot be said of the LHC. To elaborate, the $\mathcal{O}(10~\mathrm{TeV})$ bounds reported in Table~\ref{tab:BNV_bounds} are obtained under the conservative assumption that the dimensionless couplings entering the Wilson coefficients take their maximal values. Relaxing this assumption can have important implications for future LHC searches. For example, taking each of the three dimensionless couplings entering the relevant Wilson coefficient to be $\mathcal{O}(0.1)$ weakens the scale of $M_{\text{BNV}}$ to approximately $3-7~\mathrm{TeV}$, firmly in LHC territory.

\section{Representative Model and Existing Phenomenology for $\mathbf{B-L/3}$}
\label{sec:Model}

The possibility of TeV-scale BNV from $B - L/3$ theories is consequential beyond the value of the bound itself. Since this lies below the energies probed at the LHC, the effective field theory description breaks down and the search must be
formulated in terms of an explicit UV completion. This is a significant advantage thanks to the suppression due to the high dimensionality of $\mathcal{O}^{\text{L}}_{\text{BNV}}$ and $\mathcal{O}^{\text{R}}_{\text{BNV}}$. In this section we introduce a minimal \emph{representative}
model realising $B -L/3$, establish the constraints that existing
BNC and LNC searches already place on it, study their implications for the predicted proton lifetime, and identify the
region of parameter space that remains open for a dedicated $B - L/3$ BNV search strategy at the LHC. 
\subsection{Minimal Model for $\mathbf{B - L/3}$}
\begin{table}[htbp]
  \centering
  \small
  \renewcommand{\arraystretch}{1.3}
  \begin{tabular}{c c c c c c}
    \toprule
    Particle & $SU(3)_{c}$ & $SU(2)_{L}$ & $U(1)_{Y}$ & $U(1)_{B}$ & $U(1)_{L}$ \\
    \midrule
    $\chi_{uu}$ & $\mathbf{3}$ & $\mathbf{1}$ & $-\tfrac{4}{3}$ & $-\tfrac{2}{3}$ & $0$ \\
    $\chi_{LQ}$ & $\mathbf{3}$ & $\mathbf{1}$ & $-\tfrac{1}{3}$ & $+\tfrac{1}{3}$ & $+1$ \\
    $\chi_{LL}$ & $\mathbf{1}$ & $\mathbf{1}$ & $+1$ & $0$ & $-2$ \\
    \bottomrule
  \end{tabular}
  \caption{Quantum numbers of the three new scalars of the representative model for $B - L/3$ considered in this work.}
  \label{tab:ParticleContent13}
\end{table}
We add three new scalar fields to the Standard Model of Particle Physics (SM). The first is a diquark, $\chi_{_{uu}}$, that couples to two quarks and arises in the context of Grand Unification models based on the $\text{E}_{6}$ \cite{Hewett:1988xc} or Pati-Salam \cite{Pati:1974yy,Mohapatra:2007af} gauge groups. In our theory, it couples to two right handed up-type quarks of different generations
\begin{equation}\label{eq:LSDiquark1}
     \mathcal{L}_{uu} = y^{ij}_{_{uu}}\epsilon^{c_{1}c_{2}c_{3}} \ \overline{u}^{c}_{ic_{1}}P_{R}u_{jc_{3}}\chi_{uu c_{2}} + \text{h.c}.
\end{equation}
In the equation above $i$ and $j$ label quark flavours, $i \neq j$  due to fermionic statistics, and the $c$ indices label QCD color.\\ 

The second particle we add is a charge 1/3 leptoquark, $\chi_{_{LQ}}$, also common in many GUT theories --- particularly the ones based on $SU(5)$ \cite{Buchmuller:1986iq} or $ SO(10)$ \cite{Senjanovic:1982ex}. It couples either a left-handed quark doublet to a left-handed lepton doublet or a right-handed up type quark to a right-handed charged lepton,
\begin{equation}\label{eq:LLeptoquark1}
\begin{split}
\mathcal{L}_{LQ} =& y^{ij*}_{LQ}\, \epsilon^{\alpha\beta}\, \overline{Q}^{c}_{\alpha ic_{1}} \chi^{\dagger c_{1}}_{LQ} L_{\beta j} + y^{ij*}_{ul}\, \overline{u}^{c}_{ic_{1}} P_{R}\, l_{j}\, \chi^{\dagger c_{1}}_{LQ} + \text{h.c.} \\
= &  \overline{u}^{c}_{ic_{1}} y^{ij*}_{LQ}P_{L}\, l_{j} - \overline{d}^{c}_{ic_{1}} \left(V^{T}y^{*}_{LQ}U\right)^{ij} P_{L}\, \nu_{j}  \chi^{\dagger c_{1}}_{LQ} + \, \overline{u}^{c}_{ic_{1}} y^{ij*}_{ul} P_{R}\, l_{j}\, \chi^{\dagger c_{1}}_{LQ} + \text{h.c.},
\end{split}
\end{equation}
where $Q,L$ are the SM left-handed $SU(2)_L$ quark and lepton doublets respectively, $l$ is a right-handed charged lepton, and $\alpha,\beta$ are the weak isospin indices. The second line of Equation~\eqref{eq:LLeptoquark1} is written in the mass basis following the convention adopted in Section~\ref{sec:ModelIndpConstraints}. 
The leptoquark couplings play an important role in determining which BNV effective operators in \cref{eq:BNVOpt} are generated since the two Yukawa matrices generate them separately: $y_{LQ}$ alone produces $\mathcal{O}^{\text{L}}_{\text{BNV}}$ while $y_{ul}$ alone produces $\mathcal{O}^{\text{R}}_{\text{BNV}}$. In a theory with both of them non-zero, both are generated simultaneously. Such  leptoquarks are, however, severely constrained by baryon number conserving low energy probes that we will survey next. Hence, we will restrict ourselves to the ``maximally chiral'' version of $\chi_{_{LQ}}$\footnote{For notational simplicity, we use $\chi_{_{LQ}}$ to denote both types of leptoquarks.} for the remainder of this work, the one that generates either $\mathcal{O}^{\text{L}}_{\text{BNV}}$ or $\mathcal{O}^{\text{R}}_{\text{BNV}}$, but not both. \\

Finally, we add a Zee-Babu \cite{Zee:1980ai,Babu:1988ki} singly charged dilepton singlet, $\chi_{_{LL}}$, coupling two left handed leptons of different generations,
\begin{equation}\label{eq:LSDilepton1}
\begin{split}
    \mathcal{L}_{_{LL}} &= y^{ij}_{_{LL}} \ \epsilon^{\alpha\beta} \overline{L}^{c}_{\alpha i}\chi_{_{LL}}L_{\beta j}  + \  \text{h.c}.\\
    &=  \overline{\nu}^{c}_{i}\left(U^{T}y_{_{LL}}\right)^{ij} P_{L}l_{j}\chi_{_{LL}} - \ \overline{l}^{c}_{i}\left(y_{_{LL}}U\right)^{ij}P_{L}\nu_{j}\chi_{_{LL}} + \  \text{h.c.},
\end{split}
\end{equation}
where, as with our presentation of the leptoquark couplings, we write the second line in the mass basis. \\

Since these Yukawa interactions represent a \emph{hard} breaking of $B$ and $L$, we must assign quantum numbers to the new scalars such that both global symmetries are conserved. These are listed in Table~\ref{tab:ParticleContent13}. Therefore, these interactions \emph{cannot} induce BNV even at loop level since they are all $B$ and $L$ conserving and nothing about these interactions are different from those of the three scalars individually studied in multiple other works \cite{Pascual-Dias:2020hxo,Giudice:2011ak,Dorsner:2016wpm,Crivellin:2021ejk,Zee:1980ai,Babu:1988ki}. \\

To achieve BNV, we introduce the \emph{only} baryon and lepton number violating interaction of this model, one that occurs in the scalar sector:
\begin{equation}\label{eq:BViolatingInteraction}
\mathcal{L}_{\text{BNV}} =  \Lambda_{\text{BNV}} \ \chi^{\dagger}_{_{LQ}}\chi_{_{uu}}\chi_{_{LL}} + \ \text{h.c}.  
\end{equation}
This trilinear interaction breaks $B$ and $L$ softly but conserves $B - L/3$, as can be checked using the quantum numbers listed in Table~\ref{tab:ParticleContent13}.  \\

The other interactions in the scalar sector can be written compactly as:
\begin{equation}\label{eq:BViolatingInteraction}
    \mathcal{L}_{S} =  \frac{\mu^{2}_{j}}{2}\chi^{\dagger}_{j}\chi_{j} + \lambda_{jj} \ (\chi^{\dagger}_{j}\chi_{j})^{2} + \lambda_{jk} \ \chi^{\dagger}_{j}\chi_{j}\chi^{\dagger}_{k}\chi_{k}+ \lambda_{jH} \ \chi^{\dagger}_{j}\chi_{j}H^{\dagger}H,
\end{equation}    
where $j=LQ,LL,uu$ and, without loss of generality, we assume that $\mu^{2}_{j}$ > 0. This imposes $\lambda_{jj} > 0$ and $\lambda_{jk} > 0$ since the new scalars all carry non-zero electromagnetic charge and therefore cannot acquire vacuum expectation values. Mass positivity imposes a bound on their mixing with the Higgs boson 
\begin{equation}
  \lambda_{jH} > - \frac{\mu^{2}_{j}}{v^{2}}, 
\end{equation}
where $v = 246$ GeV is the Higgs VEV. \\
 
Finally, we close this presentation of our model by matching its particular couplings to the ones from the low energy effective operators discussed previously. Denoting the dimensionless part of the Wilson coefficient for $\mathcal{O}^{\text{L}}_{\text{BNV}}$ and $\mathcal{O}^{\text{R}}_{\text{proton}}$ as $c^{\text{L} \ ijklmn}_{9}$ and $c^{\text{R} \ ijklmn}_{9}$ respectively, 
\begin{equation}
\begin{gathered}
    M^{5}_{\text{BNV}} = \frac{\left(m_{\chi_{LQ}}\, m_{\chi_{uu}}\, m_{\chi_{LL}}\right)^{2}}{\Lambda_{\text{BNV}}} , \\[1ex]
    c^{\text{L}\,ijklmn}_{9} = y^{ij}_{LQ}\, y^{kl}_{uu}\, y^{mn}_{LL} , \qquad
    c^{\text{R}\,ijklmn}_{9} = y^{ij}_{ul}\, y^{kl}_{uu}\, y^{mn}_{LL} .
\end{gathered}
\end{equation}
Here, we mention that $M_{\text{BNV}}$ cannot be made arbitrarily small by increasing $\Lambda_{\text{BNV}}$ since imposing perturbative unitarity -- without which all our computations in this work lose their validity -- places important theoretical constraints on the latter, as discussed in Appendix~\ref{sec:PerturbativeUnitarity}. \\

\subsection{Existing Baryon Number Conserving Constraints}
\label{sec:BNC_constraints}
Let us now discuss the relevant BNC and LNC constraints on the new scalar fields.
A novel feature is that there are some processes for which the leptoquark and the dilepton singlet contribute together at the amplitude level. For example, measurements of the anomalous magnetic moment of leptons are sensitive to contributions from both of them. Leptoquarks with both ``$LQ$-type'' and ``$ul$-type'' couplings are in fact strongly constrained by these measurements \cite{Bigaran:2020jil}, motivating us to consider the maximally chiral nature of the Yukawas. In this limit, anomalous magnetic moments of leptons are suppressed by $m^{2}_{l}/(m_{\chis{j}}m_{\chis{k}})$, with $j,k=LL,LQ$ and, for $\mathcal{O}$(TeV) masses, the corresponding constraints are easily satisfied. \\

Another strong probe comes from the contributions of the leptoquark and the dilepton singlet to the decay of a muon into an electron and a photon, $\mu \rightarrow e\gamma$. This process has a predicted SM rate that renders it unobservable for all practical purposes and can serve as a sensitive background-free probe of new physics. Expressions for the $\mu\rightarrow e\gamma$ decay rate from theories with only one leptoquark or only one dilepton singlet have been computed in \cite{Bigaran:2020jil,Dorsner:2016wpm,McLaughlin:1999rr}, but the novelty in our case is that they must be added together at the amplitude level. Making the simplifying assumption that $m_{\rm SM}/m_{\chi} \ll$ 1, where $m_{\rm SM}$ is the mass of \emph{any} SM fermion and $m_{\chi}$ is the mass of the dilepton singlet or leptoquark, the branching ratio for this decay mode can be written as
\begin{equation}\label{eq:mutoegamma}
\begin{split}
    \text{BR}(\mu \rightarrow e\gamma) = \frac{12\pi\alpha_{\text{EM}}}{G^{2}_{F}\, m^{4}_{\chi_{LQ}}}
    \biggl\vert\, & N^{2}_{c} \left(\frac{Q_{LQ}}{4}-\frac{1}{6}\right) \sum^{3}_{j=1} y^{\mu j*}_{X}\, y^{ej}_{X} \\
    &+ \frac{m^{2}_{\chi_{LQ}}}{m^{2}_{\chi_{LL}}} \left(\frac{Q_{LL}}{4}-\frac{1}{6}\right) y^{e\tau*}_{LL}\, y^{\mu\tau}_{LL} \,\biggr\vert^{2} ,
\end{split}
\end{equation}
where $Q_{LQ} = -1/3$ and $Q_{LL} = +1$ are the electromagnetic charges of the leptoquark and the dilepton respectively, $N_{c} = 3$ is the colour factor, and $X = LQ$ or $X = ul$ depending on the chirality of the leptoquark coupling.\footnote{Here, we remind the reader that we do not consider leptoquarks with simultaneous $LQ$-type and $ul$-type couplings.} From this expression, it can be checked that if the Yukawa couplings of the leptoquark and/or the dilepton to muons and electrons are set to unity, the combined limit of 
\begin{equation*}
   \text{BR}({\mu\rightarrow e\gamma}) < 3\times10^{-13} 
\end{equation*}
reported by the MEG \cite{MEG:2016leq} and MEG II \cite{MEGII:2023ltw} experiments implies that these two scalars must be heavier than $\mathcal{O}$(10) TeV. 
Furthermore, the lowest order diagram for the $\mu\rightarrow e\gamma$ process via the dilepton singlet must involve a $\tau$ neutrino that couples to both $\mu$ and $e$. Accordingly, we fix the $y^{e\tau}_{_{LL}}$ coupling to values consistent with MEG limits for an $\mathcal{O}$(TeV) mass dilepton singlet. 
For the leptoquark, we assume that they couple diagonally across the lepton and quark families, except for the top quark for which we allow, in addition to its coupling to the $\tau$ lepton, an $\mathcal{O}$(1) coupling to the muons. This choice of parameters prohibits lepton flavour violation (LFV) at tree level and satisfies the corresponding MEG limits.\footnote{We note the possibility -- which we do not pursue in this work -- that fine tuning between the sum of leptoquark Yukawas and that of the dilepton singlet to the $\tau$ generation in the limit of nearly degenerate scalar masses can lead to cancellations that provide a further avenue to avoid the stringent constraints from observed limits on the rate for the $\mu \rightarrow e \gamma$ process.} This is also a common choice in the LHC searches for leptoquark \cite{ATLAS:2021oiz,CMS:2022nty,ATLAS:2020xov}.\\ 

Next, we discuss the remaining constraints on the leptoquark, $\chis{LQ}$.
The tree-level interactions between the leptoquark with ``$LQ$-type'' couplings to SM fermions produce dimension six low energy effective operators such as 
\begin{equation*}
    \mathcal{L}_{\text{eff}} = \frac{c_{ijkm}}{m^{2}_{\chi_{_{LQ}}}}\left(\bar{u}^{i}_{L}\gamma^{\mu}d^{j}_{L}\right)\left(\bar{l}^{k}_{L}\gamma_{\mu} \nu^{m}_{L}\right),
\end{equation*}
that contribute to leptonic and semi-leptonic meson decays (only through the $y_{_{LQ}}$ couplings) as well as $\tau$ lepton decays (through both $y_{_{LQ}}$ and $y_{ul}$). In the expression above, $i,j$ and $k,m$ label quark and lepton flavours respectively, and $c_{ijkm}$ is the Wilson coefficient that results from integrating out the leptoquark. An up-to-date review of these various limits can be found in \cite{ParticleDataGroupLeptoquarks:2024cfk}, which we summarize in Table~\ref{tab:leptoquarklimits}.

\begin{table}[htbp]
    \centering
    \small
    \renewcommand{\arraystretch}{1.2}
    \begin{tabular}{@{}lc@{}}
        \toprule
        \textbf{Yukawa coupling} & \textbf{Constraint} \\
        \midrule
        $y^{ue}_{X}$ &
        $y^{ue}_{X} < 0.3 m_{\chi_{LQ}}/\mathrm{TeV}$ \\

        $y^{c\mu}_{X}$ &
        $y^{c\mu}_{X} < 0.8 m_{\chi_{LQ}}/\mathrm{TeV}$ \\
        $y^{t\mu}_{X}$ &
        $y^{t\mu}_{X} = 1 \ \text{for } m_{\chi_{LQ}} > 1.2 \ \text{TeV}$ \\
        $y^{t\tau}_{X}$ &
        $y^{t\tau}_{X} = 1 \ \text{for } m_{\chi_{LQ}} > 1.2 \ \text{TeV}$ \\
        \bottomrule
    \end{tabular}
    \caption{Summary of existing constraints on leptoquark Yukawa couplings, reproduced from Ref.~\cite{ParticleDataGroupLeptoquarks:2024cfk}. Here, $X=LQ$ or $X = ul$ denotes purely left-handed or right-handed couplings, respectively. Scenarios in which both types of couplings are present simultaneously are substantially more constrained by lepton anomalous magnetic moment searches and are therefore not considered. The limits on the couplings of the top quark to muons and $\tau$ leptons are taken from ATLAS and CMS searches for leptoquarks decaying primarily to second or third generation SM fermions \cite{ATLAS:2021oiz,CMS:2022nty,ATLAS:2020xov}. 
    }
    \label{tab:leptoquarklimits}
\end{table}

For the remainder of this work, we focus on those with preferential couplings to the third generation quarks and/or leptons, particularly the top quark. In contrast to the leptoquark couplings to the lighter quarks, those to the top quark can only be efficiently probed at the LHC. Such leptoquarks can be produced in pairs via gluon fusion, a channel that is independent of any Yukawa couplings (the production part) and has been searched for by both the ATLAS and CMS experiments \cite{ATLAS:2021oiz,CMS:2022nty}. In particular, ATLAS has set lower limits of 1.48, 1.47, and 1.43~TeV on the masses of scalar leptoquarks decaying exclusively into a top quark and an electron, a muon, or a tau lepton, respectively~\cite{ATLAS:2020xov,ATLAS:2021oiz} with unit Yukawa couplings. \\

Then, we have the dilepton singlet, $\chis{LL}$. 
This scalar field \emph{necessarily} violates lepton flavor at tree level, making it particularly sensitive to experiments searching for lepton flavor violation or testing lepton flavour universality, particularly in $\tau$ decays. 
Using measurements of leptonic and hadronic universality in meson decays and $e/\mu$ universality in $\tau$ decays reported in \cite{Pich:2013lsa}, the authors of \cite{Herrero-Garcia:2014hfa} obtained limits on $y^{e\mu}_{LL}$ and $y^{\mu\tau}_{LL}$, respectively.
We list these limits and the limits from $\mu\to e\gamma$ in Table~\ref{tab:dileptonsingletlimits}. \\

\begin{table}[htbp]
    \centering
    \small
    \renewcommand{\arraystretch}{1.2}
    \begin{tabular}{@{}lc@{}}
        \toprule
        \textbf{Yukawa coupling} & \textbf{Constraint} \\
        \midrule
        $|y^{e\mu}_{LL}|$
        & $< 0.08 \ m_{\chi_{LL}}/ \ \mathrm{TeV}$ \\

        $|y^{\mu\tau}_{LL}|$
        & $\sqrt{\left||y^{\mu\tau}_{LL}|^{2}
        - |y^{e\mu}_{LL}|^{2}\right|}
        < 0.2 \ m_{\chi_{LL}}/ \ \mathrm{TeV}$ \\

        $|y^{e\tau}_{LL}|$
        & $< 0.001$ \quad (see $\mu\rightarrow e\gamma$) \\
        \bottomrule
    \end{tabular}
    \caption{
        Summary of existing constraints on the dilepton singlet
        from Ref.~\cite{Herrero-Garcia:2014hfa}.
        See the text for further details.
    }
    \label{tab:dileptonsingletlimits}
\end{table}

Finally, we discuss the constraints on diquark, $\chis{uu}$. There are 3 Yukawa couplings, $y_{uu}^{uc}$, $y^{ut}_{uu}$ and $y^{ct}_{uu}$. The latter two are constrained by $\Delta F = 2$ flavour violation probes, specifically $D_{0} \rightarrow \overline{D}_{0}$ mixing. As explained in \cite{Giudice:2011ak}, such processes occur at the one-loop level and, due to the chirality and anti-symmetric nature of the couplings, necessarily involve a box diagram of two diquarks and two top quarks. The authors of \cite{Giudice:2011ak,Pascual-Dias:2020hxo} find particularly tight constraints on the up-top and charm-top Yukawas. These bounds do not affect the main production channel for diquarks at the LHC, the $s$-channel process $u \ c \rightarrow \chis{uu}^\dagger$, but favor dijet final states instead of top and jet ones. The former has been searched for by both the ATLAS and CMS experiments~\cite{ATLAS-CONF-2019-007,CMS:2019gwf}, providing strong constraints on $\chi_{_{uu}}$ that are relevant to our model.
These constraints, however, would be weakened by the extra BNV decay mode $\chis{uu}^\dagger\to\chis{LL}^\dagger \chis{LQ}$ if it is kinematically allowed.
We will apply the dijet constraints to our studies of collider phenomenology, to be discussed in the next section. For diquark masses above 3 TeV, Yukawa couplings $\vert y^{uc}_{uu}\vert $ = 0.25 are allowed by existing dijet searches, and we adopt this benchmark for the remainder of this work. The constraints on diquark Yukawa couplings are summarized in Table~\ref{tab:diquarklimits}.
\begin{table}[htbp]
    \centering
    \small
    \renewcommand{\arraystretch}{1.2}
    \begin{tabular}{@{}lc@{}}
        \toprule
        \textbf{Yukawa coupling} & \textbf{Constraint} \\
        \midrule
        $|y^{u c}_{uu}|$
        &  $|y^{u c}_{uu}|$ < 0.25  \\

          $\vert y^{ut}_{uu}y^{ct}_{uu}\vert\times \frac{m_{\chi_{_{uu}}}}{\text{TeV}}$ 
        & $< 1.5\times10^{-2}$ \\
        \bottomrule
    \end{tabular}
    \caption{
        Summary of existing constraints on the diquark
        from Ref.~\cite{Giudice:2011ak,Pascual-Dias:2020hxo} and LHC dijet searches \cite{ATLAS-CONF-2019-007,CMS:2019gwf}.
        See the text for further details.
    }
    \label{tab:diquarklimits}
\end{table}
\subsection{Implications of BNC constraints for proton decay.}\label{sec:BNC_constraints}
\begin{figure}
 \raisebox{2.5mm}{\input{tikz/proton-decay-q}}~~
 \input{tikz/proton-decay-l}
\caption{Feynman diagrams for loop level proton decay, where $j,k,l,m,n$ are flavour indices. A chirality flip along the $u^k$ propagator is necessary to ensure the correct chirality on both vertices, which is induced by a mass insertion denoted by the $\times$. When the two antineutrinos are of the same flavour, $l = m$, each of these diagrams has a companion with $k_{1} \leftrightarrow k_{2}$.
In the left panel, $\chis{LQ}^\dagger$ couples one of the outgoing antineutrinos with an intermediate $d$ quark, which can only be mediated when the leptoquark has ``$LQ$-type'' couplings.
In the right panel, $\chis{LQ}^\dagger$ couples one of the $u$ quarks in the proton with an intermediate antilepton, which can be mediated by either ``$LQ$-type'' or ``$ul$-type'' couplings. For simplicity, we do not consider the case when both types of couplings are present in this work.  A chirality flip by mass insertion is necessary on the intermediate $\bar{\ell}^j$ propagator for the $ul$-type coupling, as denoted by the red cross on the diagram.
For the $LQ$-type coupling, because of the $SU(2)$ index contraction of the leptoquark couplings, these two diagrams carry a relative minus sign between them and interfere destructively.}
\label{fig:ProtonDecayUVComplete}
\end{figure}
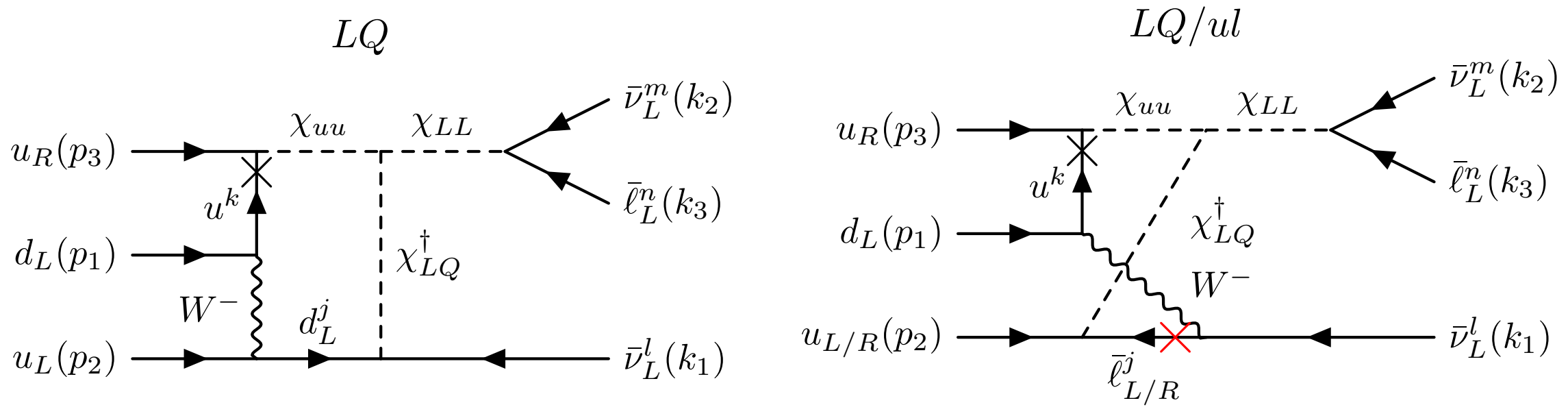

\begin{figure}[htbp]
    \centering
    \includegraphics[width=0.99\textwidth]
        {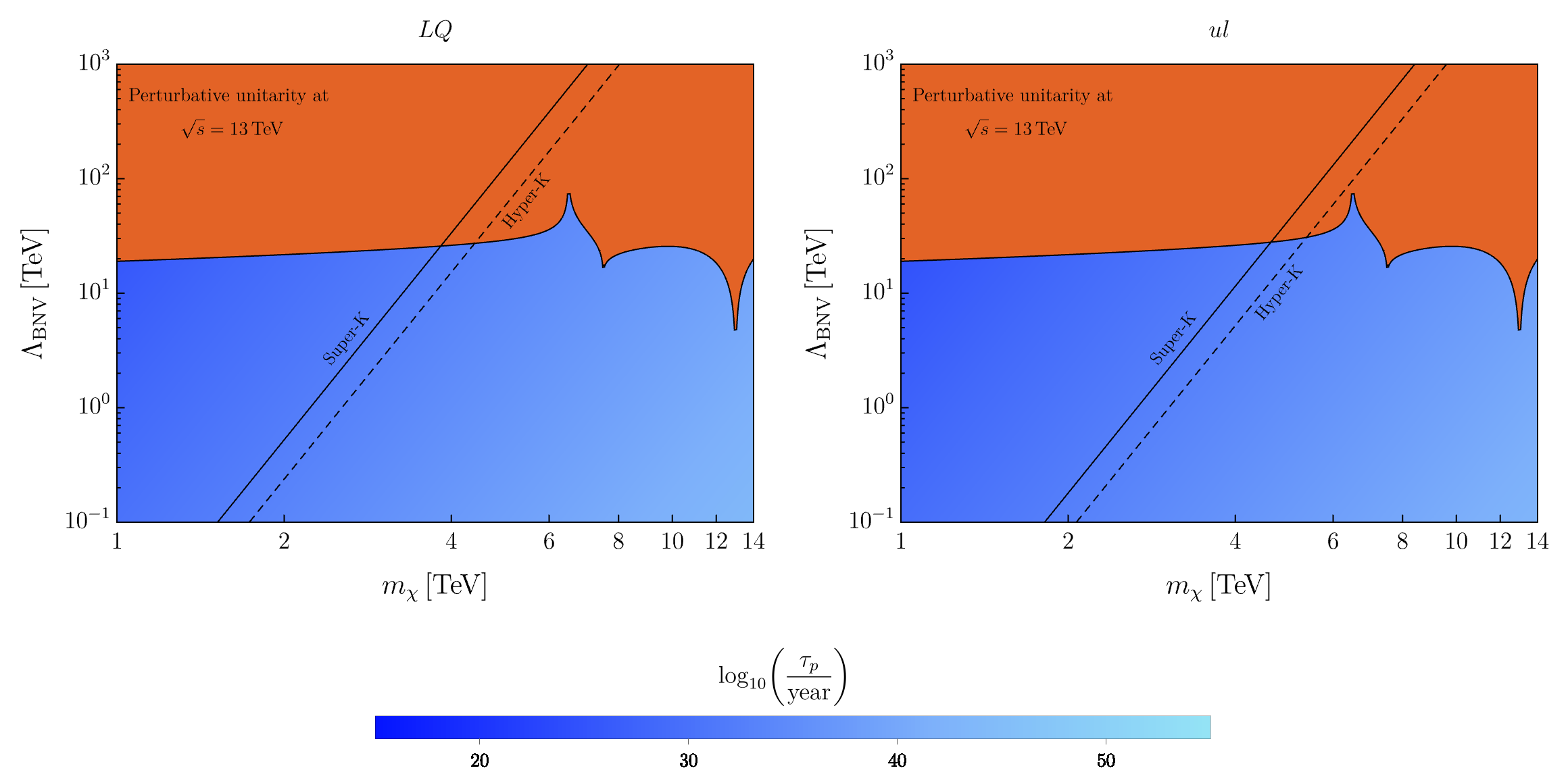}
    \caption{
        Proton decay constraints considering $LQ$-type (left) and
        $ul$-type (right)
        couplings for the leptoquark in our representative model.
        The color maps show the predicted proton lifetime
        for $p\rightarrow\mu^{+}\bar{\nu}\bar{\nu}$
        as a function of the degenerate scalar mass $m_\chi$
        and $\Lambda_{\mathrm{BNV}}$.
        Solid black contours correspond to the existing
        Super-Kamiokande limits, while dashed black contours
        indicate the projected Hyper-Kamiokande sensitivity
        of 10.7$\times 10^{32}$ years, taken from Table VII of \cite{Dev:2022jbf}.
        The corresponding electron-channel contours are
        omitted for clarity and 
        the orange regions indicate values of $\Lambda_{\text{BNV}}$ that violate perturbative unitarity at
        $\sqrt{s}=13~\mathrm{TeV}$ for a given $m_{\chi}$, as discussed in Appendix~\ref{sec:PerturbativeUnitarity}.}
    \label{fig:PDLimits}
\end{figure}
Taking the BNC constraints into account, we re-evaluate the constraints on $M_{\rm BNV}$ from proton decay experiments for the representative model, induced by the diagrams shown in Figure \ref{fig:ProtonDecayUVComplete}.
To illustrate the implications of the survey of phenomenological constraints for proton decay, in Fig.~\ref{fig:PDLimits} we show the predicted proton lifetime as a function of the scalar masses\footnote{For simplicity, we assume that all three new scalars are degenerate in mass in this section} and the BNV coupling with the Yukawa couplings fixed at saturation values fixed by the BNC constraints. Since these constraints suppress the relevant combinations of Wilson coefficients they increase the proton lifetime and weaken the lower limits on $M_{\mathrm{BNV}}$ to energies easily accessible at the LHC, as can be seen by comparing the limits on  $M_{\mathrm{BNV}}$ in Tables~\ref{tab:BNV_bounds} and \ref{tab:BNV_model_dependent_bounds}.

These results highlight the potential for proton decay experiments and collider searches to serve as complementary probes of $B - L/3$. Trilepton nucleon-decay searches at Super-Kamiokande and Hyper-Kamiokande provide exceptional sensitivity to interactions involving light quarks, while remaining sensitive to heavier-generation couplings through flavor mixing by the CKM and PMNS matrices. The LHC can directly probe the production and decay of new particles coupled preferentially to heavier quark and lepton generations, thereby providing access to flavor structures that are difficult to disentangle through nucleon decay alone. Crucially, in UV-complete realizations, the production of BNV mediators and their subsequent decays do not necessarily depend on the same combinations of Yukawa couplings that govern proton decay. 

\begin{table}[htbp]
    \centering
    \small
    \renewcommand{\arraystretch}{1.2}

    \begin{tabular}{l c c c}
        \toprule
        Proton decay mode &
        Experimental bound &
        $\chis{LQ}$ coupling &
        Lower limit on $M_{\mathrm{BNV}}$ [GeV] \\
        \midrule

        $p \rightarrow e^{+}\bar{\nu}\bar{\nu}$
        & $\tau \geq 1.7 \times 10^{32}\,\mathrm{yr}$
        & %
        $LQ$
        & $2.1 \times 10^{3}$ \\

        $p \rightarrow e^{+}\bar{\nu}\bar{\nu}$
        & $\tau \geq 1.7 \times 10^{32}\,\mathrm{yr}$
        & %
        $ul$
        & $2.6 \times 10^{3}$ \\

        $p \rightarrow \mu^{+}\bar{\nu}\bar{\nu}$
        & $\tau \geq 2.2 \times 10^{32}\,\mathrm{yr}$
        & %
        $LQ$
        & $2.7 \times 10^{3}$ \\

        $p \rightarrow \mu^{+}\bar{\nu}\bar{\nu}$
        & $\tau \geq 2.2 \times 10^{32}\,\mathrm{yr}$
        & %
        $ul$
        & $3.3 \times 10^{3}$ \\

        \bottomrule
    \end{tabular}

    \caption{
        Model-dependent lower limits on the BNV scale
        $M_{\mathrm{BNV}}$ obtained from the experimental
        proton-decay constraints. The Yukawa couplings are
        fixed to their saturation values, as determined by
        the phenomenological constraints discussed in the previous section.
        These limits illustrate the proton-decay sensitivity
        within the specified UV completion, in contrast to the
        operator-level bounds presented in
        Table~\ref{tab:BNV_bounds}.
    }
    \label{tab:BNV_model_dependent_bounds}
\end{table}

\section{Optimal process selection for probing $\mathbf{B-L/3}$ at the LHC}\label{sec:OptimalProcessSelection}

While collider signatures of BNV have been considered previously \cite{Dong:2011rh,CMS:2024dzv}, dedicated experimental searches remain limited, in part due to the very high lower limits implied by proton decay experiments. However, the results of the previous section have shown that the latter does not always preclude testing BNV at colliders. Using the benchmark presented at the end of Section~\ref{sec:Model}, we now examine the various processes that can be used at the LHC to probe the BNV interactions of $B - L/3$ theories and choose the optimal one.  \\
 
Compared to proton decay experiments, colliders offer the important advantage of directly probing BNV interactions involving heavier fermion generations or, for coloured new particles such as the diquark and leptoquark in our representative model, Yukawa independent production channels. However, establishing BNV at a hadron collider presents distinct experimental challenges. In particular, identifying whether a reconstructed jet originated from a quark or anti-quark is challenging, complicating the determination of the baryon number carried by the final-state particles. This difficulty is compounded by additional hadronic activity from initial and final state radiation (ISR and FSR respectively), which can obscure the identification of particles originating from the BNV process. In this section, we establish three guiding principles for selecting final states that minimize these challenges and facilitate the experimental identification of BNV. We then systematically compare the possible mass hierarchies of the new particles in our representative model to identify the most promising one for the LHC. This motivates the dedicated search presented in Section~\ref{sec:collidermain}, whose applicability to other ultraviolet completions is briefly discussed in Section~\ref{sec:Generalizations}. \\

Let us now turn to the three guiding principles. We will focus on our representative model for concreteness but the latter should also be adaptable to other models conserving $B - L/3$, or more general $B + aL$  symmetries compatible with TeV-scale BNV. Firstly, we favor processes with large production cross sections and minimal phase-space suppression. In particular, a mass hierarchy allowing one of the new scalars to decay into the two lighter scalars via the BNV trilinear vertex is preferable. The resulting LHC relevant BNV process then proceeds through the production of a heavy resonance from proton-proton collisions and its decay into daughter particles who, in turn, decay into SM fermions whose counting indicate $B$ and $L$ violation. \\

Secondly, to establish BNV, we favor final states for which the baryon number of the outgoing particles can be reliably identified. This is challenging for light-quark jets, since determining whether a jet originates from a quark or anti-quark is experimentally difficult. We therefore favor (anti) top quarks, whose charges can be identified through their leptonic decays. Here, we must carefully take into account the impact of ISR and FSR since they introduce baryon number into the final states unrelated to the primary BNV decay chain. However, the large momenta and isolation of particles produced through heavy-resonance decays can help distinguish them from additional radiation. In Section~\ref{sec:collidermain}, we demonstrate that the $b$-jet originating from the boosted anti-top quark is typically much more energetic compared to ISR and FSR jets and that the latter are well isolated from the former, $\Delta R \gtrsim 0.8$. \\

Thirdly, to determine the amount of lepton-number violation, we favor final states containing as many charged leptons and as few neutrinos as possible. It is important here to mention that the latter are sometimes unavoidable, as is the case for $B - L/3$. This follows from $SU(2)$ invariance, as can be seen by inspecting $\mathcal{O}^{\text{L}}_{\text{BNV}}$ and $\mathcal{O}^{\text{R}}_{\text{BNV}}$. Since neutrinos and anti-neutrinos cannot be distinguished using missing-energy measurements, their presence introduces an unavoidable ambiguity in determining $\Delta L$ experimentally, although compelling theory arguments can be made in favour of the true $\Delta L$ value in some cases. \\

\begin{figure}[h!]
    \centering
    \includegraphics[width=0.9\textwidth]{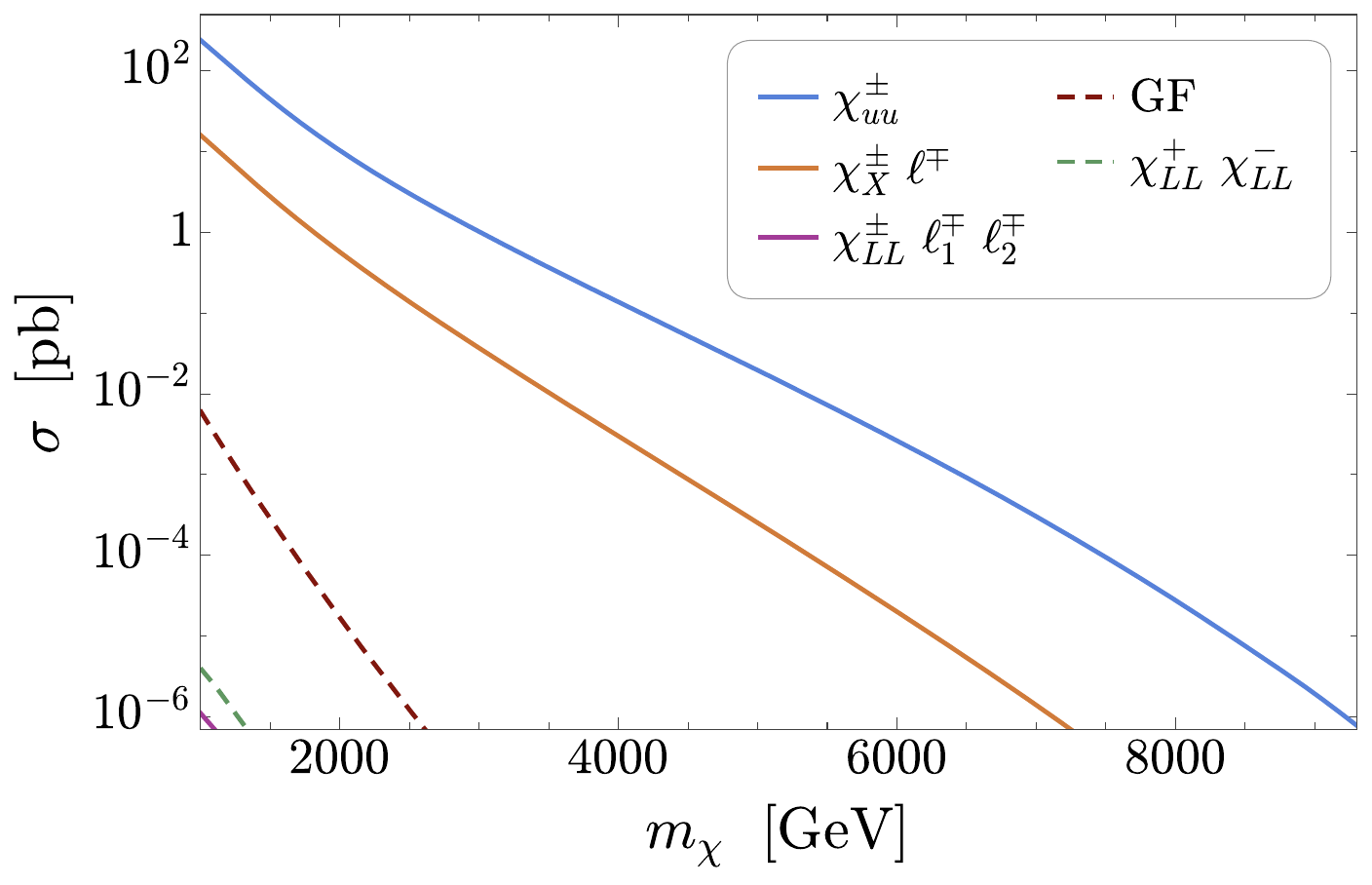}
    \caption{Production cross sections at the LHC with unit couplings and $\Lambda_{\rm BNV}$ fixed to 1 TeV. Pair production to diquarks and leptoquarks is identical since both are produced via gluon fusion (GF). BNV is most easily observed in pair production if one scalar decays through the trilinear coupling and the other decays directly to SM since otherwise if both decay through the BNV mode the net baryon and lepton number of all final states is zero. Single production of leptoquarks and dileptons requires additional SM particles in the final state, which complicates baryon and lepton number counting.}
    \label{fig:prodCS} 
\end{figure}

Using these guiding principles, we now discuss the three possible mass hierarchies between the different scalars and examine their different signals at the LHC, evaluating their suitability for establishing $B$ and $L$ violations by taking into account both experimental limitations and fundamental theoretical arguments. The relevant production cross-sections at 13 TeV, assuming unit Yukakwa couplings where appropriate, are summarized in Figure~\ref{fig:prodCS}. 

\subsection*{\textbf{I:} $\mathbf{m_{\chi_{_{LL}}} > m_{\chi_{_{LQ}}} + m_{\chi_{_{uu}}}}$.} 
    
When the dilepton singlet is the heaviest of the three scalars and is kinematically allowed to decay to the leptoquark and the diquark, the relevant production processes at the LHC to probe its BNV decay mode are pair and single production,
\begin{equation}
\begin{split}
    p \ p  & \rightarrow \gamma^{*}/Z^{*} \rightarrow \chi^{\dagger}_{_{LL}}\chi_{_{LL}} \\
    p \ p  & \rightarrow \chi^{\dagger}_{_{LL}}  \ l^{+} \ \bar{\nu}
\end{split}
\end{equation}
where, $l^{-}$ is any lepton and, for single production, without loss of generality, we only consider the dilepton singlet and not its conjugate. For pair production we require that only one dilepton singlet, either itself or its conjugate, decay via its BNV mode,
\begin{align}
    \chi^{\dagger}_{_{LL}} & \rightarrow \chi_{_{uu}}\chi^{\dagger}_{_{LQ}}, \nonumber\\ 
        \chi_{_{LL}} & \rightarrow \bar{l}\bar{\nu}.
\end{align}
This is because if they both decay via their BNV mode, the net baryon and lepton number of all final states will be zero, adding an additional complication. 
Since the existing constraints from gluon fusion require both leptoquark and diquark to be heavier than $1$ TeV, the dilepton singlet has to be heavier than $2$ TeV for this mass hierarchy to work.
However, as shown in Figure~\ref{fig:prodCS}, the cross sections for both pair and single production of dilepton are very suppressed above this mass scale, ruling out the LHC as an optimal probe of this particular mass hierarchy.
We therefore do not consider it further here but offer some remarks on the prospects for probing it at future lepton colliders in our conclusions. \\
    
    \subsection*{II: $\mathbf{m_{\chi_{_{LQ}}} > m_{\chi_{_{LL}}} + m_{\chi_{_{uu}}}}$.} 
    
    When the leptoquark is the heaviest of the three new scalars and is kinematically allowed to decay to the two other there are two relevant processes for the LHC:
    \begin{equation}
    \begin{split}
        g \ g & \rightarrow \chi^{\dagger}_{_{LQ}}\chi_{_{LQ}} \\
        p \ p & \rightarrow \chi^{\dagger}_{_{LQ}} \ l^{+}, \\
    \end{split}
    \end{equation}
    Since pair production for leptoquarks occurs via gluon fusion, it can be larger than one might naively anticipate since it depends only on the strong coupling and not on the Yukawa ones. However, as we can see from Figure~\ref{fig:prodCS}, at the masses of interest pair production imposes a heavy kinematic suppression compared to single production. Focusing on the latter, the optimal visible final states are then a charged anti-lepton, an anti-top quark, a jet (unavoidable due to the leptoquark decaying into a diquark), missing energy due to the neutrino produced from the decay of the dilepton singlet, and an anti-lepton from the single conjugate leptoquark production process (see \cite{ATLAS:2025upm} for more details). \\
    
To determine $\Delta B$ and $\Delta L$, we offer some idealized assumptions to show that given enough statistics, it should be possible to determine both. Whether or not the level of required statistics is achieveable at future collider experiments is, of course, an entirely different question. Using our knowledge of the single production process\footnote{Using observations of its BNC decay mode from single production should be particularly helpful here, particularly since the decay products of the latter can be fully visible.}\label{fn:LQBNC}, with these final states we would naively conclude that either $\Delta B = -1$, assuming the jet from the diquark is from an anti-quark, or $\Delta B = -1/3$, assuming the jet from the diquark is from an quark. However, the latter is forbidden by $SU(3)$ invariance in the absence of exotic coloured fermions that are charged under a different global symmetry (neither $B$ nor $L$). This is a fundamental theory requirement, which enables us to conclude that since these new coloured fermions have not been observed, $\Delta B = -1$. \\
    
For $\Delta L$, since we know one of the anti-leptons comes from the production of the leptoquark and is not related to its BNV interactions, we can discount it from our final state count and determine, naively, that either $\Delta L = 0$ (outgoing neutrino) or $\Delta L = -2$ (outgoing anti-neutrino). However, since this vertex has $\Delta B = -1$ both of these possibilities are forbidden by Lorentz invariance in the absence of exotic fermions -- not necessarily coloured -- charged under a global symmetry other than $L$. Non observation of the latter leads us to conclude that the intermediate heavy resonances must carry the correct baryon and lepton numbers to make this set of final states consistent with $SU(3)$ and Lorentz invariance, which leads us to conclude that $\Delta L$ must be odd. The simplest possibility,  $\Delta L = -1$, which implies $\Delta B = \Delta L = -1$, is constrained by proton decay experiments to the GUT scale, and should therefore be rejected in favour of the latter. The next possibility, $\Delta L = -3$, since $SU(2)$ invariance imposes that larger odd numbered violations of $\Delta L$ must contain one or more charged leptons or occur association with Higgs (either the doublet or triplet) or electroweak bosons. The situation can be different if right handed neutrinos are added to the theory, which we do not consider this possibility here. 
    
\subsection*{III: $\mathbf{m_{\chi_{_{uu}}} > m_{\chi_{_{LQ}}} + m_{\chi_{_{LL}}}}$.} Finally, we examine the scenario when the diquark is the heaviest of the three scalars and can decay into the other two. Since it is coloured, like for the previous benchmark the two LHC relevant processes are gluon fusion pair production and resonant single production, 
    \begin{equation}
    \begin{split}
    g \ g & \rightarrow \chi_{uu} \ \chi^{\dagger}_{uu}, \\ 
    p \ p & \rightarrow \chis{uu}^\dagger,
    \end{split}
    \label{eq:LHCsignal}
\end{equation} 
while the BNV decay mode is
\begin{equation}
   \chi_{uu}^\dagger \rightarrow \chis{LQ}^{\dagger} \chis{LL}.
\end{equation}
The single production process here involves the fusion of an up and a charm parton to produce the diquark via an $s$-channel resonance. The corresponding cross-section is therefore substantially larger than pair production and offers a much broader diquark mass reach, as can be seen in Figure~\ref{fig:prodCS}. Furthermore, since the diquark is now an intermediate state, final states can be chosen that are free of light jets, which greatly helps baryon number counting. The selected anti-leptoquark decay products for this analysis are an anti-top quark and a charged anti-lepton. The dilepton singlet produces another anti-lepton and an anti-neutrino. Here, assuming the anti-top quark is reconstructed from its decay products, $\Delta B = -1$ can be reliably determined. For lepton number counting, due to the unavoidable anti-neutrino final state, unfortunately, experimentally we cannot distinguish between $\Delta  L = -1$  and $\Delta L = -3$. However, from a theory perspective, the same argument we presented for the determination of $\Delta L$ for the leptoquark regarding proton decay constraints on $\Delta B =  \Delta L = -1$ theories leads us to $\Delta L = -3$ as the next, and most likely, possibility. \\ 

\section{BNV through resonantly produced diquarks at the LHC}\label{sec:collidermain}
\begin{figure}[h!]
\centering
\input{tikz/collider_signature}
\caption{Parton level description of the distinctive hadron collider signature we propose for $B - L/3$. An up and a charm quark from the protons annihilate to produce a diquark, a process which benefits from an $s$-channel enhancement. It then decays to a leptoquark and a dilepton singlet. These in turn decay to an anti-top quark and an anti-lepton, and an anti-lepton and an anti-neutrino.}
\label{fig:ColliderSignatureFinal}
\end{figure}
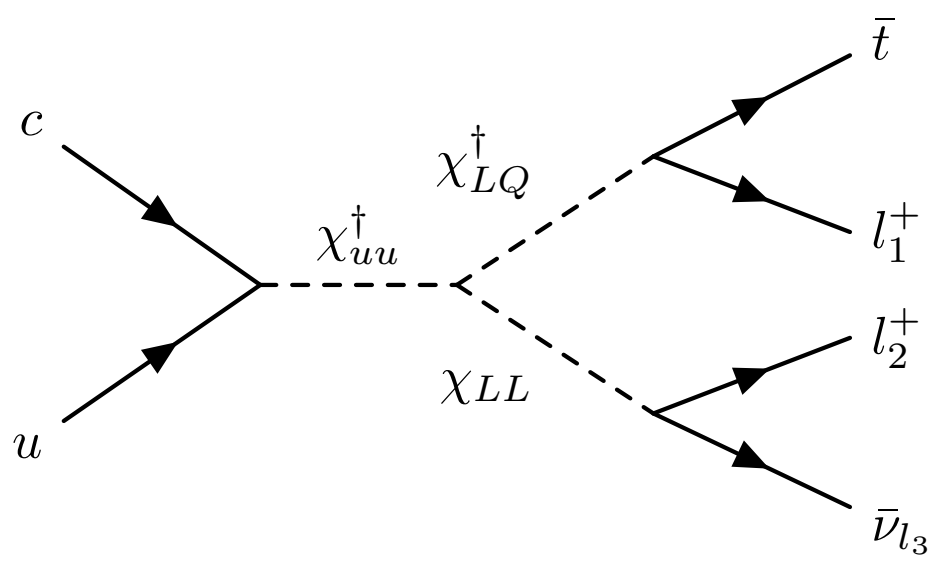

Of the three mass hierarchies considered in Section~\ref{sec:OptimalProcessSelection}, the one with the diquark as the heaviest scalar is the most promising for an LHC search. It is resonantly produced via an \(s\)-channel process, providing a large production cross-section, and its subsequent BNV decay results in final states with minimal hadronic activity, facilitating baryon-number identification. Although the outgoing antineutrino introduces an experimental ambiguity in determining $\Delta L$, theory arguments and results from proton decay experiments lead us to identify $\Delta L = -3$ as a strong possibility. In this Section, we present an analysis of the kinematics for this benchmark. For simplicity, we focus the discussion on ``$ul$-type'' leptoquarks, which has a larger branching ratio to our chosen final states compared to the ``$LQ$-type'' one. Since the chirality of the leptoquark couplings has very little impact on the kinematics of the final state particles, we adopt the same selection cuts for both.

\subsection{Simulation details.}
Our proposed search is based, at the parton level, on the Feynman diagram shown in Figure~\ref{fig:ColliderSignatureFinal}. The diquark is produced via the $u c \rightarrow \chi^{\dagger}_{_{uu}}$  annihilation process. It then undergoes its BNV decay chain, with the chosen SM final states for the analysis being the ones discussed in Section~\ref{sec:OptimalProcessSelection}. This signature shares many similarities with previous same-sign dilepton searches performed by both the ATLAS \cite{ATLAS:2011izm,ATLAS:2023afl} and CMS \cite{CMS:2016mku,CMS:2020cpy} experiments. While these provide indirect constraints on our signal process, they are sub-optimal given that they do not require top-tagging and require several high $p_{\text{T}}$ jets. More importantly, these two requirements are not well suited to our central goal of demonstrating BNV as unambiguously as possible. Therefore, while we will show that the results from \cite{ATLAS:2023afl}, in particular, \emph{indirectly} constrain some of the parameter space of interest for our model, much of it remains open and could be explored with the dedicated search we propose below using \emph{both} Run 2 and Run 3 data. \\

The dominant SM background process with \emph{almost} the same final states is 
 \begin{equation}
 \begin{split}
     p \ p \rightarrow t\overline{t}W^{+}, W^{+} \rightarrow l_{1}^{+}\nu_{l_{1}},t \rightarrow b W^{+}, W^{+} \rightarrow l_{2}^{+}\nu_{l_{2}}, \\
     p \ p \rightarrow t \overline{t} Z, Z \rightarrow l_{1}^{+} l_{1}^{-}, t \rightarrow b W^{+}, W^{+} \rightarrow l_{2}^{+}\nu_{l_{2}}.\  
 \end{split} 
 \end{equation}
They both come with an additional $b$ jet not present for the signal and can be further suppressed using a jet veto. However, we still expect the signal to carry missing energy in the form of neutrinos and anti-neutrinos produced from the leptonic decays of hadronization mesons. For this reason, in our simulations we do not enforce the requirement that for the $t\overline{t}Z$ background, the lepton from the $Z$ decay is misidentified. \\

To determine efficient selection cuts for our search, we study various kinematic quantities,  specified below, for two benchmark points broadly representative of the mass range that the (HL) LHC can potentially explore, 
\begin{enumerate}
    \item Benchmark I (BP I): $m_{\chi_{uu}}$ = 3.5 TeV, $m_{\chi_{LQ}} = m_{\chi_{LL}}$ = 1.5 TeV, with $\Lambda_{\text{BNV}}$ fixed at 10 TeV giving an $20\%$ BNV branching fraction for the diquark. This is potentially searchable with currently available LHC data and presents an interesting target for the HL-LHC.  
    \item Benchmark II (BP II): $m_{\chi_{uu}}$ = 5.5 TeV, $m_{\chi_{LQ}} = m_{\chi_{LL}}$ = 2.5 TeV, with $\Lambda_{\text{BNV}}$ fixed at 10 TeV giving an $8\%$ BNV branching fraction for the diquark. This represents a soft upper limit on what is potentially searchable at the HL-LHC assuming an integrated luminosity of 3000 fb$^{-1}$. 
\end{enumerate}
For both these benchmarks, we consider a simple scenario where the diquark Yukawa couplings are left to their current saturation value and multiply the couplings of the leptoquark and the dilepton by an overall common factor, $y_{\text{SK}}$,
\begin{align}
    y_{uu} &= \begin{pmatrix}
                0 & 0.25 & 0.12 \\
                -0.25  & 0 & 0.12 \\
                -0.12  & -0.12 & 0
    \end{pmatrix}, \\ 
    y_{LL} &= y_{\text{SK}}\begin{pmatrix}
                0 & 0.125 & 10^{-3} \\
                -0.125 & 0 & 0.325 \\
                -10^{-3} & -0.325 & 0
    \end{pmatrix} \text{and} \ \
    y_{X} = y_{\text{SK}} \begin{pmatrix}
                0.397 & 0 & 0 \\
                0 & 0.949 & 0 \\
                0 & 1 & 1
    \end{pmatrix},
\end{align}
where $X = LQ \ \text{or} \ ul$. To fix $y_{\text{SK}}$, we consider the values consistent with the Super-K bound for when the heaviest of the three scalars has a mass of 3 TeV and the other two are degenerate at 1.5 TeV and fix $\Lambda_{\text{BNV}}$ = 10 TeV, which gives $M_{\text{BNV}} = 1.4$ TeV. After making the appropriate rotations to the mass basis, we then take $y_{\text{SK}}$ to be the smallest one consistent with Super-K bounds on trilepton nucleon decay to positrons or anti-muons, finding that when  $X = LQ$, $y_{\text{SK}}$ = 0.1 and that when $X = ul$, $y_{\text{SK}}$ = 0.06. \\

We fixed the leptoquark and dilepton singlet masses to be degenerate for simplicity. In the second benchmark, since it is kinematically allowed due to the new production mode, we increase their masses to exceed the current limits from searches based on direct production.  The differential cross-section for the production of the diquark for both benchmark points are shown in Figure~\ref{fig:kinematics_unitarity}, along with the perturbative unitarity constraints (see Appendix~\ref{sec:PerturbativeUnitarity}) on $\Lambda_{\text{BNV}}$ at 13 and 14 TeV respectively. 

\begin{figure}[h!]
    \centering
    \includegraphics[width=0.487\textwidth]{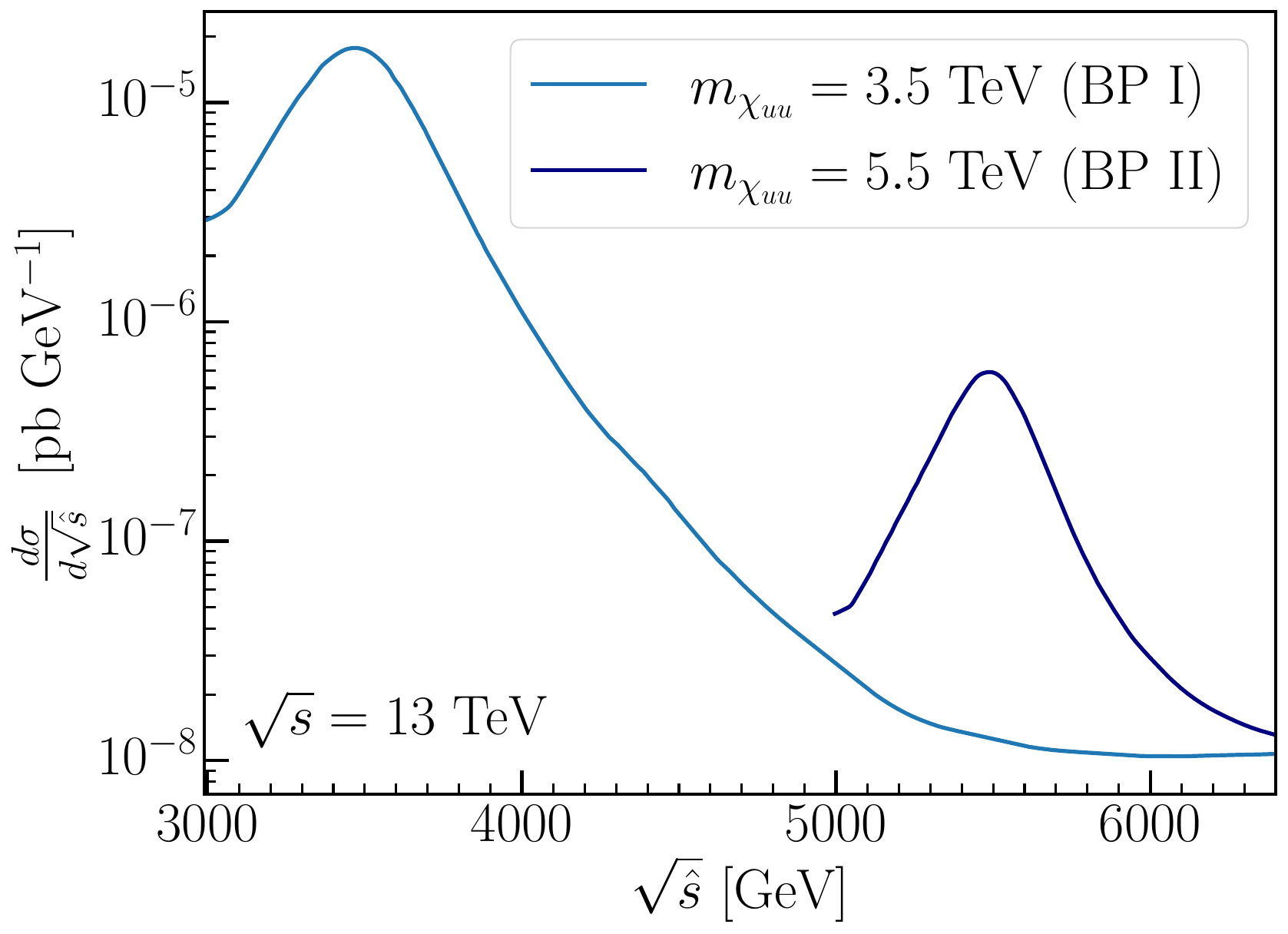}
    \includegraphics[width=0.482\textwidth]{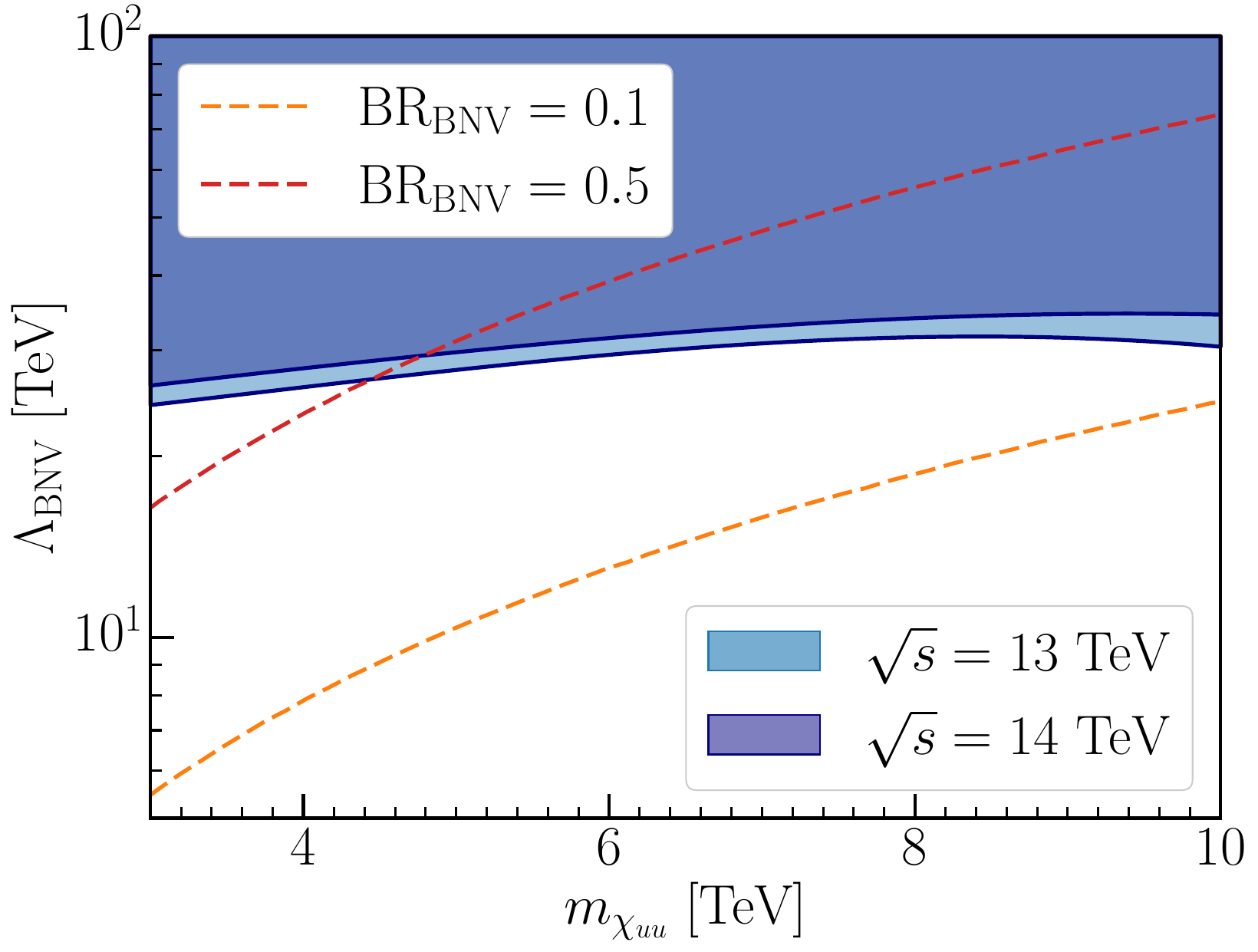}
    \caption{\textit{Left:} Differential cross section with respect to parton center of mass energy, $d\sigma/d\sqrt{\hat{s}}$, at $\sqrt{s} = 13$~TeV for $p \ p \rightarrow \chi^{\dagger}_{uu}, \ \chi^{\dagger}_{uu} \rightarrow \chi_{_{LQ}}\chi_{_{LL}}$ showing an $s$-channel resonance at the diquark mass at 3.5 and 5.5 TeV.  \textit{Right:} Perturbative unitarity constraints at $\sqrt{s} = 13$ (light blue shaded) and $14$~TeV (dark blue shaded), with contours of the diquark BNV branching fraction, $\mathrm{BR}_{\mathrm{BNV}}$ at 10$\%$ (dashed orange) and $50\%$ (dashed red overlaid). The shaded regions violate perturbative unitarity limits. 
    } 
    \label{fig:kinematics_unitarity}
\end{figure}

As these benchmarks points will illustrate, the signal final states possess several distinctive kinematical differences with their background conter-parts that can be used for selection cuts, namely: 
\begin{enumerate}
 \item Transverse momentum, $p_{\text{T}}$, of the final state anti-leptons.  Both signal anti-leptons come from the on-shell decay of $\mathcal{O}$(TeV) resonances and are therefore expected to peak at more than $\mathcal{O}$(500) GeV, much larger than the SM background ones that are produced from the decays of (boosted) $W$ bosons.
    \item Missing Transverse Energy (MET), defined as,
    \begin{equation}\label{eq:ETmissVis}
    \slashed{E}_{\text{T}} \equiv \vert \sum_{\text{vis}} \vec{p}_{\text{T}} \vert, 
\end{equation} 
    where the $p_{\text{T}}$ sum is over all visible final states. MET is also expected to be peak at much larger values for the signal because it is dominantly carried by the single anti-neutrino produced in the decay of the heavy, $\mathcal{O}$(TeV), mass dilepton singlet. In comparison, the SM $t\overline{t}W$ background MET is carried by neutrino daughter particles produced in the decay of (boosted) $W$ bosons, which means it peaks at a much lower value. When the anti-top decays leptonically, there is an additional, sub-dominant, contribution to MET for both the signal and the background.  
    \item Transverse cluster mass of all visible final states, 
    \begin{equation}\label{eq:MCVis}
        \text{M}^{2}_{\text{C}}\left(\text{vis}\right) = \left(\sqrt{p^{2}_{\text{T,vis}}+m^{2}_{\text{vis}}} + \slashed{p}_{\text{T}}\right)^{2} - \left(\vec{p}_{\text{T,vis}}+\slashed{p}_{\text{T}}\right)^{2},
    \end{equation}
    where  $p_{\text{T,vis}}$ and $m_{\text{vis}}$ are the transverse momentum and invariant mass of all visible final state particles respectively. The 2 $\rightarrow$ 1 $\rightarrow$ 2 $\rightarrow$ (2,2) topology of our signal motivates using $\text{M}_{\text{C}}\left(\text{vis}\right)$ as a selection cut. Since all signal final state particles originate from the decay of the same ``grand-parent'' diquark particle, we expect it to show a smeared peak near its mass. The smearing is due to a combination of neutrinos in the final state, ISR, FSR, and detector effects but these do not change the fact that the signal $\text{M}_{\text{C}}\left(\text{vis}\right)$ peaks at a much higher value compared to the background which does not share the same topology.  
\end{enumerate}
We also examine our choice of final state particles in more detail. A particularly important criteria in deciding whether or not we want to include anti-$\tau$ lepton final states is isolation. Given that the latter decays hadronically 64.79$\%$ of the time \cite{HeavyFlavorAveragingGroupHFLAV:2024ctg}, its inclusion depends on if it is well isolated, particularly from the anti-top daughter $b$-jet. If it is not, the reconstruction becomes challenging for its dominant hadronic decay mode and we must focus instead on only $e^{+}$ and $\mu^{+}$ final states. Another important consideration with $\tau^{+}$ final states is that the final state particles are either a secondary $e^{+}$ or $\mu^{+}$ when it decays leptonically or a positively charged $\tau$-tagged jet when it decays hadronically. The emission of neutrino(s) from the parent $\tau^{+}$ lowers the $p_{\text{T}}$ of the final states, which we must take into account when selecting a cutoff for anti-lepton $p_{\text{T}}$ as a selection cut. \\

A related issue is whether or not to consider $\tau$ leptons for the leptonic decay of the anti-top quark. For simplicity, we follow conventional ATLAS and CMS measurements of semileptonic single-top production \cite{Pintucci:2024ude} and consider only $e^{-}$ or $\mu^{-}$ final states, again due to the challenges associated with reconstructing a candidate hadronically decaying $\tau$ lepton close to a $b$-jet. However, even with only $e^{-}$ or $\mu^{-}$ final state leptons from the anti-top decay,  reconstructing its charge remains challenging. A signal anti-top quark originates from the decay of a heavy leptoquark whose mass is constrained by existing direct searches to exceed 1.2 TeV. Consequently, it typically carries a transverse momentum of $\mathcal{O}(600)$ GeV, resulting in a characteristic angular separation between its decay products of order
\[
    \Delta R(\ell^-,\bar b)\simeq \frac{2m_t}{p_{T,\bar t}}
    \simeq 0.6
    \left(\frac{600\,\mathrm{GeV}}{p_{T,\bar t}}\right),
    \qquad \ell^-=e^-,\mu^-.
\]
Although this relation provides only a characteristic angular scale rather than an exact lepton--jet separation, it illustrates how the charged lepton can lie within the radius of a large-radius AK8 jet, complicating conventional lepton isolation and the reconstruction of the leptonic anti-top decay. This is the well-known ``lepton-in-jet'' problem, arising when the charged lepton becomes collimated with the accompanying $b$-initiated jet. In our case, this problem can be even worse than the naive estimate above suggests since the leptoquark is allowed to be heavier than the lower limits implied by direct searches due to the new, BNV, production mechanism. While a detailed analysis of the anti-top reconstruction in this highly boosted regime is beyond the scope of this work, the two leptoquark masses in our benchmarks, 1.5 TeV for BP I and 2.5 TeV for BP II, are chosen specifically to quantify the scope of this problem. \\

To study the above mentioned issues and kinematic distributions we simulate  \(10^5\) events for each sample using \texttt{MadGraph5\_aMC@NLO} ~\cite{Alwall:2014hca}, considering the leptonic and hadronic anti-top decay channels separately, for the signal benchmark points above. We compare their distributions with those of the SM \(t\bar t W\) and \(t\bar t Z\) backgrounds at 13 and 14 TeV respectively. Parton showering and hadronization are simulated using \textsc{Pythia}~8~\cite{Bierlich:2022pfr} and detector effects are studied using fast detector simulation with \textsc{Delphes}~3~\cite{deFavereau:2013fsa}. Because the highly boosted anti-top decay products cannot always be resolved by the standard reconstruction procedure, we adopt a hybrid approach combining detector-level objects with select generator-level information. The relevant simulation prescriptions are summarized below.

\begin{enumerate}
\item \textbf{Basic acceptance requirements.} For both signal and background, we impose \(|\eta_{\ell,j,b}|<2.5\), approximating the central detector acceptance of ATLAS and CMS, and require reconstructed jets considered in the analysis to satisfy \(p_{\text{T}j}>20\) GeV. 

\item \textbf{Treatment of boosted anti-top decays.} The strong collimation of the anti-top decay products complicates their individual reconstruction, particularly for the ``lepton-in-jet'' and  ``merged-jet'' topologies for leptonic and hadronic decays respectively. Since a dedicated reconstruction algorithm for these boosted configurations, including their charge identification, lies beyond the scope of this present study, we adopt the following simplified hybrid prescription:

\begin{enumerate}    \item \textit{Leptonic decays.} The two positively charged lepton candidates are reconstructed using \textsc{Delphes}~3, including hadronically decaying $\tau^+$ candidates. For the $\overline{b}$ quark and the negatively charged lepton originating from the anti-top, we instead use the \textsc{Pythia}~8 event record. The \(\bar b\) is identified as the highest $p_\text{T}$ parton with status code 23 and PDG ID \(-5\), while the charged lepton is selected as the highest-$p_\text{T}$ electron or muon. This prescription is motivated by the expectation that other negatively charged leptons, predominantly arising from secondary meson decays, are typically softer. 

   \item \textit{Hadronic decays.} The positively charged lepton candidates are reconstructed as in the leptonic case. The three partons originating from the anti-top decay are identified using the \textsc{Pythia}~8 event record, since the strong collimation of their decay products prevents a straightforward resolved-jet reconstruction. We select the highest $p_{\text{T}}$ status-23 $\overline{b}$ parton, together with the two highest $p_{\text{T}}$ eligible light-quark candidates from the $W$ boson decay, corresponding to PDG IDs $(-2,-4)$ for up-type antiquarks and $(1,3)$ for down-type quarks. Their four-momenta are used to study the kinematics of the hadronically decaying anti-top.

   \item \textit{Additional jets.} For both the leptonic and hadronic anti-top decay channels, ISR and FSR jets are identified using the \textsc{Delphes}~3 reconstructed jet collection, with reference to the corresponding \textsc{Pythia}~8 event records. We exclude jets associated with the status-23 hard-processes but retain all additional ones. This allows us to study their kinematic distributions and angular separation from the hard-process decay products.

   \item \textit{Missing transverse energy and transverse cluster mass.} Unlike the explicit reconstruction of individual anti-top decay products, these observables depend primarily on the overall missing transverse momentum and the combined visible final-state four-momentum. Consequently, they are less sensitive to whether nearby decay products are individually resolved. We examine these observables using the detector-level information provided by \textsc{Delphes}~3 and compare them with their generator-level counterparts where available. 

   \item \textit{Comparison of generator and detector-level results.} Wherever possible, we present both generator-level and detector-level distributions to assess the impact of detector effects on the proposed kinematic selection criteria. Distributions involving the explicitly identified anti-top decay partons are presented at generator level and labeled accordingly.
   \end{enumerate}

\end{enumerate}

Our resulting study is therefore intended to establish a realistic starting point for a dedicated experimental analysis, while leaving the development and validation of a fully reconstructed boosted anti-top selection to future work. 

\subsection{Leptonic Anti-Top Decay}
\label{sec:tbar_leptonic}
With these clarifications, we now present our detailed kinematic analysis, starting with when the anti-top decays leptonically. The signal process we search for is then
\begin{equation}
p p \rightarrow j_{b} \ e^{-}/\mu^{-} \ l_{1}^{+}l_{2}^{+} + \text{Missing Energy},   
\end{equation}
where the $j_{b}$ denotes a $b$ jet originating from the $b$ quark daughter of the anti-top quark, and the electron or muon is produced along with the corresponding anti-neutrino $\overline{\nu}_{e}/\overline{\nu}_{\mu}$ from the decaying $W$ boson. 

\subsubsection*{Final state kinematics}
\begin{figure}[h!]
   \centering
    \includegraphics[width=0.99\textwidth]{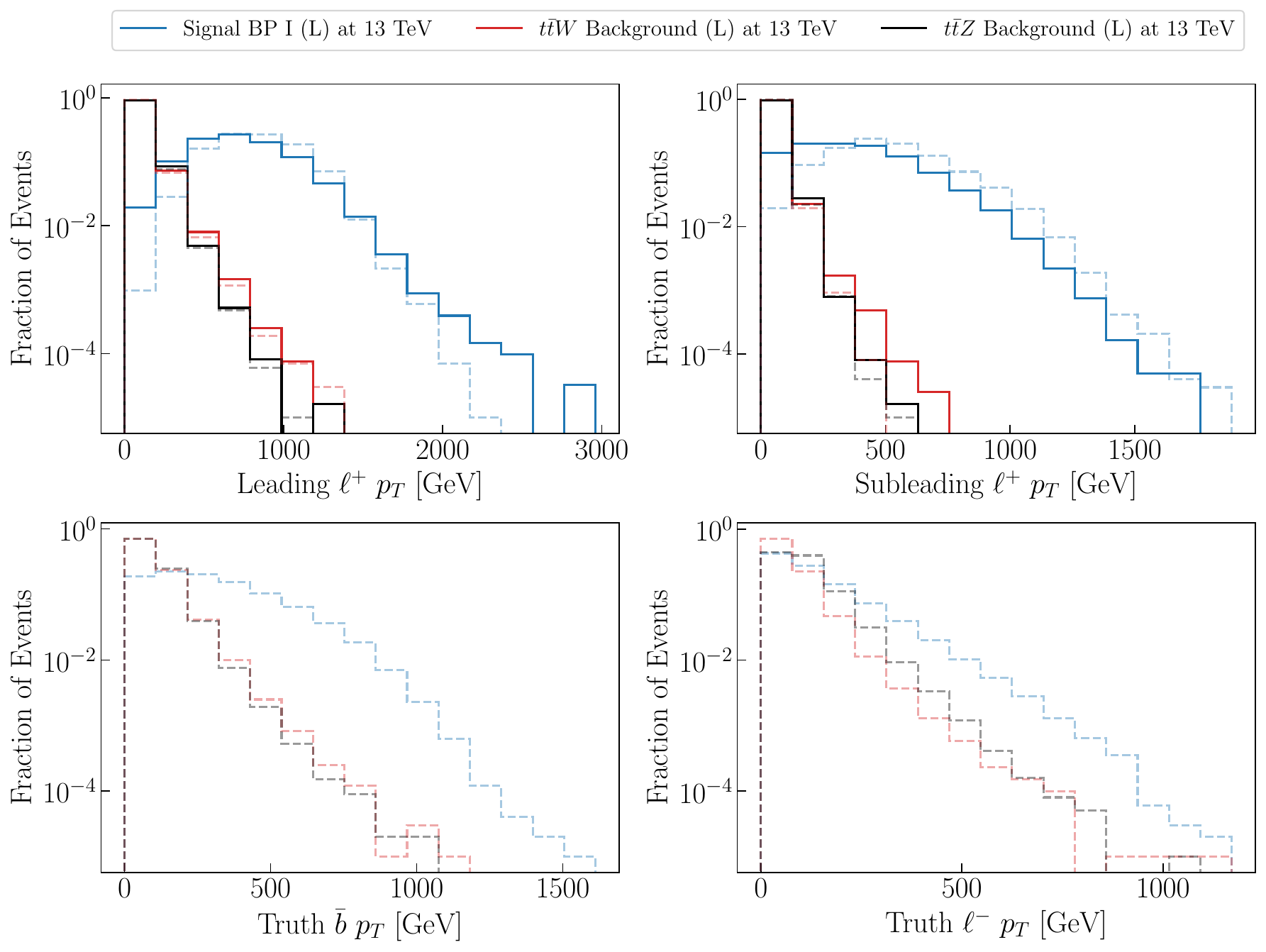}
    \caption{Transverse momentum distributions of all visible final states for the signal BP I, with the corresponding backgrounds shown in red and black. The top left and right panels for each benchmark correspond to the anti-leptons sorted by $p_{\text{T}}$, with the solid and dashed lines corresponding to detector and generator level distributions respectively. The signal anti-lepton transverse momentum distribution, in particular, illustrate the effectiveness of a large $p_{\text{T}}$ cut on the final state anti-leptons. There are much more detector level anti-lepton events at lower $p_{\text{T}}$s compared with the generator level ones since in the former case $\tau^{+}$ particles are taken as the visible decay products of the parent particle rather than the parent particle itself, see the discussion in \cref{sec:tbar_leptonic} for more details. The lower right panel shows the $p_{\text{T}}$ distribution (generator level) of the lepton produced from the leptonic decay of the anti-top, and the lower left one for the generator level anti-$b$-quark.}
    \label{fig:pT_DistributionsBPI}
   \end{figure}

\begin{figure}[h!]
    \centering
    \includegraphics[width=0.99\textwidth]{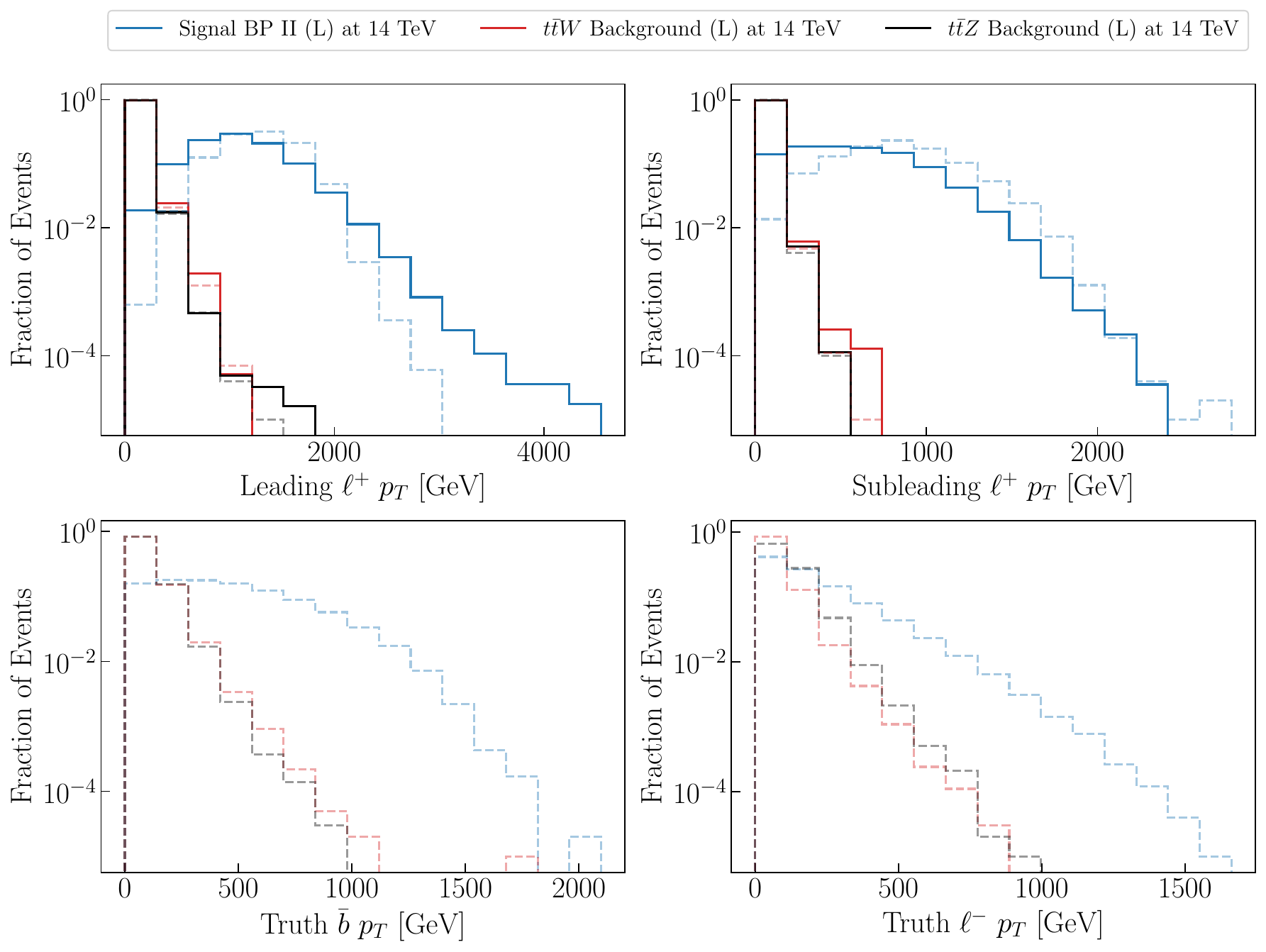}
    \caption{Same as \cref{fig:pT_DistributionsBPI} but for BP II.}   \label{fig:pT_DistributionsBPII}
\end{figure}

\textit{Transverse momentum of hard visible final states:} We start our discussion of signal kinematics with the transverse momentum of all hard visible final states in our simulated signal and background events in Figures~\ref{fig:pT_DistributionsBPI} and~\ref{fig:pT_DistributionsBPII}.  Several remarks are in order:  
\begin{enumerate}
    \item \textit{The anti-leptons:} On the top left and right panels of Figures~\ref{fig:pT_DistributionsBPI} and~\ref{fig:pT_DistributionsBPII}, we show the distribution of charged anti-lepton final states sorted by $p_{\text{T}}$. The solid lines correspond to the distribution of our  \textsc{Delphes}~3 events while the dashed ones correspond to our \textsc{Pythia}~8 ones. At generator level, we use the momentum of the parent $\tau^+$, whereas at detector level we use its decay products, either an $e^+$ or $\mu^+$ in leptonic decays or a tau-tagged jet in hadronic decays. Since these decays involve one or more neutrinos, the visible decay products carry only a fraction of the parent $\tau^{+}$ momentum. This contributes to the shift towards lower
$p_{\text{T}}$ seen in Figures~\ref{fig:pT_DistributionsBPI} and~\ref{fig:pT_DistributionsBPII} for the detector-level distributions. However, for both benchmarks the signal dileptons still peak at a $p_{\text{T}}$ > 500 GeV, consistent with their origin from the decay of a TeV scale particle. The background ones, on the other hand, peak much lower, around $\mathcal{O}$(100) GeV, consistent with their origins from (boosted) $W$ bosons. We therefore impose a  $p_{\text{T}}$ > 50 GeV cut on all candidate anti-lepton final states. If only $e^{+}$ or $\mu^{+}$ candidates are considered, this cutoff can be raised higher. 
    \item \textit{The electron or muon from the anti-top decay:} On the lower right panels of Figures~\ref{fig:pT_DistributionsBPI}  and~\ref{fig:pT_DistributionsBPII}, we show the distribution of the $e^{-}/\mu^{-}$ $p_{\text{T}}$ produced from the decay of the boosted anti-tops. Here, we remind the reader that these are generator level events from our \textsc{Pythia}~8 records. Since the energy of the latter is shared  with the $b$-jet and the anti-neutrino, the $e^{-}/\mu^{-}$ are softer than the anti-leptons. This is more pronounced in the signal case. For both benchmarks, the signal peaks roughly in the same region as the background, although it has a much larger spread. We will therefore not consider selection cuts based on the final state electrons or muons. 
    \item \textit{The b-jet from the anti-top decay:} In the lower left panel of Figures~\ref{fig:pT_DistributionsBPI} and~\ref{fig:pT_DistributionsBPII}, we show the distribution of the initial $b$ quark from the anti-top decay that produces the $b$-tagged jet. 
    Following the same observation as in the anti-lepton case, a selection cut with $p_{\text{T},b}$ > 50 GeV would be applied to suppress the background which peaks at lower $p_{\rm T}$.\\
\end{enumerate}

\begin{figure}[h!]
    \centering
    \includegraphics[width=0.49\textwidth]{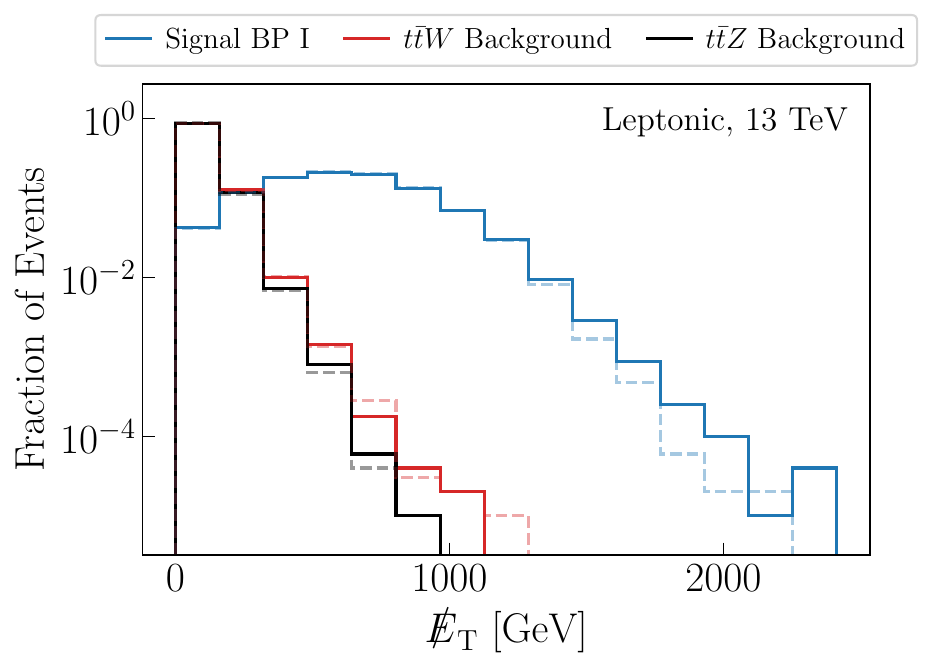}
    \includegraphics[width=0.49\textwidth]{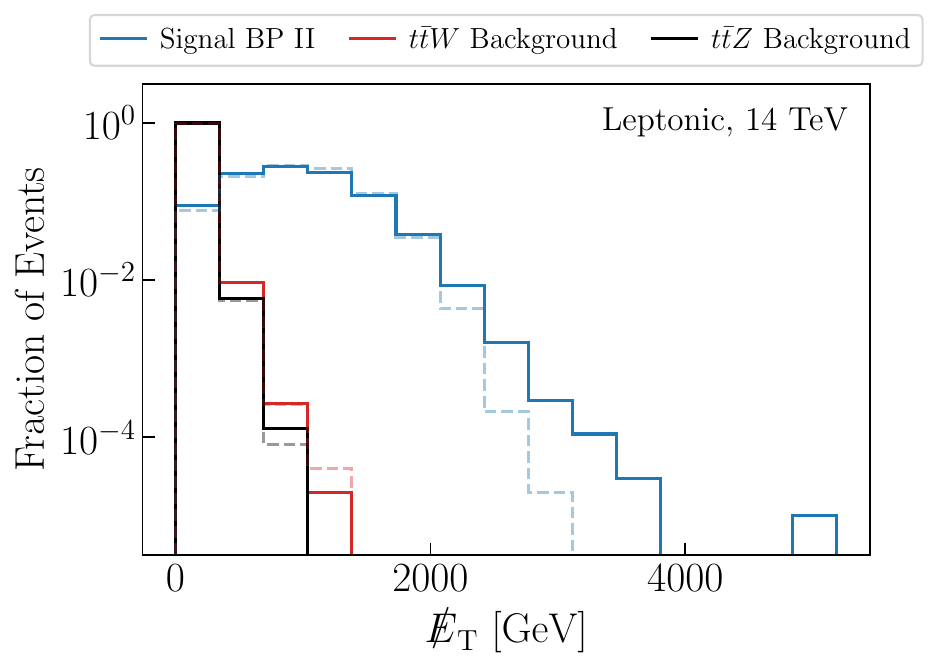}

    \caption{MET distribution for the background compared to signal benchmarks BP I (left) and BP II (right) for leptonically decaying anti-top. Solid lines correspond to detector level distributions while dashed ones correspond to generator level ones. The signal missing energy, dominated by the neutrino produced from the dilepton singlet (1.5 TeV for BP I and 2.5 TeV for BP II) decay, peaks at much higher values compared to the SM backgrounds, the latter originating from neutrinos produced via the decay of (boosted) $W$ bosons.}
    \label{fig:ET_Distributions}
\end{figure}

\noindent\textit{Missing Transverse Energy:} Next, we consider MET, as presented in in Figure~\ref{fig:ET_Distributions}. In the signal case, MET is carried by two anti-neutrinos, one from the decay of a TeV mass dilepton singlet and the other from that of the boosted anti-top. Its MET therefore peaks around $\mathcal{O}(m_{\chi_{LL}}/2) \gtrsim 500$ GeV. In contrast, the background MET is carried by a trio of neutrinos originating from the decays of (boosted) $W$ bosons and carrying a combined energy of roughly $\mathcal{O}$(100) GeV.\footnote{For both background and signal, there is also a small contribution from neutrinos emitted from mesons decays originating from hadronic activity but these are typically much softer.} The signal MET peaks at a much higher value than the SM backgrounds, motivating its use as a selection cut. \\

\begin{figure}[h!]
    \centering
    \includegraphics[width=0.48\textwidth]{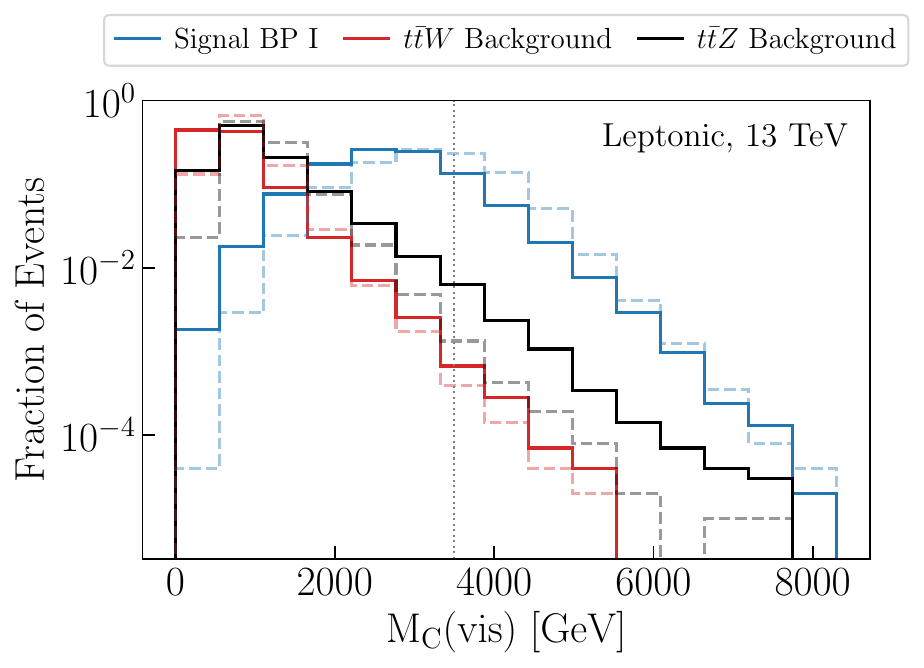}
    \includegraphics[width=0.486\textwidth]{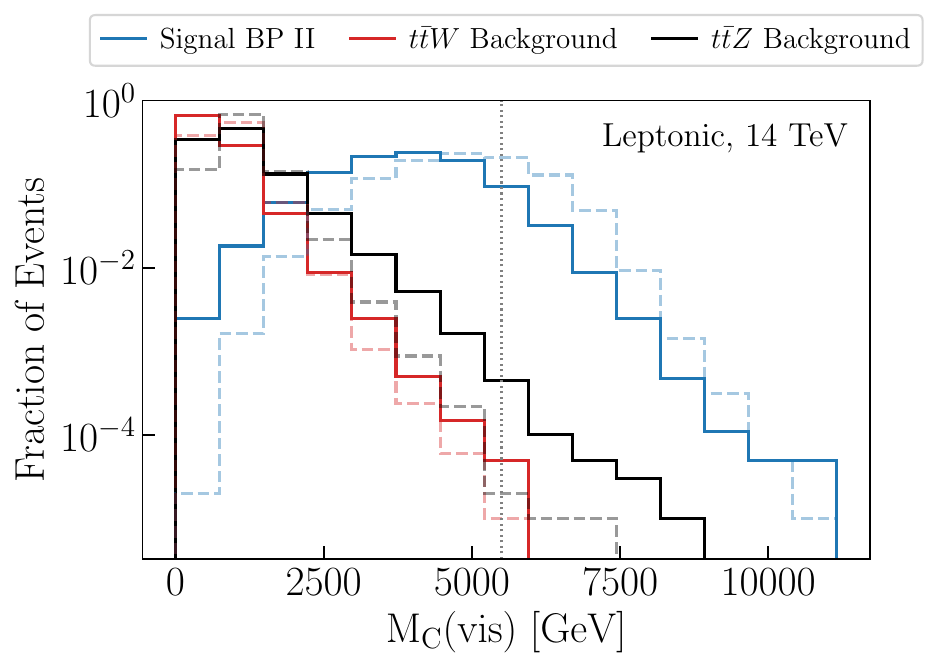}
    \caption{Transverse cluster mass of all visible final states for BP I (left) and BP II (right) compared to the corresponding SM backgrounds. Solid lines correspond to detector level distributions while dotted lines correspond to generator level ones. In both cases, the peak is smeared away from the corresponding diquark mass due to the outgoing neutrinos. The signal $\text{M}_{\text{C}}\left(\text{vis}\right)$ peaks at a much higher value compared to the backgrounds, motivating its use as a selection cut.}
    \label{fig:M_Cluster_Distributions}
\end{figure}

\noindent\textit{Transverse cluster mass of all visible final states:} Finally, since, at the parton level, all four final state particles are produced from the decay of the diquark ``grand-parent'' particle, we examine the kinematics of the former to search for any peaked structures around the diquark mass. To this end, we compute the transverse cluster mass of all visible final states, which, with respect to our simulations, refers to all stable final state from our \textsc{Delphes}~3 event records. The resulting distribution of $\text{M}_{\text{C}}\left(\text{vis}\right)$ for our two signal benchmark points and their corresponding backgrounds are shown in Figure~\ref{fig:M_Cluster_Distributions}, with the distribution from our \textsc{Pythia}~8 event records shown in faint dashed lines for comparison. For both benchmark cases, the $\text{M}_{\text{C}}$(vis) signal dominates over the backgrounds for $\text{M}_{\text{C}}(\text{vis.)}\gtrsim 2000$ GeV.

\subsubsection*{Separation between final state particles}
\begin{figure}[h!]
    \centering

    \includegraphics[width=\textwidth]{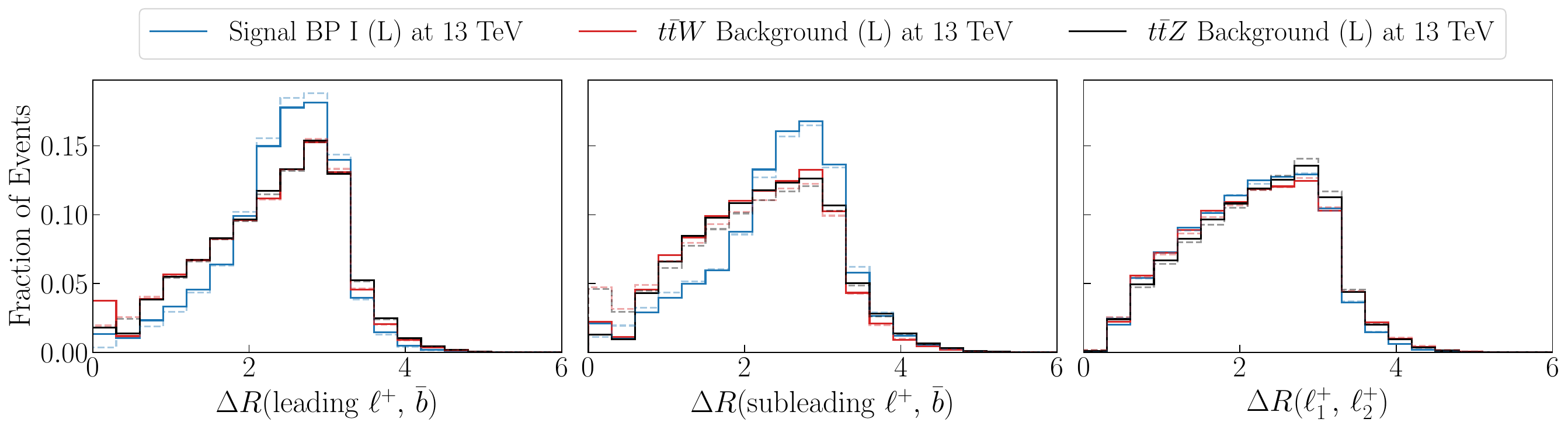}
    \vfill
    \includegraphics[width=\textwidth]{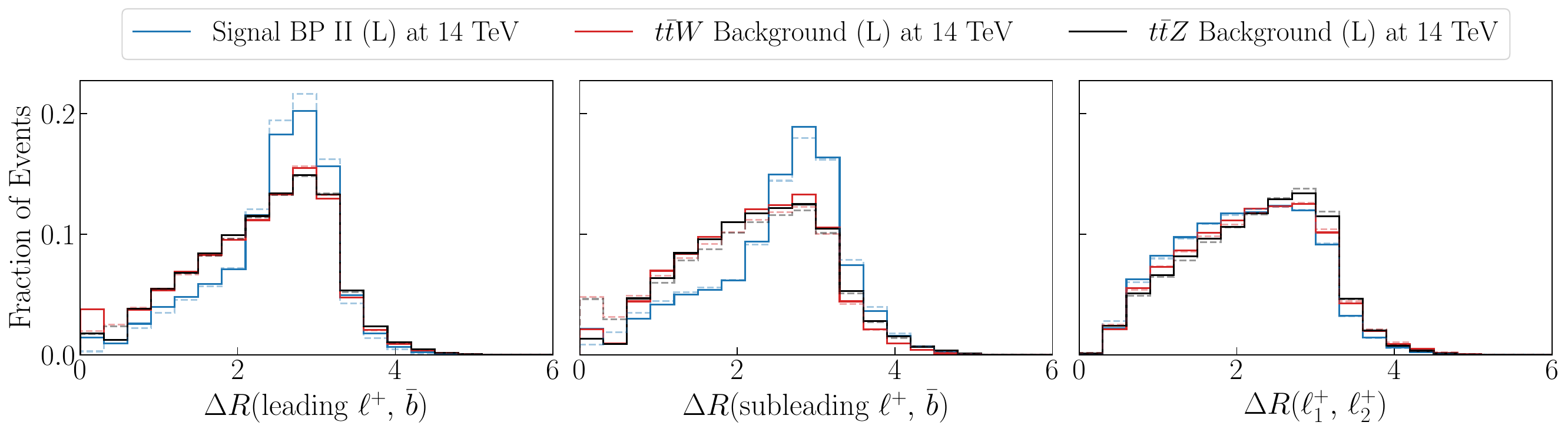}

    \caption{Top: Separation, $\Delta R$, between the highest $p_{T}$ $b$-quark and the leading (left) and sub-leading (middle) anti-leptons as well as that between the two anti-leptons (right panel) for BP I and the background. This is measured between the generator-level anti-$b$-quark between the generator-level (dash line) and detector-level (solid line) antileptons respectively. Bottom: Same as top, but for BP II. See the text for an explanation for the small bump at low $\Delta R$ for detector level anti-leptons. In our simulated events for BP I, the separation between the lepton and the $b$-tagged jet is $\Delta R > 0.8$ for 97.9\% of leading leptons and 94.5\% for the sub-leading leptons. For BP II, $\Delta R > 0.8$ for 97.4\% and 95.3\% of the leading and sub-leading leptons respectively. Regarding the separation between the two anti-leptons themselves, $\Delta R > 0.8$ for 90.3\% of the simulated events for BP I and 89.7\% for BP II. This relative isolation motivates including $\tau^{+}$ final states for our signal process.}
    \label{fig:lep_b_dR_distributions}
\end{figure}
\textit{Separation between anti-leptons and the b-quark:} The separation between the final state anti-leptons, especially the one originating from the decay of the leptoquark, and the $b$-quark jet is, as mentioned previously, important to inform our choice of which of the former to include in an inclusive search. As shown in Figure~\ref{fig:lep_b_dR_distributions}, the anti-leptons are always well isolated from the $b$-tagged jet and from each other, justifying our decision to include $\tau$ final states in our analysis. We note the small bumps at low $\Delta R$. This affects both the background and signal, and is due to the fact sometimes an anti-lepton emitted from the showering of the initial $b$-quark can have higher $p_{\text{T}}$ compared to the anti-lepton from the decays of the heavy scalar, especially if the anti-lepton is itself a positron, anti-muon, or $\tau$-tagged jet daughter of a parent $\tau^{+}$ particle.  Since our selection criteria at the moment only involves choosing the two hardest anti-leptons, this misidentification is possible and we take this into account with a proper isolation requirement for our final state selection cuts. \\

\begin{figure}[h!]
    \centering
    \includegraphics[width=0.99\textwidth]{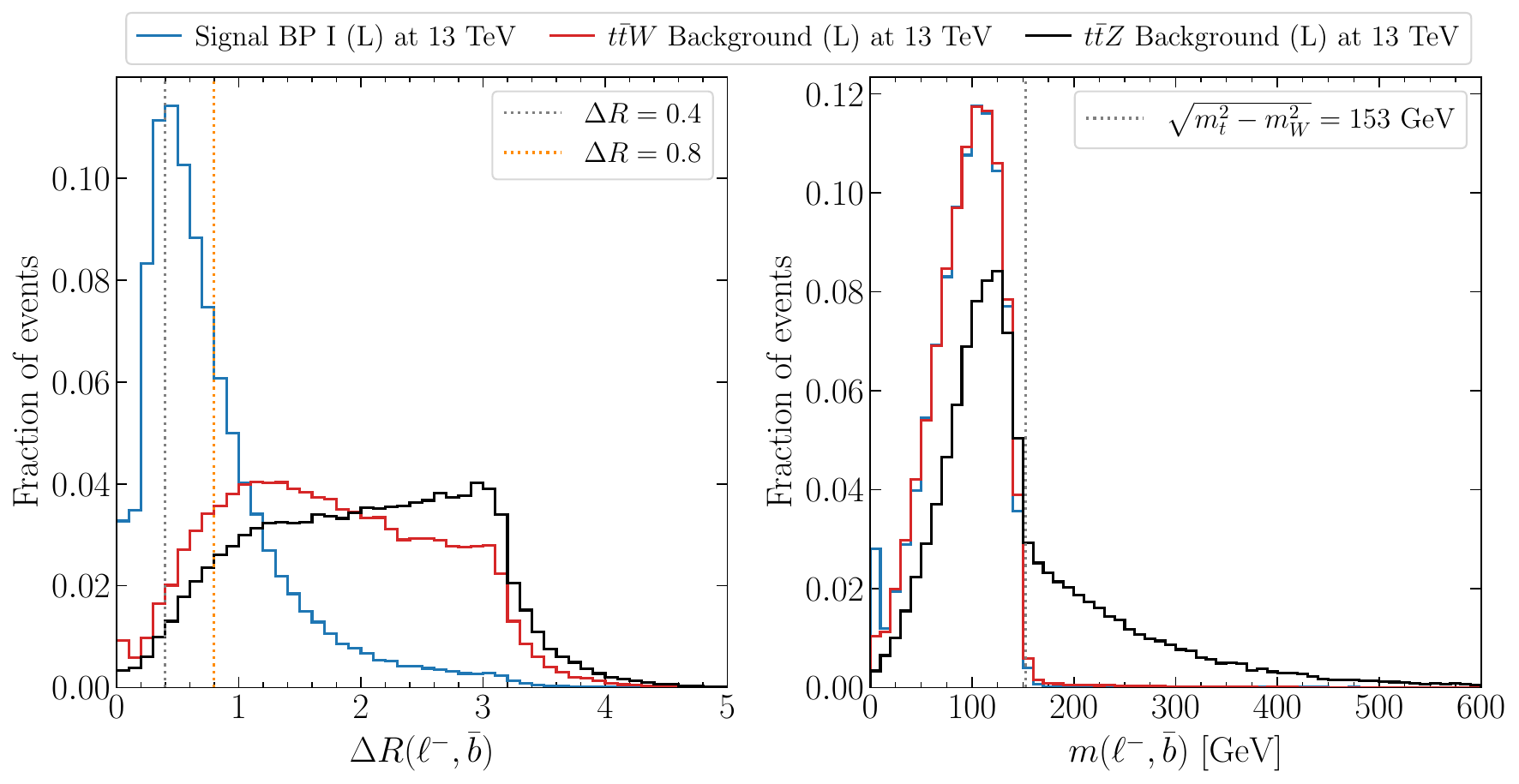}
    \hfill
    \includegraphics[width=0.99\textwidth]{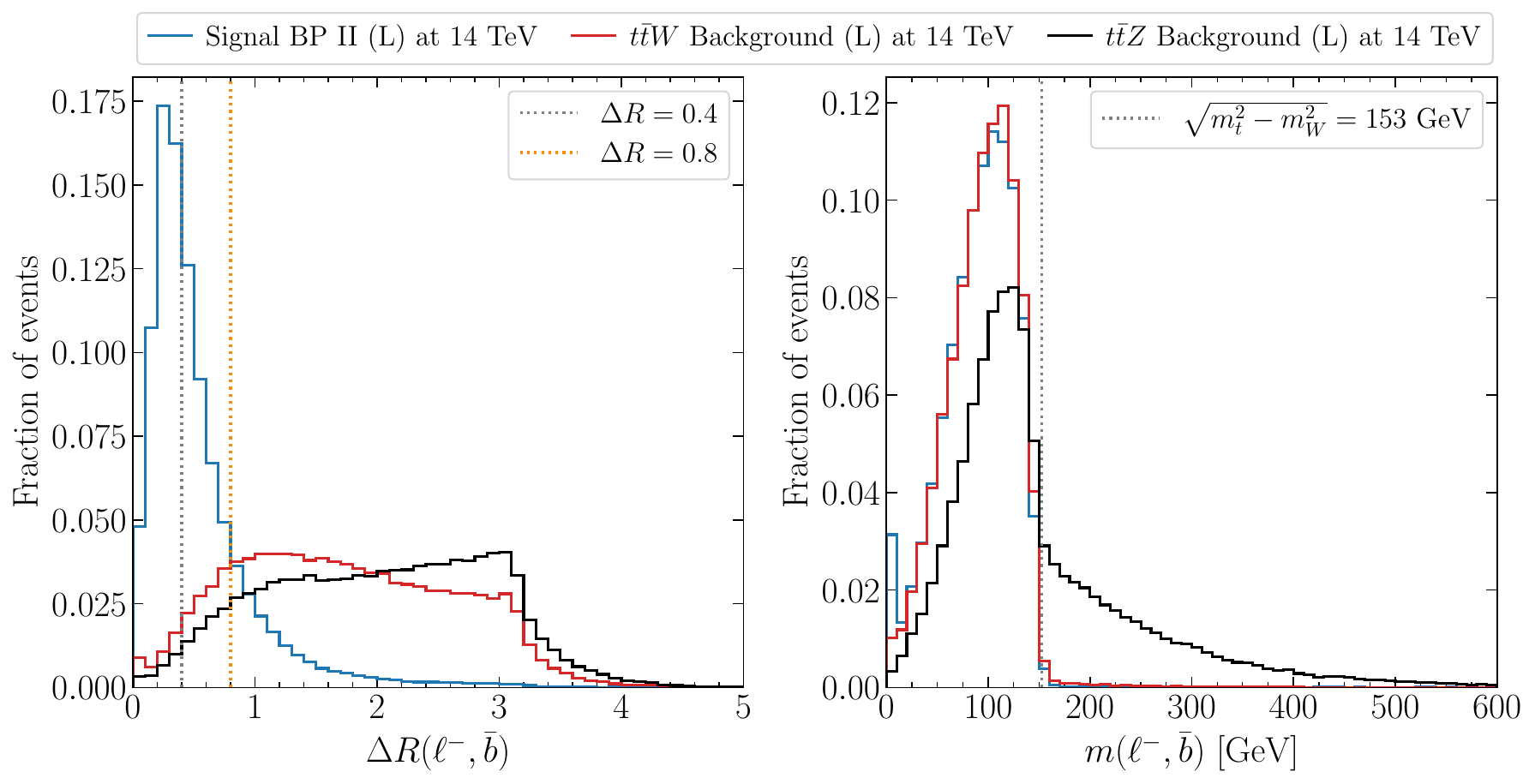}
    \caption{Left panels: Angular separation between the truth level $\overline{b}$ quark and lepton $l^{-}= e^{-}, \mu^{-}$, both produced from the decay of the boosted anti-top quark to illustrate the ``lepton in jet'' inherent to our BNV search given existing constraints on our leptoquark. For our signal BPI (top left), $\Delta R\left(l^{-},\overline{b}\right) \leq $ 0.4 (0.8) for 27 (64) $\%$ of events while for BPII $\Delta R \left(l^{-},\overline{b}\right) \leq $ 0.4 (0.8) for 48 (82) of events. As a helpful comparison, for the $t\overline{t}W$ background only 15 (16) $\%$ of events have $\Delta R\left(l^{-},\overline{b}\right) \leq $ 0.4 (0.8). The difference for both separation and invariant mass between the $t\overline{t}W$ and $t\overline{t}Z$ background can be explained by how we choose the \emph{lepton}: the latter is chosen as the highest $p_{\text{T}} > 10$ GeV electron or muon, and, for the $t\overline{t}Z$ process, this is dominantly the lepton coming from the decay of the $Z$ boson instead of the one from the anti-top decay. Right panels: invariant mass distribution of the lepton and anti-bottom quark showing consistency with the kinematic end-point of $m(\ell^{-}b) < \sqrt{m_{t}^{2}-m_{W}^{2}} \simeq 153$~GeV for anti-top decay for background, signal BPI (top right) and BPII (bottom right).}
    \label{fig:Lepton_In_Jet}
\end{figure}

\noindent\textit{Separation between electron or muon and $b$-tagged jet:}
Next we consider in more detail the angular separation between the electron or muon produced from the anti-top decay and the $b$-jet to quantify the ``lepton in jet'' problem. For a proper search, one should ideally remain agnostic about the mass of the leptoquark, with the only requirement being that lower limits from direct and indirect searches for both the leptoquark and dilepton singlet and the kinematical requirement $m_{\chi_{LQ}} + m_{\chi_{LL}} < m_{\chi_{uu}}$ be satisfied. Here we illustrate the scope of the problem through our benchmark choices. In going from BP I, which involves a 1.5 TeV leptoquark and $\Delta R\left(l^{-},\bar{b}\right) \simeq 0.5$, to BP II, which involves a 2.5 TeV leptoquark and $\Delta R\left(l^{-},\bar{b}\right) \simeq 0.3$, we illustrate how this effect is made worse for heavier leptoquarks in Figure~\ref{fig:Lepton_In_Jet}. For our signal BP I, 27 (64) $\%$ of the $\bar{b}$ quark and $l^{-}$ have angular separation $\Delta R\left(l^{-},\bar{b}\right) \leq $ 0.4 (0.8). This increases sharply in our signal BP II, with 48 (82)$\%$ of the $\bar{b}$ quark and $l^{-}$ having angular separation $\Delta R\left(l^{-},\bar{b}\right) \leq $ 0.4 (0.8). In comparison, the background case only 4.0 (15.3) $\%$ of events have $\Delta R\left(l^{-},\bar{b}\right) \leq $ 0.4 (0.8) at 13 TeV, with no appreciable change at 14 TeV (see Table~\ref{tab:leptoninjetsummary}). For completeness, we also show the invariant mass distributions of the lepton and anti-$b$ quark for the background, signal BP I, and signal BP II, all of which are consistent with the kinematic end-point $m(\ell^{-}b)  \simeq 153$~GeV. 

\begin{table}[htbp]
  \centering
  \small
  \renewcommand{\arraystretch}{1.2}
  \begin{tabular}{@{}lrrrrrr@{}}
    \toprule
    & \multicolumn{2}{c}{\quad\quad$13$ TeV}
    & &\multicolumn{2}{c}{\quad~~$14$ TeV} \\
    \cmidrule(lr){2-4} \cmidrule(l){5-7}
    Fraction of events [\%] & BP I & $t\bar{t}W$ & $t\bar{t}Z$ & BP II & $t\bar{t}W$ & $t\bar{t}Z$ \\
    \midrule
    $\Delta R(\ell^{-},\bar{b})\leq 0.4$
      & 27 & 4.0 & 2.3 & 49 & 4.3 & 2.3 \\
    $\Delta R(\ell^{-},\bar{b})\leq 0.8$
      & 65 & 15 & 9.8 & 82 & 16 & 9.9  \\
    \bottomrule
  \end{tabular}
  \caption{Percentage of simulated signal and background events in
  which the charged lepton $\ell^{-}$ from the leptonic anti-top decay
  lies within the given angular separation of the $\bar{b}$-jet,
  quantifying the ``lepton in jet'' problem. BP I and BP II, defined
  in the text, correspond to leptoquark masses of 1.5 and 2.5 TeV,
  respectively.}
  \label{tab:leptoninjetsummary}
\end{table}

\subsection{Hadronic Anti-Top Decay}
When the anti-top decays hadronically, we search for 
\begin{equation}
p p \rightarrow j_{b} \ j_{2} \ j_{3} \ l_{1}^{+}l_{2}^{+} + \text{Missing Energy},  
\end{equation}
where $j_{2,3}$ are jets of any type originating from the hadronic decay of the $W$ boson, itself a bi-product of the anti-top decay. Given the aforementioned issue about our signal anti-tops being highly boosted, we expect that for a large fraction of events, the two jets will be difficult to resolve individually. \\

Before moving on to an analysis of the final state kinematics, we spell out our motivation for analysing the hadronic decay events. Even though we favour the leptonic decay of the anti-top quark since it allows us to determine its charge, and hence that baryon number violation by one unit has occurred, even when the decay is hadronic charge conservation can be used to make an indirect argument for $\Delta B = -1$. To elaborate, the final state for the full decay of the anti-top quark contains two anti-leptons, missing energy, and three jets, one of which is $b$-tagged. Even after successfully reconstructing the top final state using the three jets, we are still left with the ambiguity of whether or not it was a particle or an anti-particle. To resolve this, we consider the two cases separately: 

\begin{enumerate}
    \item Assuming it is a top quark, the total charge of all final state particles is then +$\frac{8}{3}$. By charge conservation, the initial states must also have the same charge. However, the simplest possibility within the SM for such an initial state is a four up-type quark process that is extremely suppressed and does not yield the final states required by our signal\footnote{There could also be processes in which two undetected (soft or forward) up-type jets are in the final state. No such process appears at tree level in the SM, and while our model does contain such processes, their cross sections are subdominant to our signal process.}. 
    \item Assuming, on the other hand, that the fourth final state is an anti-top quark, the total charge of all final states is then +$\frac{4}{3}$. Using charge conservation arguments once again, we can then see that if we insist on two initial state partons, they must be two up-type ones. This can be further corroborated via the BNC dijet searches for the scalar diquark.
    \end{enumerate} 
Since this charge conservation argument presents a compelling argument for asserting $\Delta B = -1$, we also consider hadronically decaying anti-top quarks in our analysis.

\subsubsection*{Final state kinematics}
\begin{figure}[htbp]
    \centering

    \subfloat[BP I\label{fig:pT_Distributions_H_BPI}]{
        \includegraphics[width=\textwidth]{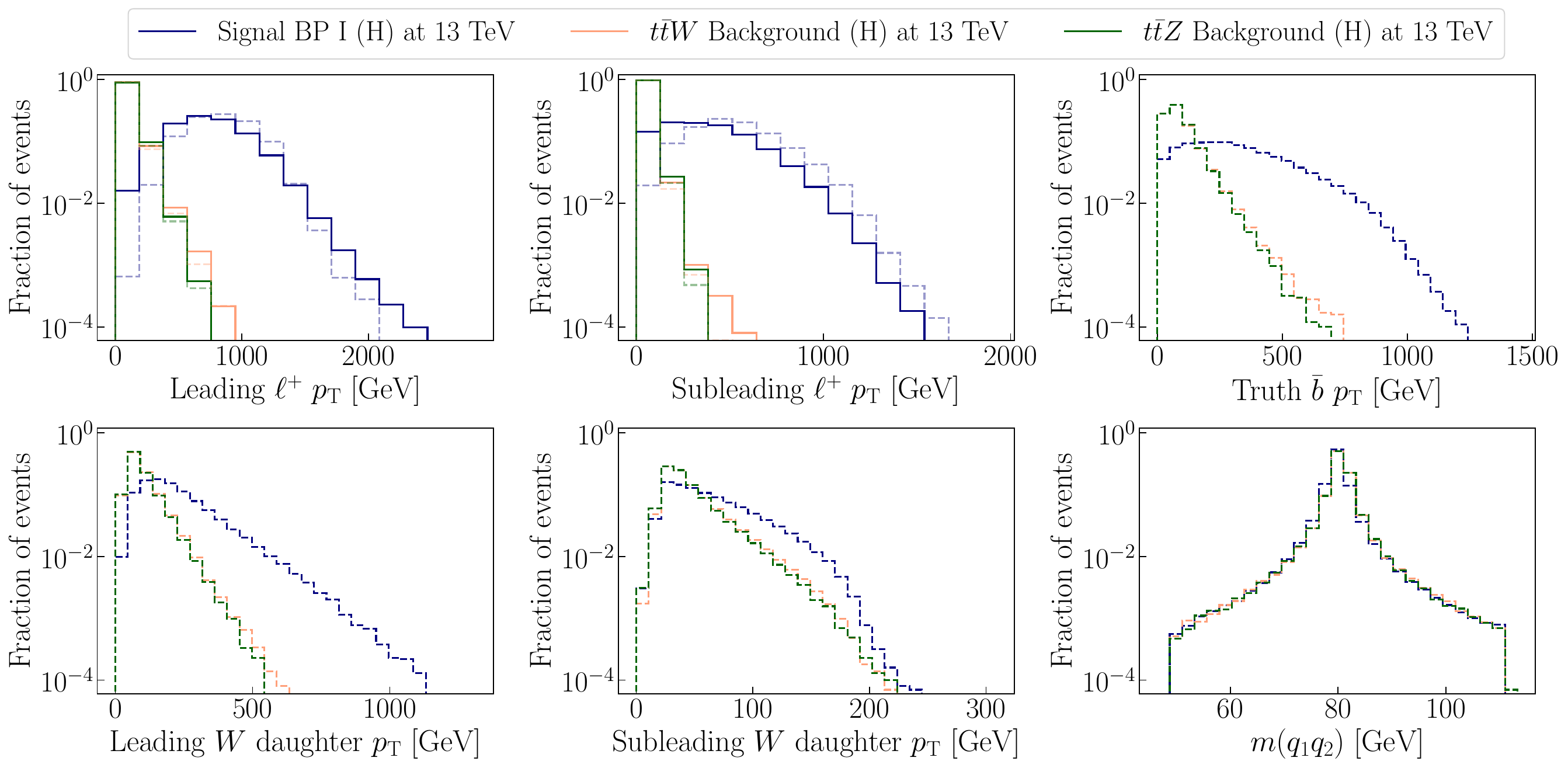}
    }
    
    \subfloat[BP II\label{fig:pT_Distributions_H_BPII}]{
        \includegraphics[width=\textwidth]{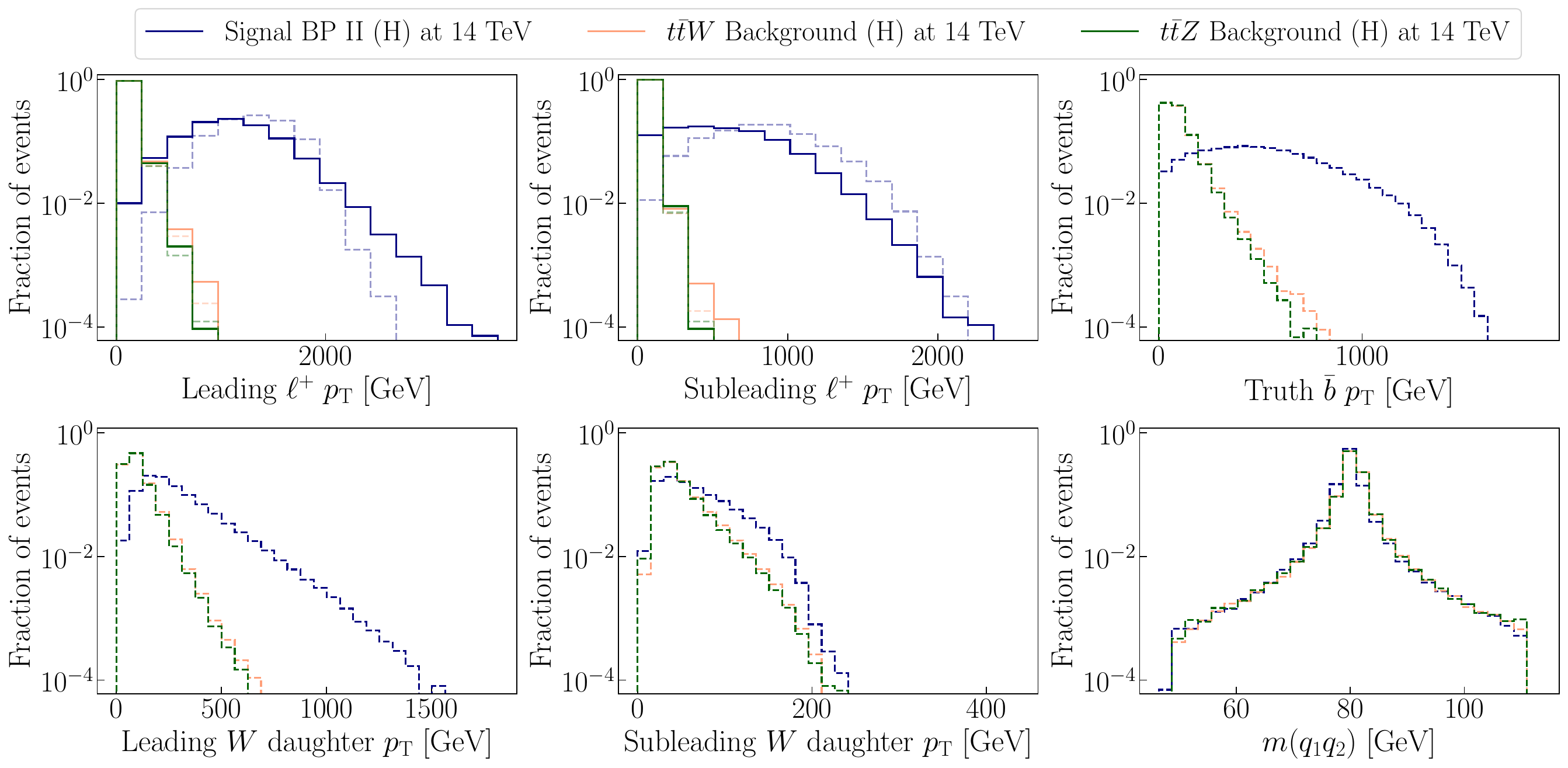}
    }
    \caption{Transverse momentum of all visible final state particles when the anti-top quark decays hadronically. The top two rows correspond to our signal BP I and the SM backgrounds at 13 TeV while the bottom two correspond to our signal BP II and the SM  backgrounds at 14 TeV. On the upper panels of each of these, we show the distribution of charged anti-lepton final states sorted by $p_{\text{T}}$ (left and middle) and of the truth-level $b$-jet produced from the anti-top decay (right). In bottom left and middle panels, we show the $p_{\text{T}}$ distribution of the truth level quark jets produced from the decay of the $W$ boson, itself a daughter particle of the anti-top decay, sorted according to their $p_{\text{T}}$. For completeness, we show the invariant mass distribution of these jets (using their truth level information), with a visible peak at the $W$ mass for both signal benchmarks and the background.}
    \label{fig:pT_DistributionsBPI_BPII_Hadronic}
\end{figure}

\textit{Transverse momentum of hard visible final states:} The main difference with the leptonic case is that there is now one more hard visible final state since, instead of the lepton and the neutrino from the anti-top decay, we have two jets. The general conclusions reached for the leptonic case hold: visible final states carry more transverse momentum compared to background ones, to varying degrees. To illustrate this, we plot the $p_{\text{T}}$ distributions of all our visible final states from our simulated signal and background events in Figure~\ref{fig:pT_Distributions_H_BPI} (signal BP I and backgrounds at 13 TeV) and Figure~\ref{fig:pT_Distributions_H_BPII} (signal BP II and backgrounds at 14 TeV). On the upper left and middle panels of each of these figures, we show the distribution of charged anti-lepton final states sorted by $p_{\text{T}}$. The general situation is unchanged from the leptonic case, meaning that our proposed selection cut, $p_{\text{T}}$ > 50 GeV for the two signal anti-leptons, remains valid. The distribution of the truth-level $b$-jet produced from the anti-top decay is shown in the upper right panel of each figure, with the distribution again remaining fairly similar to the leptonic decay case. The major difference is that we now have two hard jets instead of a single lepton. These are shown in the bottom left and middle panels of each figure. For completeness, we show the invariant mass distribution of these jets (using their truth level information), with a visible peak at the $W$ mass for both signal benchmarks and the background, on the bottom right panels of each figure. \\

\begin{figure}[htbp]
    \centering
    \includegraphics[width=0.486\textwidth]{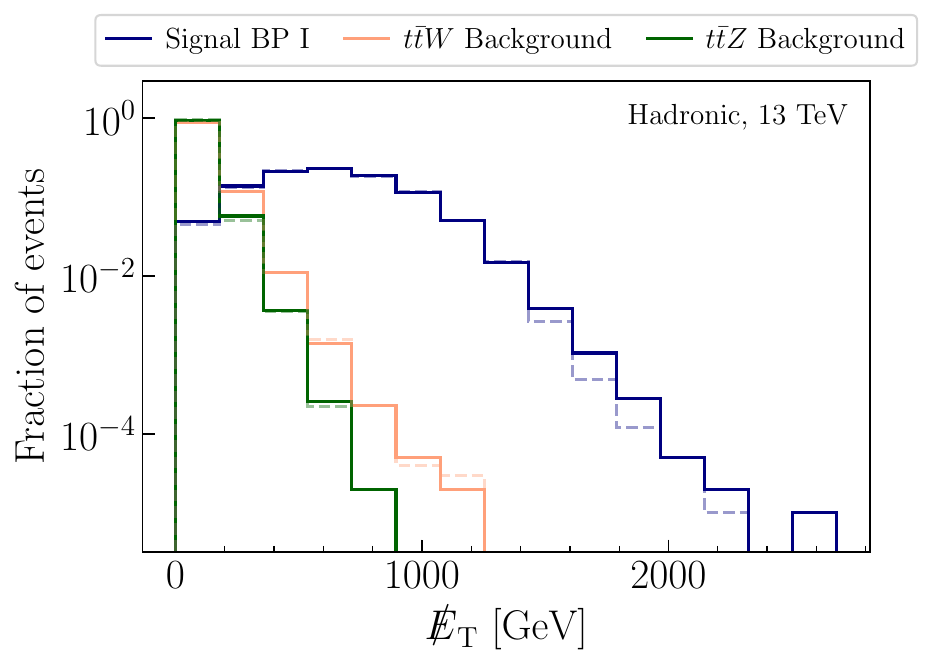}
    \hfill
    \includegraphics[width=0.48\textwidth]{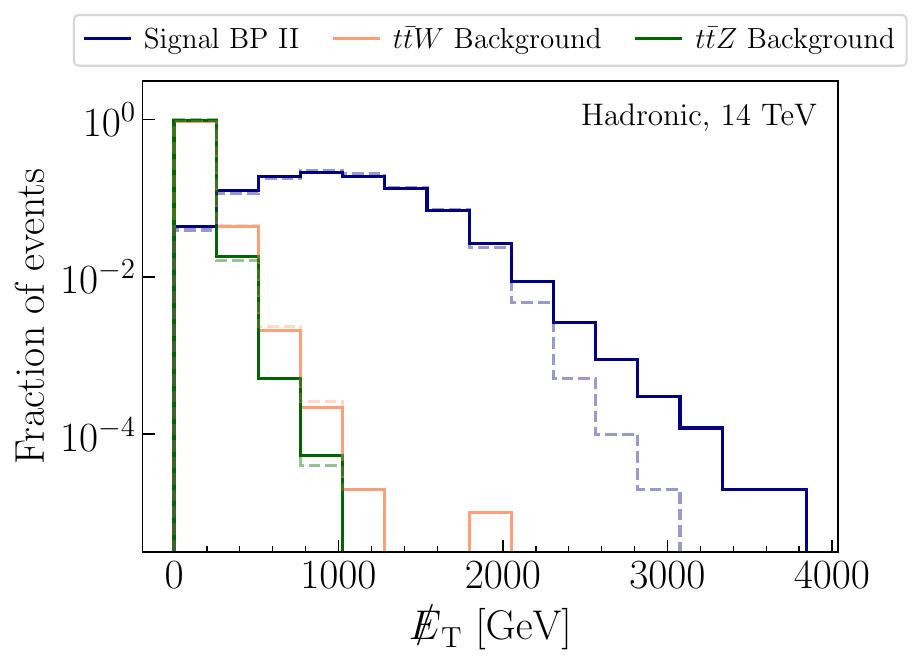}
    \caption{MET distribution for the background compared to signal benchmarks BP I (left) and BP II (right) when the anti-top decays hadronically. There is one fewer neutrino in the final state but the result is essentially unchanged from the leptonic case since the extra neutrino from the anti-top decay in that scenario is much less energetic compared to the one produced from the decay of the dilepton singlet. For easy comparison, we overlay the MET distribution computed from detector (solid) and generator level (dashed) events.}
    \label{fig:ET_Distributions_H}
\end{figure}

\noindent\textit{Missing Transverse Energy:} Since the anti-top decays hadronically, there is one fewer neutrino in the final state, meaning that they carry less energy. However, since the dominant contribution to the latter came from the neutrino produced from the decay of $\chi_{_{LL}}$, this has no significant bearing on the use of MET as a selection cut, as can be seen from Figure~\ref{fig:ET_Distributions_H}. \\

\begin{figure}[htbp]
    \centering
    \includegraphics[width=0.48\textwidth]{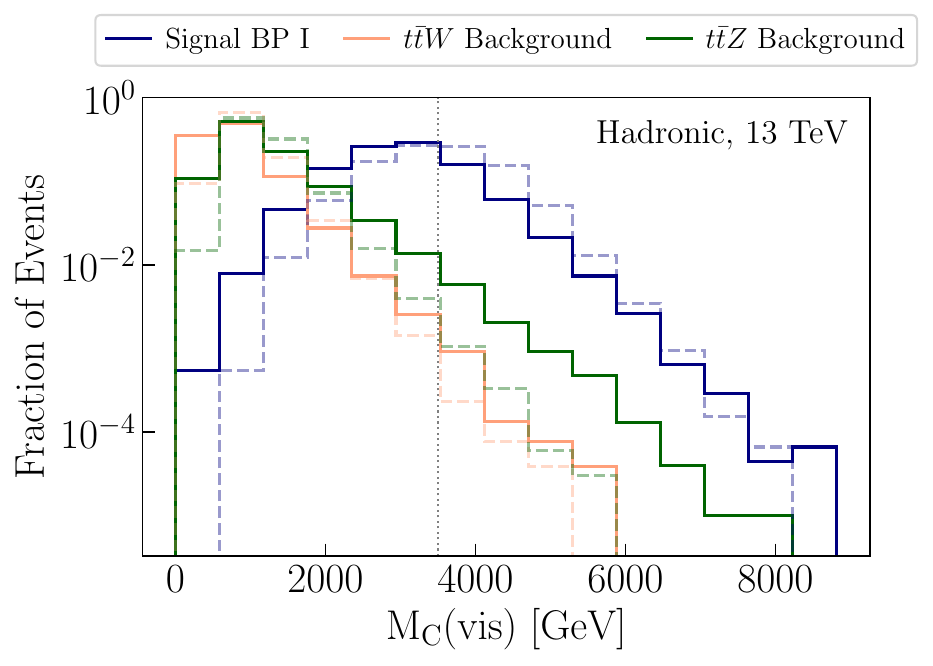}
    \hfill
    \includegraphics[width=0.48\textwidth]{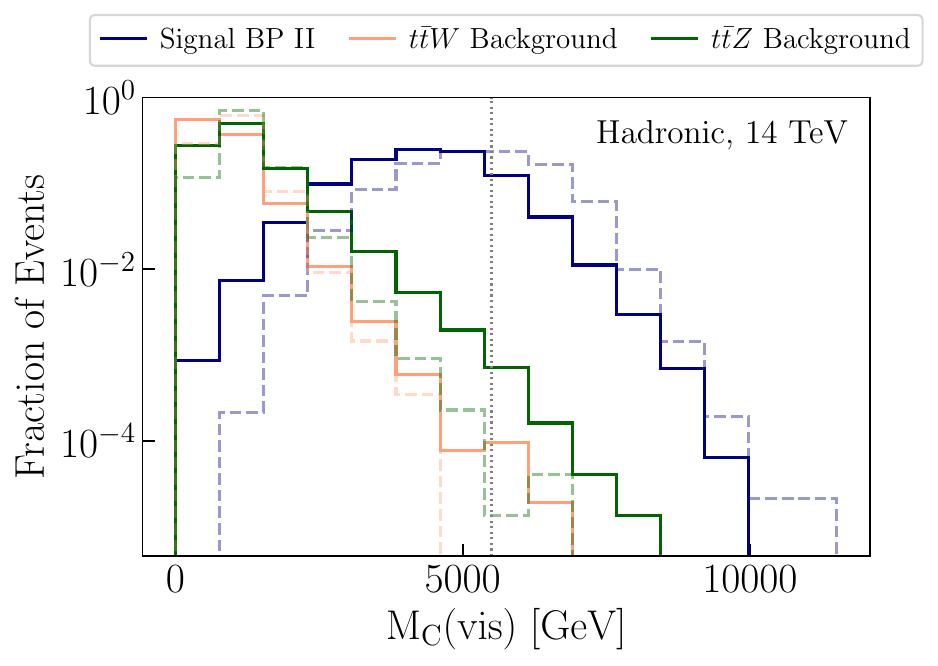}
    \caption{Transverse cluster mass, of all visible final states for BP I (left) and BP II (right) compared to the corresponding SM backgrounds when the anti-top decays hadronically. In both cases, the peak is smeared away from the corresponding diquark mass due to the outgoing neutrinos. It can again be seen that the signal $\text{M}_{\text{C}}\left(\text{vis}\right)$ peaks at a much higher value compared to the background, motivating its use as a selection cut. For comparison purposes, we include both $\text{M}_{\text{C}}\left(\text{vis}\right)$ computed from detector (solid) and generator level (dashed) distributions.}
    \label{fig:M_Cluster_Distributions_Hadronic}
\end{figure}

\noindent\textit{Transverse cluster mass of all visible final states:} As presented in Figure \ref{fig:M_Cluster_Distributions_Hadronic}, the distribution is similar to that of the leptonically decaying case, with the only change being that since the hadronic decay produces two jets instead of a lepton and an anti-neutrino, there is one less neutrino in the final state. However, as with MET, given that the extra neutrino in the leptonic case carried a sub-dominant amount of energy compared to the one produced in the decay of the dilepton singlet, the two $\text{M}_{\text{C}}\left(\text{vis}\right)$ distributions are not noticeably different. 
\subsubsection*{Separation between final state particles} 
\begin{figure}[h!]
    \centering
    \includegraphics[width=0.99\textwidth]{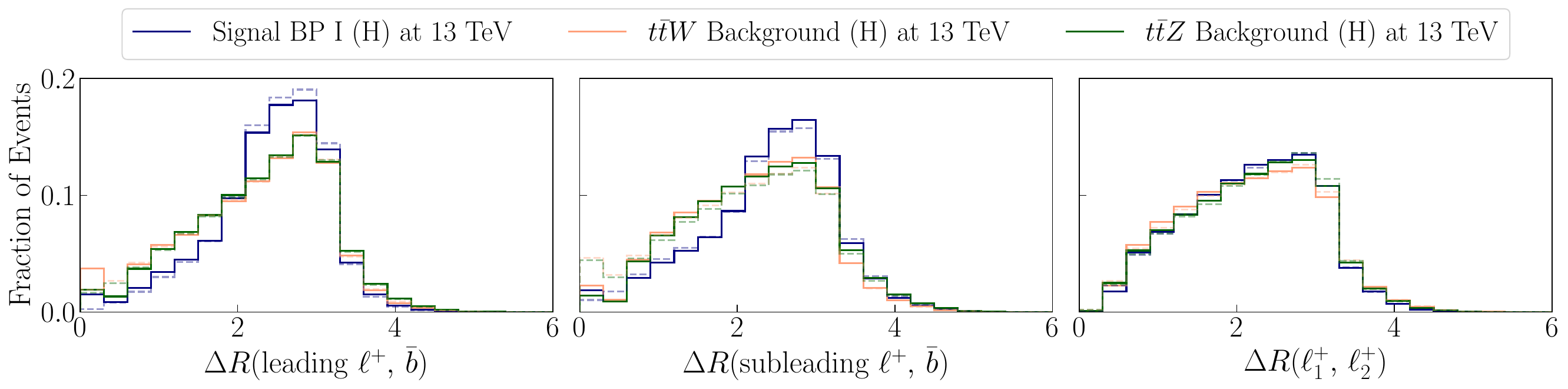}
    \includegraphics[width=0.99\textwidth]{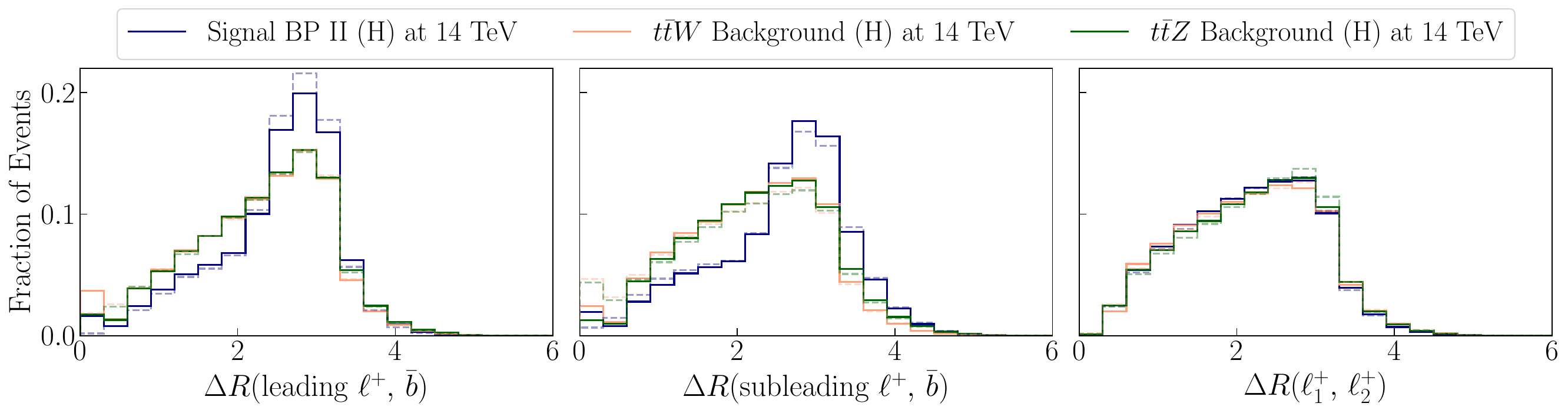}
    \caption{Isolation of the hard anti-leptons in the signal case. Top: $\Delta R$ between the leading and sub-leading anti-lepton (left and middle respectively) and anti-bottom quark. Right: separation between the anti-leptons themselves. The top panels are for our signal BP I and SM $t\overline{t}W$ background at 13 TeV while the bottom ones are for our signal BP II and SM $t\overline{t}W$ background at 14 TeV. The anti-leptons remain very isolated from the anti-top quark and from each other. }
    \label{fig:lep_b_dR_distributions_H}
\end{figure}

\textit{Separation between anti-leptons and the b-tagged jet:} In contrast to the leptonic case, when considering whether the anti-lepton final states are isolated enough from the decay products of the anti-top, we now have, in principle, three jets to consider. In particular, any $\tau^{+}$ particle that is too collimated with one of these jets can be challenging to reconstruct if it decays hadronically. Fortunately, as can be seen from Figure~\ref{fig:lep_b_dR_distributions_H}, the angular separation between the anti-leptons and the reconstructed anti-top, which we take to be the vector sum of the three hard daughter jets (at truth level), is large.  \\

\begin{figure}[h!]
    \centering

    \subfloat[BP I\label{fig:merged_jet_problem_BPI}]{
        \includegraphics[width=0.99\textwidth]{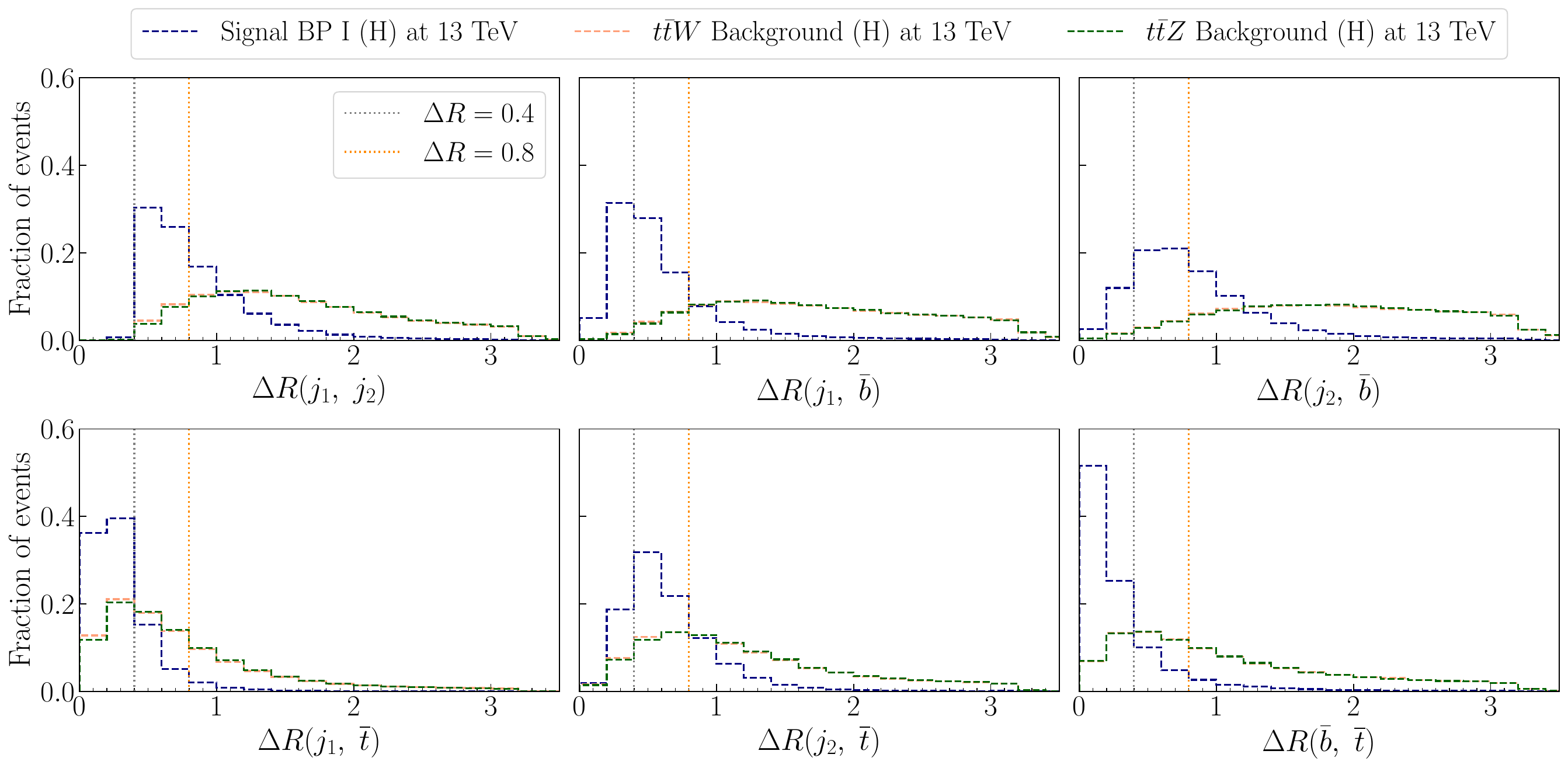}
    }
    \hfill
    \subfloat[BP II\label{fig:merged_jet_problem_BPII}]{
        \includegraphics[width=0.99\textwidth]{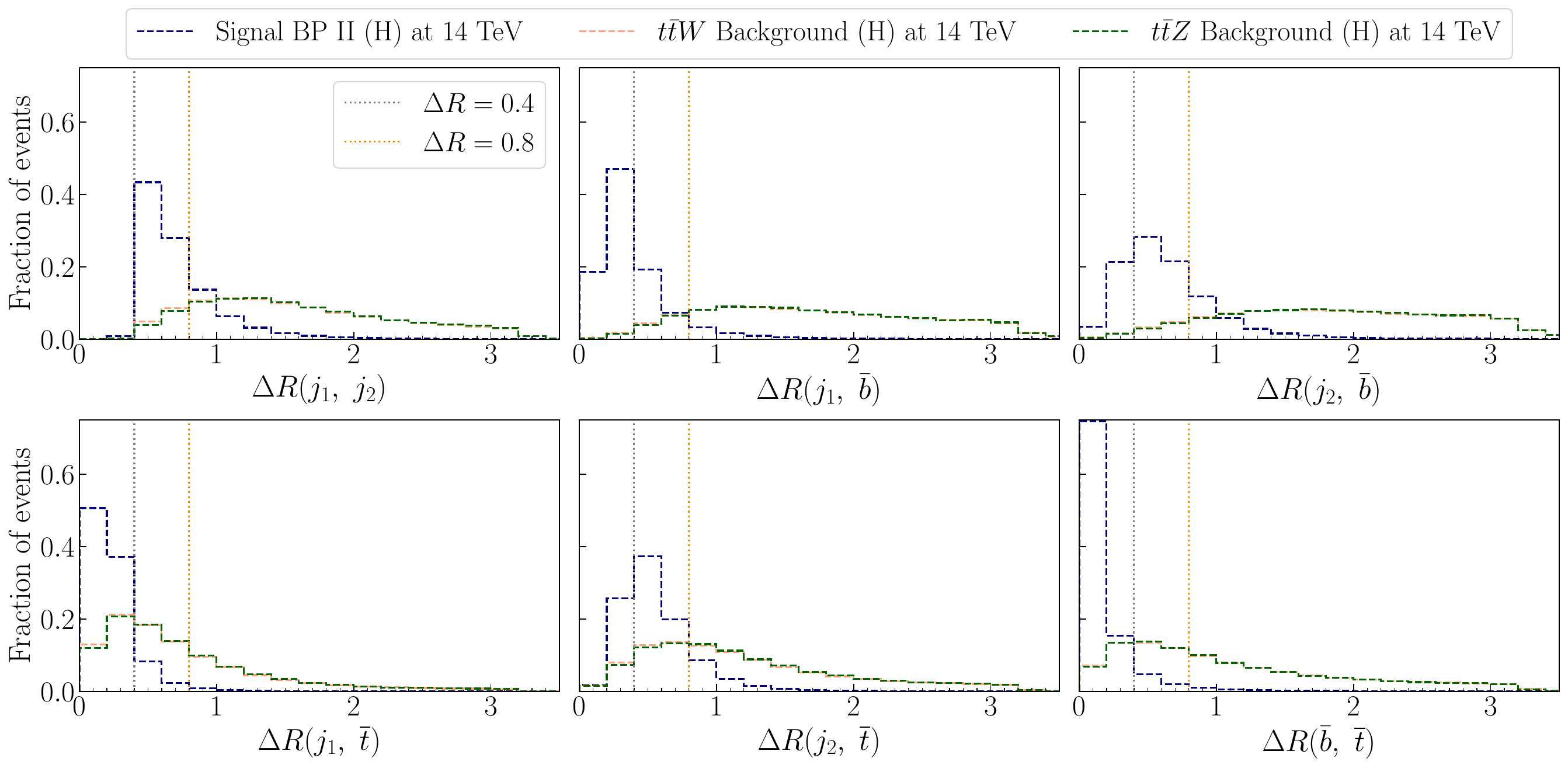}
    }
    \caption{Angular separation of the three jets originating from the hadronic decay of the anti-top. The two upper rows are for the signal BP I and SM $t\overline{t}W$ background at 13 TeV while the lower two are for the signal BP II and SM $t\overline{t}W$ background at 14 TeV. In each sub-figure, the upper left, middle, and right panels are, the separation between the two jets originating from the decay of the $W$ boson $\Delta R\left(j_{1},j_{2}\right)$, the leading jet from the $W$ boson decay and the $b$-jet, $\Delta R \left(j_{1},j_{b}\right)$, and the the sub-leading jet from the $W$ boson decay and the $b$-jet, $\Delta R\left(j_{2},j_{b}\right)$. The lower left, middle, and right panels show the angular separation between the leading jet from the $W$ boson decay and the reconstructed $\bar{t}$, $\Delta R\left(j_{1},\bar{t}\right)$, the sub-leading jet from the $W$ boson decay and the reconstructed $\bar{t}$, $\Delta R\left(j_{2},\bar{t}\right)$, and the $b$-jet and the reconstructed $\bar{t}$, $\Delta R\left(\overline{b},\bar{t}\right)$, respectively.   The merged jet problem can be easily inferred from the observation that the for all the different combinations of $\Delta R$ between the different hadronic decay products of the anti-top quark, the separation in the signal case (for both benchmarks) is much smaller than the background. Additionally, the merging problem worsens when going from BP I (1.5 TeV leptoquark) to BP II (2.5 TeV leptoquark).}
    \label{fig:merged_jet_problem}
\end{figure}

\noindent\textit{The merged jet problem:} Finally, we illustrate the merged jet problem discussed previously. Since our highly boosted anti-top now decays hadronically, instead of a single lepton, we now have two jets that are highly collimated with the $b$-jet, which we label as $j_{1}$ and $j_{2}$ for the leading and sub-leading jets from the decay of the $W$ boson and $j_{b}$ the $b$-jet. We quantify this in Table~\ref{tab:mergedjetsummary}, where we report the fraction of signal and background three jets from the hadronic decay of the anti-top within a cone $\Delta R$ < 0.4 from the reconstructed anti-top axis. The full distributions of the angular separations of the anti-top hadronic decay products are shown in Figures~\ref{fig:merged_jet_problem_BPI} and~\ref{fig:merged_jet_problem_BPII}. 

\begin{table}[htbp]
  \centering
  \small
  \renewcommand{\arraystretch}{1.2}
  \begin{tabular}{@{}lrrrrrr@{}}
    \toprule
    & \multicolumn{2}{c}{\quad\quad$13$ TeV}
    & &\multicolumn{2}{c}{\quad~~$14$ TeV} \\
    \cmidrule(lr){2-4} \cmidrule(l){5-7}
    Fraction of events [\%] & BP I & $t\bar{t}W$ & $t\bar{t}Z$ & BP II & $t\bar{t}W$ & $t\bar{t}Z$\\
    \midrule
    $\Delta R(j_{1},\bar{t})\leq 0.4$
      & 76 & 34 & 32 & 87 & 34 & 33 \\
    $\Delta R(j_{2},\bar{t})\leq 0.4$
      & 20 & 10 & 9 & 27 & 10 & 9 \\
    $\Delta R(j_{b},\bar{t})\leq 0.4$
      & 76 & 20 & 20 & 89 & 22 & 20 \\
    \bottomrule
  \end{tabular}
  \caption{Percentage of simulated signal and background events in
  which each hadronic decay product of the anti-top quark lies within
  $\Delta R\leq 0.4$ of the reconstructed anti-top axis. The jets
  $j_{1}$ and $j_{2}$ are the leading and sub-leading jets from the
  hadronic decay of the $W$ boson, and $j_{b}$ is the $b$-jet.}
  \label{tab:mergedjetsummary}
\end{table}

\subsection{Impact of initial and final state radiation}\label{ssec:ExperimentalChallengers}
\begin{figure}[htbp]
\centering
\includegraphics[width=0.99\textwidth]{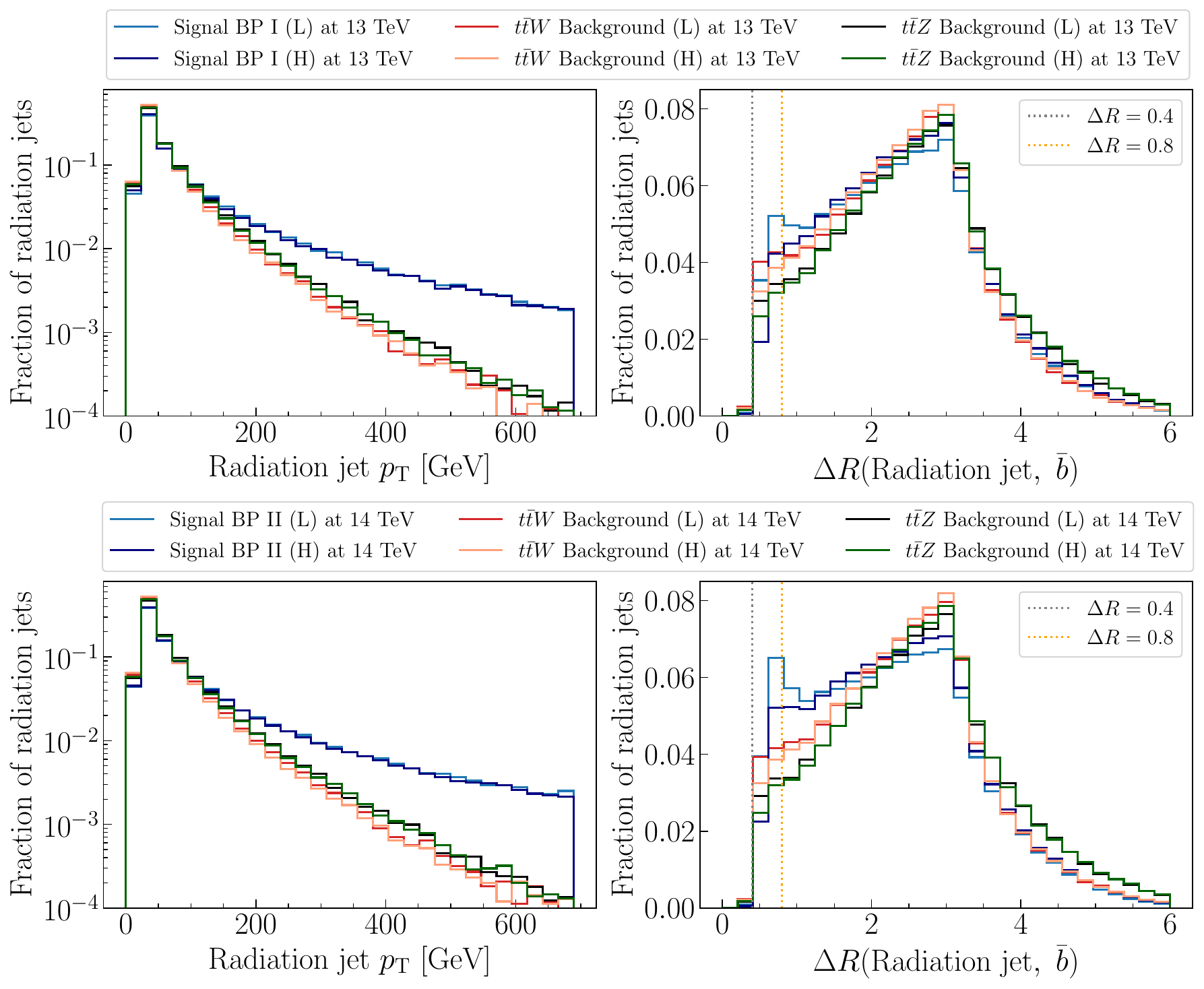}
\caption{Top: $p_{\text{T}}$ distribution of radiation jets for the signal BP I and $t\bar{t}W$ background (left) and the angular separation between the $b$-jet and ISR/FSR jets for both the leptonic and hadronic decay modes of the anti-top.  Bottom: Same, but for BP II. For BP I, 94.1$\%$ (94.4$\%$) of radiation jets are separated from the $b$-jet by $\Delta R$ > 0.8, with the mean being 2.4 (2.4) for the leptonic (hadronic) decay mode while for BP II, 92.2$\%$ (93.4$\%$) of radiation jets are located at $\Delta R$ > 0.8, with the mean separation being 2.3 (2.4) for the leptonic (hadronic) decay mode. The background is similar, with 93.4$\%$ (93.7$\%$) of radiation jets separated by $\Delta R$ > 0.8 from the $b$-jet, with the mean being 2.4 (2.4) at 13 TeV for the leptonic (hadronic) decay mode. At 14 TeV, 93.2$\%$ (93.5$\%$) of radiation jets separated by $\Delta R$ > 0.8 from the $b$-jet, with the mean being 2.5 (2.4) for the leptonic (hadronic) decay mode. The baryon number for the final state of interest, the boosted $b$-jet is therefore not contaminated, in its immediate angular vicinity, by radiation jets. It should be noted here that we have not considered the presence of additional soft jets due to pileup effects and simply assumed that these can be reliably identified.}
\label{fig:ISR_FSR_distributions_BPI_BP2}
\end{figure}

As already mentioned, ISR and FSR contaminate any BNV signal since they affect the baryon number counting of the final states and care must be taken in dealing with them. As already alluded to, we find that the most promising approach is not to impose a larger $p_{\text{T}}$ cut on jets since this risks significantly reducing the signal yield. Instead, we propose to make use of the fact that the final state particles involved in the BNV process originate from heavy resonances are firstly much more boosted compared to the SM backgrounds and, secondly, relatively isolated from ISR and FSR.   This is illustrated in Figure~\ref{fig:ISR_FSR_distributions_BPI_BP2}, where we show the $p_{\text{T}}$ distribution of all radiation jets from our simulations (a single event can have many of them) as well as their angular separation from the anti-bottom quark produced in the anti-top decay. Inspecting the upper and lower left panels, it is evident that any large jet $p_{\text{T}}$ cut will significantly reduce the signal yield. However, from the upper and lower right panels, we see that for both BP I and BP II, ISR and FSR are very isolated from the leading $b$-quark jet. From our simulation results, in BP I, 94.1$\%$ of radiation jets are separated from the $b$ quark by $\Delta R$ > 0.8, with the mean being 2.4. For BP II, 92.2$\%$ of radiation jets are located at $\Delta R$ > 0.8, with the mean separation being 2.3. This is not drastically different from the background. For example, for $t\overline{t}W$, 93.4$\%$ (93.2$\%$) of radiation jets separated by $\Delta R$ > 0.8 from the $b$-quark at 13 TeV (14 TeV). While this offers no avenue for selection cuts, it does provide clarity on an important point for final state baryon number counting: the baryon number of the final state of interest is carried by the boosted $b$-jet and is not contaminated, in its immediate angular vicinity, by radiation jets. However, we note that a real detector will contain, in addition to radiation jets, soft jets due to pileup effects that will need to be efficiently identified and categorized as such.

\newcommand\minisec[1]{\emph{\textbf{#1}}.}
\newcommand{\mr}{m_{\Phi_r}}

\subsection{Final selection cuts}
\label{sec:finalselection}

\begin{table}[htbp]
  \centering
  \footnotesize
  \setlength{\tabcolsep}{4pt}
  \renewcommand{\arraystretch}{1.2}
  \begin{tabular}{@{}lrrrrrr@{}}
    \toprule
    & \multicolumn{3}{c}{$13$ TeV, $139\,\mathrm{fb}^{-1}$}
    & \multicolumn{3}{c}{$14$ TeV, $3000\,\mathrm{fb}^{-1}$} \\
    \cmidrule(lr){2-4} \cmidrule(l){5-7}
    Cut
      & BP I & $t\bar{t}W$ & $t\bar{t}Z$
      & BP II & $t\bar{t}W$ & $t\bar{t}Z$ \\
    \midrule
    Initial events
      & 54.8 & 403 & 34.3
      & 29.6 & $9.64\times10^3$ & 878 \\

    $p_T^{\bar b}>50$ GeV
      & 50.7 & 293 & 25.2
      & 28.2 & $7.03\times10^3$ & 643 \\

    No additional jet within $\Delta R<0.8$
      & 43.6 & 266 & 22.6
      & 23.4 & $6.38\times10^3$ & 577 \\

    $p_T^{\ell^-}>10$ GeV
      & 43.4 & 265 & 22.6
      & 23.3 & $6.36\times10^3$ & 577 \\

    $m(\ell^-,\bar b)<160$ GeV
      & 43.2 & 262 & 14.1
      & 23.2 & $6.29\times10^3$ & 357 \\

    $N_{\ell^+}\geq 2$
      & 26.4 & 105 & 8.71
      & 13.1 & $2.48\times10^3$ & 219 \\

    $\Delta R(\ell^+,\bar b)>0.4$
      & 25.7 & $100$ & 8.58
      & 12.7 & $2.37\times10^3$ & 215 \\

    $p_T^{\ell^+}>50$ GeV
      & 24.6 & 28.0 & 3.15
      & 12.4 & 666 & 79.8 \\

    $\slashed{E}_T>350$ GeV
      & 20.1 & 0.326 & 0.0117
      & 11.4 & 8.96 & 0.369 \\

    $\text{M}_\text{C}(\mathrm{vis})>1500$ GeV
      & -- & -- & --
      & 11.4 & 6.65 & 0.272 \\

    \midrule
    Events after kinematic cuts
      & 20.1 & 0.326 & 0.0117
      & 11.4 & 6.65 & 0.272 \\

    Selection efficiency [\%]
      & 36.7 & 0.0810 & 0.0340
      & 38.4 & 0.0690 & 0.0310 \\

    Tagging weight [\%]
      & 63.0 & 35.0 & 22.5
      & 57.0 & 32.0 & 46.0 \\

    \midrule
    Final selected events (weighted)
      & 12.7 & 0.114 & 0.00262
      & 6.49 & 2.13 & 0.125 \\

    \bottomrule
  \end{tabular}
  \caption{
    Cut-flow table for the leptonic decay channel of the
    anti-top quark mediated by
    $\mathcal{O}^{\mathrm{R}}_{\mathrm{BNV}}$.
    The expected event yields are normalized to
    $139\,\mathrm{fb}^{-1}$ at $\sqrt{s}=13$ TeV
    and $3000\,\mathrm{fb}^{-1}$ at
    $\sqrt{s}=14$ TeV.
    BP I and BP II correspond to representative scalar
    masses of $3.5$ TeV and $5.5$ TeV, respectively.
    The cluster transverse mass requirement
    $\text{M}_\text{C}(\mathrm{vis})>1500$ GeV is imposed only
    for the 14 TeV analysis.
    For the signal, the final weight accounts for
    the $b$-tagging efficiency while for the backgrounds
    it accounts for exactly one of the two b quarks getting tagged as a $b$ jet.
  }
  \label{tab:LeptonicCutFlow}
\end{table}

\begin{table}[h!]
  \centering
  \small
  \renewcommand{\arraystretch}{1.2}
  \begin{tabular}{@{}lrrrrrr@{}}
    \toprule
    & \multicolumn{3}{c}{$13$ TeV, $139\,\mathrm{fb}^{-1}$}
    & \multicolumn{3}{c}{$14$ TeV, $3000\,\mathrm{fb}^{-1}$} \\
    \cmidrule(lr){2-4} \cmidrule(l){5-7}
    Cut & BP I & $t\bar{t}W$ & $t\bar{t}Z$
        & BP II & $t\bar{t}W$ & $t\bar{t}Z$ \\
    \midrule
    Initial events
      & 100 & 898 & 74.6
      & 36.4 & $2.14\times10^4$ & $1.91\times10^3$ \\

    $p_T^{\bar b}>50$ GeV
      & 94.9 & 640 & 53.4
      & 35.5 & $1.53\times10^4$ & $1.37\times10^3$ \\

    $p_T^{j_{1,2}}>10$ GeV
      & 94.7 & 639 & 53.2
      & 35.4 & $1.53\times10^4$ & $1.36\times10^3$ \\

    $p_T^{\bar t}>75$ GeV
      & 94.4 & 532 & 44.3
      & 35.4 & $1.28\times10^4$ & $1.14\times10^3$ \\

    $N_{\ell^+}\geq 2$
      & 55.8 & 189 & 25.0
      & 19.2 & $4.54\times10^3$ & 646 \\

    $\Delta R(\ell^+,\bar b)>0.4$
      & 55.7 & 189 & 25.0
      & 19.2 & $4.53\times10^3$ & 645 \\

    $p_T^{\ell^+}>50$ GeV
      & 53.4 & 59.1 & 9.39
      & 18.6 & $1.42\times10^3$ & 244 \\

    $\slashed{E}_T>350$ GeV
      & 44.2 & 1.25 & 0.0656
      & 17.2 & 32.4 & 1.29 \\

    No additional jet within $\Delta R(j,\bar t)\leq 0.8$
      & 40.3 & 1.10 & 0.0604
      & 15.2 & 30.0 & 1.11 \\

    $M_C^{\mathrm{vis}}>1500$ GeV
      & -- & -- & --
      & 15.2 & 20.8 & 0.902 \\

    \midrule
    Events after kinematic cuts
      & 40.3 & 1.10 & 0.0604
      & 15.2 & 20.8 & 0.902 \\

    Selection efficiency [\%]
      & 40.2 & 0.123 & 0.0810
      & 41.9 & 0.0970 & 0.0473 \\

    Exactly-one-$b$-tag probability [\%]
      & -- & 48.8 & 49.5
      & -- & 51.2 & 72.5 \\

    Top identification efficiency [\%]
      & 50.0 & 50.0 & 50.0
      & 50.0 & 50.0 & 50.0 \\

    \midrule
    Final selected events (weighted)
      & 20.2 & 0.269 & 0.0150
      & 7.62 & 5.33 & 0.327 \\

    \bottomrule
  \end{tabular}
  \caption{
  Cut-flow table for the Hadronic decay channel of the
  anti-top quark mediated by
  $\mathcal{O}^{\mathrm{R}}_{\mathrm{BNV}}$.
  The expected event yields are normalized to integrated
  luminosities of $139\,\mathrm{fb}^{-1}$ at
  $\sqrt{s}=13$ TeV and $3000\,\mathrm{fb}^{-1}$ at
  $\sqrt{s}=14$ TeV.
  For the signal, the assumed $50\%$ anti-top
  identification efficiency includes $b$-tagging.
  For the background, the event-dependent probability
  of tagging exactly one $b$ quark is applied separately,
  followed by a $50\%$ anti-top reconstruction efficiency
  conditional on the tagging requirement.
  The final event yields include these efficiencies.
  }
  \label{tab:HadronicCutFlow}
\end{table}

The kinematic features identified above motivate the following event selections, in addition to the previously defined basic acceptance ones. Firstly, candidate events when the anti-top quark decays leptonically must satisfy the following criteria

\begin{enumerate}
 
  \item Exactly one $b$-tagged jet with $p_{\text{T}} > 50$~GeV. 
  
  \item At least one additional charged lepton ($e^{-}$ or
  $\mu^{-}$) with $p_{\text{T}} > 10$~GeV. 
  
  \item  The \(b\)-tagged jet and negatively charged lepton must be identified as the decay products of a single anti-top quark, as determined by an appropriate anti-top reconstruction procedure.
 
  \item Two same-sign anti-leptons, $e^{+}$, $\mu^{+}$, or a positively charged $\tau$-tagged jet, all with $p_{\text{T}} > 50$~GeV. 
  
  \item To suppress contamination from secondary positively charged leptons produced in heavy-flavor decays, we require \(\Delta R(\ell_i^+,j_b)>0.4\) for both positively charged leptons, where $i= e,\mu,\tau$.

  \item Missing transverse energy:
  $\slashed{E}_{\text{T}} \geq 350$~GeV.

  \item We further veto additional radiation jets within \(\Delta R<0.8\) of the \(b\)-tagged jet. Since the primary \(b\)-jet is reconstructed using the anti-\(k_t\) algorithm with radius parameter \(R=0.4\), this requirement primarily constrains additional resolved jets in the region \(0.4<\Delta R<0.8\).
 
  \item Transverse cluster mass (only for higher luminosities):
  $\text{M}_{\text{C}}(\text{vis}) \geq 1500$~GeV. 

\end{enumerate}
When the anti-top decays hadronically we require that selected candidate events must consist of 

\begin{enumerate}
  \item A hadronic anti-top candidate, reconstructed in one of three
configurations depending on the degree of boost: three resolved jets
including a $b$-tagged one, two jets, including a
$b$-tagged jet when the $W$ decay products merge, or a single large-radius jet containing a $b$-tagged sub-jet when all three decay products merge. 

\item The reconstructed (anti) top quark must have a $p_{\text{T}}$ > 75 GeV. 
   
  \item Two same-sign anti-leptons, $e^{+}$, $\mu^{+}$, or a positively charged $\tau$-tagged jet, all with $p_{\text{T}} > 50$~GeV. 

  \item To suppress contamination from secondary positively charged leptons produced in heavy-flavor decays, we require \(\Delta R(\ell_i^+,j_b)>0.4\) for both positively charged leptons, where $i= e,\mu,\tau$.

  \item Missing transverse energy:
  $\slashed{E}_{\text{T}} \geq 350$~GeV. 

  \item We further require that no additional radiation jet be present within an angular separation of ($\Delta R\leq$ 0.8) from the reconstructed anti-top quark, excluding jets associated with its reconstructed decay products. For anti-top candidates reconstructed using the anti-($k_t$) algorithm with radius parameter $R$ = 0.4, this requirement primarily suppresses additional resolved radiation in the vicinity of the anti-top candidate, outside the jets used for its reconstruction. The precise implementation depends on whether the anti-top decay products are fully resolved or partially or completely merged. 
 
  \item Transverse cluster mass (higher luminosities only):
  $\text{M}_{\text{C}}(\text{vis}) \geq 1500$~GeV. 

\end{enumerate}

We summarize the impact of these selection cuts on our signal benchmarks for the ``$ul$-type'' leptoquark (generating the $\mathcal{O}^\text{R}_{\text{BNV}}$ operator) in Tables~\ref{tab:LeptonicCutFlow} and~\ref{tab:HadronicCutFlow} for the leptonic and hadronic anti-top decay channels respectively. Since the kinematics are not sensitive to the chiral structure of the leptoquark coupling, similar results are expected for the ``$LQ$-type'' leptoquark. Our analysis employs a hybrid approach combining generator-level information from \textsc{Pythia}~8 with detector-level reconstruction from \textsc{Delphes}~3. Specifically, the decay products of the anti-top quark are identified using generator-level event records, whereas the two positively charged leptons originating from the heavy scalar decays are selected using detector-level objects reconstructed by \textsc{Delphes}~3. The latter is also used to implement reconstruction and identification efficiencies of $e^{+}$, $\mu^{+}$, and hadronically decaying $\tau^{+}$. The $b$-tagging efficiency, on the other hand, is accounted for separately using the transverse-momentum-dependent parametrization used in \textsc{Delphes}~3 but applied to the generator level $b$-quark,
\begin{equation}\label{eq:bjettag}
\epsilon_b(p_{\text{T}})= \frac{21.25\tanh(0.0025p_{\text{T}})}
{1+0.063p_T}.    
\end{equation}

Since a full reconstruction of highly boosted, leptonically decaying anti-top quarks is beyond the scope of this work, we adopt an approximate procedure to estimate the effects of $b$-tagging and anti-top identification. For the leptonic channel, we require the negatively charged lepton and candidate $b$ quark to satisfy $m(\ell^-,b)<160$ GeV, slightly above the kinematic endpoint of the corresponding invariant-mass distribution in top-quark decay. We then assign each candidate event a weight determined by the transverse-momentum-dependent $b$-tagging efficiency, $\epsilon_b(p_T)$. \\

For the $t\bar tW$ background, both the $b$ and $\bar b$ quarks must be considered as potential tagged candidates. For the leptonic decay channel, requiring exactly one tagged $b$ quark, which must also satisfy the invariant-mass condition, gives the event weight
\[
w_{\mathrm{BG}}=
\epsilon_1 I_1(1-\epsilon_2)
+\epsilon_2 I_2(1-\epsilon_1),
\]
where $\epsilon_i=\epsilon_b(p_{T,i})\Theta(p_{T,i}-50~\mathrm{GeV})$ and $I_i=\Theta(160~\mathrm{GeV}-m(\ell^-,j_{b_i}))$. The expected event yields are obtained by summing these weights over events passing the preceding selection cuts. For the hadronic channel, we adopt a flat anti-top identification efficiency of 50$\%$ for signal events, incorporating the assumed $b$-tagging efficiency. For background events, we explicitly evaluate the event-dependent probability of tagging exactly one $b$ quark and subsequently apply a flat 50$\%$ anti-top reconstruction efficiency, treated as conditional on satisfying the $b$-tagging requirement. \\ 

\begin{figure*}[h!]
    \centering
    \includegraphics[width=0.99\textwidth]{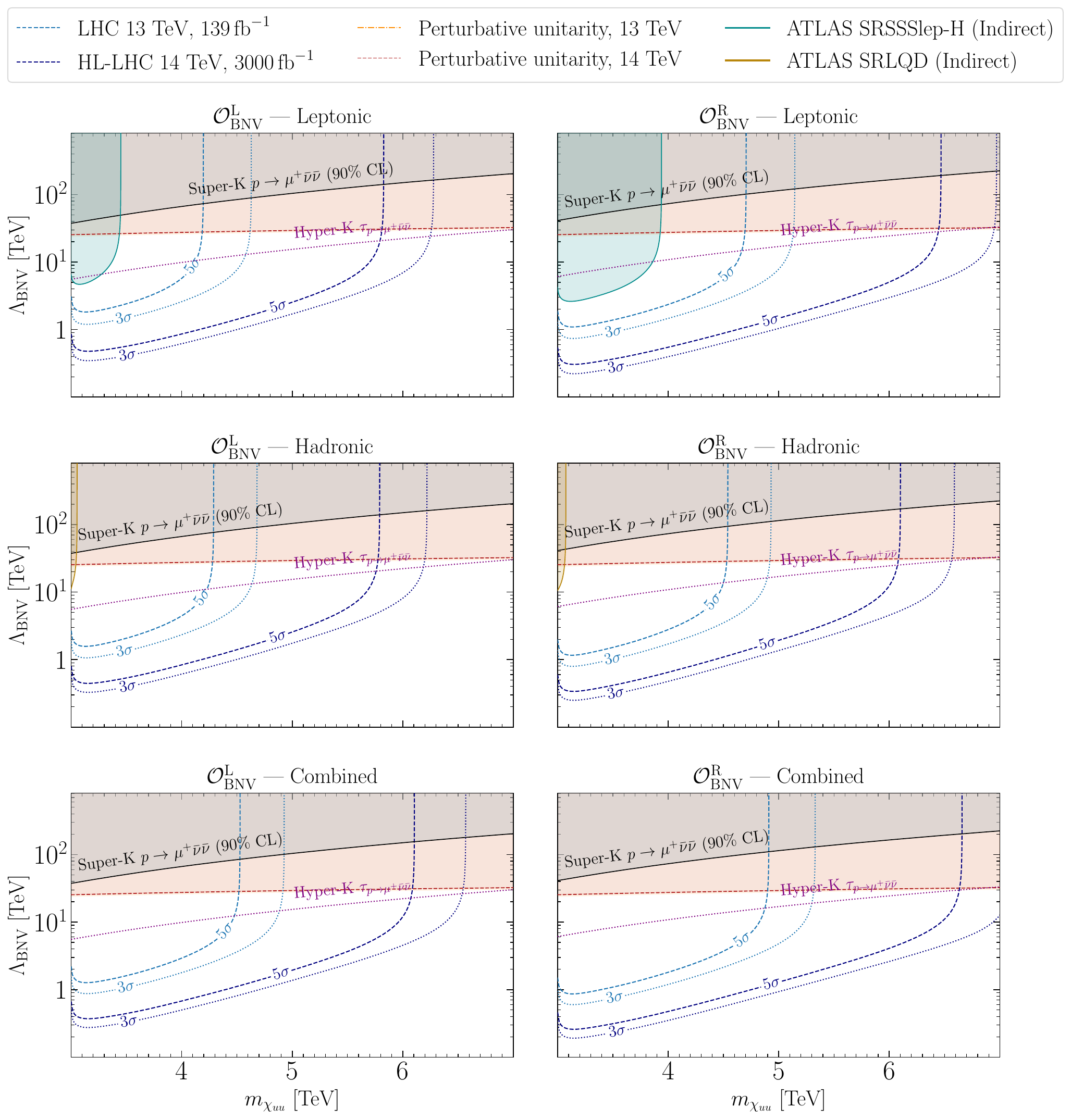}
    \caption{
    Projected LHC sensitivity to the baryon-number-violating
    operators $\mathcal{O}^{\mathrm{R}}_{\mathrm{BNV}}$
    (right column) and
    $\mathcal{O}^{\mathrm{L}}_{\mathrm{BNV}}$
    (left column) as a function of the diquark mass
    $m_{\chi_{uu}}$ and the trilinear coupling scale
    $\Lambda_{\mathrm{BNV}}$. We fix $y^{uc}_{uu} = 0.25$ and $m_{\chi_{LQ}} = m_{\chi_{LL}}$ = 1.5 TeV.
    The upper, middle, and lower rows correspond to the
    Leptonic, Hadronic, and Combined anti-top decay
    channels, respectively.
    The dotted and dashed contours indicate the
    projected $3\sigma$ and $5\sigma$ sensitivities
    at the LHC ($\sqrt{s}=13$ TeV,
    $139\,\mathrm{fb}^{-1}$) and HL-LHC
    ($\sqrt{s}=14$ TeV,
    $3000\,\mathrm{fb}^{-1}$), respectively.
    The solid curves represent the existing
    Super-Kamiokande constraints, while the dotted
    proton-decay curves indicate the projected sensitivity measurable at the
    Hyper-Kamiokande experiment from \cite{Dev:2022jbf} for muon channel only since, given our model choice to remain consistent with lepton flavour violation constraints, the electron channel constraints are always weaker.
    Perturbative unitarity constraints are shown in  orange and red for $\sqrt{s} = 13$ TeV and $\sqrt{s} = 14$ TeV respectively. 
    The Combined sensitivities are obtained by combining
    the statistical significances of the Leptonic and
    Hadronic channels, assuming statistically independent
    event samples. 
    }
    \label{fig:BNV_Collider_ProtonDecay_Reach}
\end{figure*}

The cut flow analysis for our signal benchmark points and background are studied in Tables~\ref{tab:LeptonicCutFlow} and~\ref{tab:HadronicCutFlow} respectively. Using the selection efficiencies obtained from these same selection cuts, we determine the projected sensitivities of the LHC and HL-LHC to the BNV operators $\mathcal{O}^{\mathrm{R}}_{\mathrm{BNV}}$ and $\mathcal{O}^{\mathrm{L}}_{\mathrm{BNV}}$ and show the results in Figure~\ref{fig:BNV_Collider_ProtonDecay_Reach}. \\

We consider two collider benchmarks: $\sqrt{s}=13$ TeV with an integrated luminosity of $139\,\mathrm{fb}^{-1}$, corresponding to the LHC Run 2 dataset, and $\sqrt{s}=14$ TeV with an integrated luminosity of $3000\,\mathrm{fb}^{-1}$, corresponding to the projected dataset accumulated over ten years of HL-LHC operation. For each operator and decay channel, the signal cross sections and selection efficiencies are evaluated as functions of the diquark mass $m_{\chi_{uu}}$, with the relevant tagging and reconstruction efficiencies incorporated as described above. \\

To quantify the discovery sensitivity, we employ the Asimov significance for a Poisson counting experiment with an assumed known background expectation, following Ref.~\cite{Cowan:2010js}. For a signal yield $S$ and background yield $B$, the median discovery significance is
\begin{equation}
Z_A
=
\sqrt{
2\left[
(S+B)\ln\left(1+\frac{S}{B}\right)-S
\right]
},
\label{eq:BNVAsimovSignificance}
\end{equation}
corresponding to the background-only hypothesis tested against the signal-plus-background expectation. This expression retains the Poisson nature of the event counts and does not rely on the Gaussian approximation $S/\sqrt{B}$, which is inappropriate in the low-background regime relevant to our analysis. We use the final background expectations after all selection and identification efficiencies listed in Tables \ref{tab:LeptonicCutFlow} and \ref{tab:HadronicCutFlow}, combining $t\overline{t}W$ and $t\overline{t}Z$ events in quadrature at the aforementioned center of mass energies and integrated luminosities. \\

We also show a combined sensitivity by treating the Leptonic and Hadronic selections as statistically independent event categories. Under this assumption, the joint likelihood is the product of the individual Poisson likelihoods, and their Asimov log-likelihood-ratio test statistics are additive. Consequently,
\begin{equation}
Z_A^{\mathrm{Combined}}
=
\sqrt{
\left(Z_A^{\mathrm{Lep}}\right)^2
+
\left(Z_A^{\mathrm{Had}}\right)^2
}.
\label{eq:BNVCombinedSignificance}
\end{equation}

It is important to mention here that the sensitivities we present are statistical projections with fixed background expectations. In particular, uncertainties associated with the simulated backgrounds, detector modeling, and correlated systematic effects are not included. The resulting $3\sigma$ and $5\sigma$ sensitivity contours are shown in Figure~\ref{fig:BNV_Collider_ProtonDecay_Reach} as a function of  $(m_{\chi_{uu}}$ and $ \Lambda_{\mathrm{BNV}})$. The upper two rows present the individual Leptonic and Hadronic channels, while the lower row shows their combined sensitivity, separately for $\mathcal{O}^{\mathrm{R}}_{\mathrm{BNV}}$ and $\mathcal{O}^{\mathrm{L}}_{\mathrm{BNV}}$. We also overlay the current Super-Kamiokande constraints (black) and projected Hyper-Kamiokande sensitivities (purple) from the three-lepton proton-decay modes, together with the perturbative-unitarity boundaries. The green and gold shaded regions are excluded using indirect constraints from the same sign dilepton search performed by the ATLAS experiment \cite{ATLAS:2023afl}, where, for consistency with their analysis, we consider only muons and electron final states and adopt their prescribed selection cuts. Specifically, for the leptonically decaying anti-top, we use their SRSSSlep-H limit, with the constraints shown in the green contour, while for the hadronically decaying limit we take their SRLQD limit, with the constraints shown in gold. We do not show the combined results since the two signal regions are not independent. Given that both regions explicitly require zero $b$-tagged jets, in both cases we multiply the cross-section after selection cuts with the weighted mis-tagging probability using \cref{eq:bjettag}. Assuming our mistagged $b$-jets is one of the two $p_{\text{T}} >$ 60 GeV jets required for the SRSSSlep-H region, the other one must come from an ISR or an FSR jet, which significantly reduces the signal efficiency. In the SRQLD region, the requirement of five jets with $p_{\text{T}} > 50$ GeV significantly reduces the efficiency of our hadronic signal much more compared to the leptonic case since, even assuming that three of them come from an untagged anti-top quark, we still require two high $p_{\text{T}}$ ISR or FSR jets. \\

These comparisons illustrate the complementarity between direct searches for $B - L/3$ type BNV interactions at hadron colliders and proton decay experiments. In particular, as our benchmark illustrates, even after taking into account indirect constraints from the latest ATLAS same sign dilepton searches, there are complementary regions of parameter space where the LHC and Hyper-K could be used in tandem. For diquark masses in the 3.5 -- 4 TeV mass range and $\Lambda_{\text{BNV}}$ in the 5 -- 20 TeV range, the benchmark model can be probed at both the LHC and Hyper-K. Notably, its BNV signals can be observed at 5$\sigma$ sensitivity even using the LHC Run 2 data, with the latter then predicting complementary trilepton proton decay events at an observable rate over a ten year period at Hyper-K. For diquark masses in the same range as above and $\Lambda_{\text{BNV}}$ in the 1 -- 5 TeV range, this model is beyond the sensitivity of Hyper-K, while staying within reach of the LHC.

\section{Generalizations Beyond Our Representative Model}\label{sec:Generalizations}

The selection cuts of Section~\ref{sec:collidermain}, together with the tagging and reconstruction efficiencies, suppress the SM background to an overall selection efficiency of approximately $0.016-0.042\%$ across the leptonic and hadronic channels. In contrast, the signal efficiencies remain approximately $21-25\%$, demonstrating the effectiveness of our selection criteria in discriminating our BNV signal from the SM background. While our study focused on a representative minimal model, these cuts, particularly those on the anti-lepton transverse momenta, MET, and transverse cluster mass, exploit general features of the event topology rather than the specific details of our model,
\begin{equation*}
 pp \to X \to A\,B \to (f_1 f_2)(f_3 f_4),   
\end{equation*}
where an $s$-channel resonance $X$ 
decays to two heavy states $A$ and $B$, each of which decays to a pair of SM 
fermions. In the $B - L/3$ case this topology corresponds to 
the dimension-9 operators with six external fermions that are UV-completed by 
three heavy bosons meeting at a single vertex (for more details, see Tables~19 and~20 of~\cite{Heeck:2026dmh}, with the particle quantum numbers of Tables~4 and~5). 
Requiring that $X$ be a diquark to allow for $qq$ $s$-channel production at the LHC, that $A$ be a leptoquark and $B$ be a dilepton,\footnote{$A$ and $B$ are not required to be $SU(2)_{\text{L}}$ singlets} that the 
operator be hypercharge neutral, and that the central vertex be an 
$SU(2)_L$ singlet and be renormalizable ($SSS$ or $SVV$), we find that, of the seven models listed in~\cite{Heeck:2026dmh}, five 
share the same topology as that of our representative model, as listed in Table~\ref{tab:models}.

\begin{table}[htbp]
  \centering
  \small
  \renewcommand{\arraystretch}{1.2}
  \begin{tabular}{@{}lccc@{}}
    \toprule
    Model & $X$ (diquark) & $A$ (leptoquark) & $B$ (dilepton) \\
    \midrule
    Ours              & $S_1$ & $S_2$ & $S_{14}$ \\
    Triplet extension & $S_1$ & $S_4$ & $S_9$ \\
    Vector 1          & $S_1$ & $V_2$ & $V_{12}$ \\
    Vector 2          & $V_1$ & $V_2$ & $S_{14}$ \\
    Vector 3          & $V_1$ & $V_2$ & $S_9$ \\
    \bottomrule
  \end{tabular}
  \caption{Models sharing the $s$-channel topology
  of Fig.~\ref{fig:topology}, with $S_i$ and $V_i$'s being new particles denoted as in \cite{Heeck:2026dmh}.}
  \label{tab:models}
\end{table}

All five involve an $s$-channel diquark resonance that 
decays to a leptoquark--dilepton pair, leading to four SM fermions. Our choice of an anti-top, two charged anti-leptons and an unavoidable anti-neutrino in the final state is then the natural set of final states to optimize the determination of $\Delta B$ and $\Delta L$. Differences between the models enter only 
through the production cross section, the branching ratios, and the 
$SU(2)_{\text{L}}$ multiplet structure, which changes the charges of the intermediate 
states and the number of charge combinations in the final state. These 
change the signal normalization but are not expected to significantly affect the kinematic distributions that 
the cuts exploit. Our selection cuts 
are therefore efficient against the background and relatively insensitive to these 
differences. In particular, the kinematic thresholds adopted for the anti-lepton $p_{\text{T}}$, MET, and transverse cluster mass cuts were set conservatively and are expected to produce similar efficiencies for TeV-scale heavy resonance masses. We therefore expect them to be applicable to the four remaining models from Table~\ref{tab:models}, although the precise signal efficiency would have to be computed for each. From an experimental perspective, a model-independent search targeting this distinctive final-state signature, with our selection cuts applied without imposing assumptions about the underlying mediator representations, offers an especially appealing approach. The resulting experimental constraints can subsequently be recast within each of the corresponding models. Let us note that two further models, $S_2V_2V_2$ and $S_4V_2V_2$ can produce the same SM final states but, since there is no diquark in these theories, the heavy states are produced 
in a VBF-like topology with different kinematics. Our cuts therefore do not carry over in their entirety and we leave a dedicated study of these topologies to future work.

\section{Conclusion}\label{sec:conclusion}
In this work, we have investigated the phenomenology of theories featuring $B-L/3$ conserving BNV interactions.
Our results show that the physics responsible for baryon number violation need not lie far beyond the reach of colliders. For $B-L/3$ conserving interactions and other similar models, it may reveal itself at the LHC and in nucleon-decay experiments alike.\\

From a low-energy perspective, we have shown, using a reasonable projected branching ratio for charm baryon decays and lower limits on the proton lifetime under trilepton decay measured by Super-K, that although the former is predicted at tree level and the latter at loop level, proton decay provides a significantly stronger probe of this class of BNV interactions. Despite the high dimensionality of the effective operators, the existing Super-Kamiokande limits on trilepton nucleon decay translate into constraints of $\mathcal{O}(1-30)$ TeV on the BNV mediator mass scale, depending on the underlying flavour structure and couplings. Importantly, these constraints leave open the possibility of directly probing the associated new particles at the LHC, highlighting a promising avenue for complementarity between collider and nucleon-decay experiments. \\

Motivated by this complementarity, we investigated the collider phenomenology of a representative ultraviolet completion, focusing on the distinctive final state consisting of an anti-top quark, two charged anti-leptons, and missing transverse energy. We developed selection strategies for both the leptonic and hadronic anti-top decay channels and evaluated their individual and combined discovery sensitivities at the LHC and HL-LHC. Our results demonstrate how direct collider searches can complement existing and projected proton-decay constraints, probing regions of parameter space consistent with perturbative unitarity and the relevant low-energy bounds. \\

Several aspects of our work deserve further experimental attention. Firstly, an updated Super-Kamiokande analysis of trilepton nucleon decay would be particularly valuable, incorporating the additional exposure accumulated since the publication of the previous limits more than a decade ago \cite{Super-Kamiokande:2014pqx}. Secondly, with the Hyper-Kamiokande experiment \cite{Hyper-Kamiokande:2018ofw} approaching its operational phase, a dedicated search for trilepton proton decay should be considered a priority. In parallel, a dedicated LHC search targeting the BNV final-state signature identified here, using the selection criteria we have developed, would provide an independent and complementary probe. More broadly, a signature-based search would allow the resulting experimental limits to be interpreted across a wider class of BNV models, extending beyond the specific ultraviolet completion studied in detail here. From a more theoretical perspective, it would be interesting to explore grand unified theories that feature our model as an intermediate theory between the electroweak and GUT scales, as originally envisioned by Weinberg \cite{Weinberg:1980bf}. 

\begin{acknowledgments}
 We are especially grateful to Takumi Aoki for identifying recent same sign dilepton searches relevant for our proposed signal, pointing us to Hyper-K projections for trilepton nucleon decay, and feedback on the draft. We would also like to thank to Lara Dodd, Claudia Merlassino, and Laura Pintucci for very helpful discussions on boosted top quark reconstruction and  acknowledge Joseph Boudreau, Tao Han, Julian Heeck, James Mueller, Diana Sokhashvili, and Anil Thapa for helpful correspondence and feedback on the draft. We acknowledge the use of AI tools, namely ChatGPT of OpenAI and Claude of Anthropic, to develop the Python routines used in deriving the selection cuts of Section \ref{sec:collidermain}, to perform supplementary consistency checks on the derivations of the proton and charm baryon decay rates, and for assistance in writing and proofreading the final draft. We have verified all the results presented in this work and are responsible for any errors. The work of DS is supported by NSF Grant No.~PHY-2210361, by the Maryland Center for Fundamental Physics, and the Mitchell Institute for Fundamental Physics and Astronomy. TO is supported by the DOE Office of Science Distinguished Scientist Fellows Award 2022. The work of AB is supported by the National Research Foundation of Korea (NRF) with Grant No. RS-2025-00562917. The work of AB, FB and DL was supported in part by the U.S. Department of Energy under grant No. DE-SC0007914 and in part by Pitt PACC. The work of DL was also partially supported by the National Natural Science Foundation of China under Grants No. 12321005 and partially supported by the Shandong Provincial Department of Science and Technology under Project No. TQ012025001.
\end{acknowledgments}

\appendix  

\section{Perturbative Unitarity Constraints}\label{sec:PerturbativeUnitarity}
An important theoretical constraint on various parameters in the scalar sector, in particular the BNV trilinear coupling, comes from perturbative unitarity. This is determined by, for example, the  $\tiny{\chi_{_{LQ}} \ \chi_{_{LL}} \rightarrow \chi_{_{LQ}} \ \chi_{_{LL}}}$ scattering process, for which there are four diagrams, one involving the mixed quartic $\lambda_{_{LLLQ}},$ and three other ones with the diquark in the mediator corresponding to the $s$, $t$, and $u$ channels. \\ 
  
To determine the perturbative unitarity constraints on $ \Lambda_{\text{BNV}}$ and $\lambda_{LQLL}$, we apply the general formulae provided in \cite{Goodsell:2018tti}, but with slight modifications to deal with poles in the propagator of the diquark. The amplitude can thus be written as

\begin{equation}\label{eq:partialwaveunitarityamplitude}
    \mathcal{M}(\chi_{LL}\chi_{LQ} \rightarrow \chi_{LL}\chi_{LQ}) = \lambda_{LLLQ} + \Lambda^{2}_{\text{BNV}} \sum_{x=s,t,u} \frac{1}{x - m^{2}_{\chi_{uu}} + i m_{\chi_{uu}} \Gamma_{uu}} ,
\end{equation}
where $\Gamma_{uu}$ is the decay width of $\chi_{_{uu}}$ and $s,t,$ and $u$ are the usual Mandelstam variables. The partial wave unitarity constraint is defined by the inequality
\begin{equation}
    -i\left(a_{j} - a^{\dagger}_{j}\right) \leq a_{j}a^{\dagger}_{j},
\end{equation}
where $a_{j}$ is the $j^{\text{th}}$ coefficient of the partial wave decomposition of the amplitude. Since this must be satisfied term by term, the authors of \cite{Goodsell:2018tti} consider only the zeroth order one and find,
\begin{equation}\label{eq:perturbativeunitaritycondition}
    \text{Re}(a_{0}) \leq \frac{1}{2}.
\end{equation}
We apply the same conditions and, in the $s \rightarrow \infty$ limit, find
\begin{equation}
\lambda_{LQLL} > -8\pi,    
\end{equation}
a constraint independent of $\Lambda_{\text{BNV}}$ that is easily satisfied by the previously imposed positivity criterion to avoid spontaneous breaking of electromagnetic gauge invariance. As this asymptotic limit implies, the perturbative unitarity constraints on $ \Lambda_{\text{BNV}}$ weakens with increasing $s$. However, it must hold for \emph{all} $s$ above the kinematic threshold for the scattering process in question. To derive the limits for for different values of $s$, we apply the formulae presented in \cite{Goodsell:2018tti}, but with the previously mentioned modification to deal with $s$ channel poles that effectively amounts to the replacement 
\begin{equation*}
m^{2}_{\chis{uu}} \rightarrow m^{2}_{\chis{uu}} + im_{\chis{uu}}\Gamma_{uu}.    
\end{equation*}

\section{Proton Decay Amplitudes}\label{sec:ProtonDEccayAmplitudes}
In this section, we give a brief presentation of the computation of the two proton decay amplitudes, particularly because they involve making some simplifying assumptions regarding the hadronic matrix elements in the case of $\mathcal{O}^{\text{L}}_{\text{BNV}}$. \\

Some manipulations of the hadronic portion of the three quark BNV operators are also necessary. Starting from the notation of \cite{Aoki:2017puj},
\begin{subequations}
 \begin{equation}\label{eq:AokiPDMatrixElement1}
    \left< 0 \ \vert (ud)_{R}u_{L} \vert p \right> = \alpha P_{R} u_{P}(p), 
\end{equation}
\begin{equation}\label{eq:AokiPDMatrixElement2}
 \left< 0 \ \vert (ud)_{L}u_{L} \vert p \right> = \beta P_{L} u_{P}(p),
\end{equation}
\end{subequations}
with the results of their lattice computations giving $\alpha$ = $-\beta = -0.014$ GeV$^{3}$ (evaluated at a scale $\mu$ = 2 GeV).
Noting that
\begin{equation*}
    \begin{split}
      \left( \overline{d}^{c}P_{L}u\right) & = -\left( \overline{d}^{c}P_{L}u\right)^{T} = u^{T}C P_{L}d 
    \end{split}
\end{equation*}
and using 
\begin{equation*}
\overline{u}^{c} = - u^{T}C^{-1},    
\end{equation*}
we apply transposition to the vacuum proton decay matrix element and perform some further algebra to find
\begin{equation}\label{eq:hadmatrixuseful}
    \langle \  0 \ \vert  \left(\overline{d}^{c}P_{L}u\right)\overline{u}^{c}_{L} \vert \ p \ \rangle = \beta P_{L} \overline{v}_{P}(p).
\end{equation}
This is consistent with the form of the matrix elements used in \cite{Hou:2005iu} (see, for example, just below equation (19) for neutron to vacuum matrix element). 

\subsection{$\mathbf{\mathcal{O}^{\text{R}}_{\text{BNV}}}$}
We start with the $\mathcal{O}^{\text{R}}_{\text{BNV}}$ operator since it is more straightforward compared to $\mathcal{O}^{\text{L}}_{\text{BNV}}$. The amplitude can be written as
\begin{align}
\mathcal{M}_{\text{R}}={}&
-\frac{e^2}{8s_W^2}
 \left(c_{LL}U\right)^{mn} \sum_{j,k} V^{T kd} c^{uk}_{uu}\, m_{u_{k}}\, m_{l_{j}}\, \mathcal{T}_{\text{R,loop}}(m_{u_{k}},m_{l_{j}})\, U^{\dagger jl} c^{uj}_{ul} ,
\end{align}

The loop integral can be written as
\begin{equation}
\begin{split}
\mathcal{T}_{\text{R,loop}} = \frac{-i}{16\pi^{2}}\int_{0}^{1} dx_{1} \int^{1-x_{1}}_{0} dx_{2}\,
\frac{\epsilon^{IJK}}{\Delta^{2}(x_{1},x_{2})}
&\left(\overline d_I^{\,c}(p_1)\gamma^\mu P_L u_J(p_3)\right) \\
\times&\left(\overline u^{\,c}_{K}(p_{2})\gamma_\mu P_R v_{l}(k_{1})\right)
\left(\overline v^{\,c}_{m}(k_{2}) P_L v_{n}(k_{3})\right),
\end{split}
\end{equation}   

where 
\begin{equation}
\begin{aligned}
\Delta^2(x_1,x_2) ={}&
m_W^2
+x_1\left(m_{u_k}^2-m_W^2-p_1^2\right)+x_2\left(m_{d_j}^2-m_W^2-p_2^2\right)
+\left(x_1p_1-x_2p_2\right)^2.
\end{aligned}
\end{equation}
Using a Fierz-Pauli transformation and Equation~\eqref{eq:hadmatrixuseful} the the numerator can be simplified and the loop integral becomes  
\begin{equation}
\mathcal{T}_{\text{R,loop}} = i\frac{2}{16\pi^{2}}\int_{0}^{1} dx_{1} \int^{1-x_{1}}_{0}  dx_{2} \ \frac{\mathcal{S}_{lmn}}{\Delta^{2}(x_{1},x_{2})}.    
\end{equation}

Continuing the computation by keeping only terms larger than $m_{P}$ or higher in $\Delta^{2}(x_{1},x_{2})$ then leads to the results of Section~\ref{ssec:protondecayamp}. 

\subsection{$\mathbf{\mathcal{O}^{\text{L}}_{\text{BNV}}}$}
The computation of the $\mathcal{O}^{\text{L}}_{\text{BNV}}$ operator is slightly more complicated due to its chirality structure that leaves a loop integral that depends on the constituent quark momentum. The amplitude is 
\begin{equation}
\begin{split}
\mathcal{M}_{\text{L}} = \left(c_{LL}U\right)^{mn} \sum_{j}\sum_{k} V^{T kd}c^{uk}_{uu}m_{u_{k}}
&\Bigl( \mathcal{T}^{jk}_{\text{L,loop}}\left(m_{u_{k}},m_{l_{j}}\right)U^{\dagger jl}c^{uj}_{LQ} \\
&\quad - \mathcal{T}_{\text{L,loop}}(m_{d_{j}},m_{u_{k}})\,V^{*uj}\left(V^{T}c_{LQ}U\right)^{jl} \Bigr)
\end{split}
\end{equation}
The loop integral is 
\begin{equation}
\begin{split}
\mathcal{T}_{\text{L,loop}} ={}& \frac{-i}{16\pi^{2}}\int_{0}^{1} dx_{1} \int^{1-x_{1}}_{0} dx_{2}\,
\frac{\epsilon^{IJK}\,N_{\nu}}{\Delta^{2}(x_{1},x_{2})}
\left(\overline d_I^{\,c}(p_1)\gamma^\mu P_L u_J(p_3)\right) \\
&\times\left(\overline u^{\,c}_{K}(p_{2})\gamma_\mu \gamma^{\nu} P_R v_{l}(k_{1})\right)
\left(\overline v^{\,c}_{m}(k_{2}) P_L v_{n}(k_{3})\right),
\end{split}
\end{equation}
where
\begin{equation}
N_\nu=x_1p_{1\nu}+(1-x_2)p_{2\nu}.
\end{equation}
Using a Fierz-Pauli transformation, the operator in the numerator of the loop integral can be written as
\begin{align}
\mathcal{O}^{\nu}_{\text{L}} =-2\epsilon^{IJK}
\left(\overline d_I^{\,c}(p_1)P_Lu_J(p_2)\right)
\left(\overline u^{\,c}_{K}(p_3)\gamma^\nu P_Rv(k_{1})\right)\left(\overline v^{\,c}_{m}(k_{2})
P_Lv_{n}(k_{3})\right)N_{\nu}.
\end{align}
Applying integration by parts and the Dirac equation, some terms vanish in the chiral limit. These are sub-dominant since they are proportional to the up and down quark masses. The dominant contribution comes from terms that do not vanish in the chiral limit,  
\begin{equation}
N_\nu\langle0|\mathcal{O}^{\nu}_{\text{L}} |p(P)\rangle
=
2\alpha\,
\overline v(P)\slashed N P_L v(k_{1}).
\end{equation}
To evaluate this, we make the simplifying assumption that all three consituent quarks are of roughly equal momentum,
\begin{equation}
p_1^\mu\simeq p_2^\mu\simeq p_3^\mu\simeq
\frac{P^\mu}{3},
\end{equation}
and apply the Dirac equation to obtain the results of Section~\ref{ssec:protondecayamp}.  

\bibliographystyle{JHEP}
\bibliography{apssamp}
\end{document}

%% file: tikz/charmbaryon_1.tex
\CharmBNV
  {d_L^{\,i}}
  {\mathcal{O}^{\text{L}}_{ucd_i}}
  {\bar{\nu}^{\,l}_{L}(k_1)}
  {B_c(P)=
   \begin{cases}
     \Lambda_c^+\ \text{or}\ \Sigma_c^+, & i=d\\
     \Xi_c^+, & i=s
   \end{cases}}

%% file: tikz/charmbaryon_2.tex
\CharmBNV
  {u_{L/R}^{\,i}}
  {\mathcal{O}^{\text{L}/\text{R}}_{ucu_i}}
  {\bar{\ell}^{\,l}_{L/R}(k_1)}
  {B_c^{++}(P)=
   \begin{cases}
     \Sigma_c^{++}, & i=u\\
     \Xi_{cc}^{++}, & i=c
   \end{cases}}

%% file: tikz/proton_decay_q_fig1.tex
\begin{tikzpicture}[x=1cm,y=1cm,thick,>=Stealth]

  \draw[rounded corners=0.25cm] (0,1.05) rectangle (1.45,-1.05);
  \node at (0.58, 0.65) {$u_R$};
  \node at (0.58,-0.15) {$d_L$};
  \node at (0.58,-0.78) {$u_L$};

  \coordinate (O) at (4.45,0.65);
  \coordinate (Ol) at (4.62, 0.65);
  \coordinate (A) at (3.80,-0.15);
  \coordinate (B) at (5.00,-0.15);

  \draw[fermion] (1.02,0.65) -- (1.45,0.65) -- (O);

  \draw[fermion] (1.02,-0.15) -- (A);

  \draw[fermion] (1.02,-0.78) -- (4.55,-0.78)
    .. controls (5.00,-0.78) and (5.00,-0.55) .. (B);

  \draw[fermion] (A) -- (O);
  \node[above left] at (4.00,-0.22) {$u^k$};
  \node[massinsert, rotate=45] at (4.2, 0.333) {$\boldsymbol{\times}$};

  \draw[fermion] (B) -- (O);
  \node[above right] at (4.81, -0.28) {$d_{L}^{j}$};

  \draw[wboson] (A) -- (B);
  \node[below] at (4.38,-0.17) {$W$};

  \node[
    draw,
    circle,
    fill=white,
    minimum size=0.34cm,
    inner sep=0pt
  ] at (O) {$\times$};

  \node[above=9pt] at (4.45,0.65)
    {$\displaystyle\sum_{k,j}\mathcal{O}^{\text{L}}_{u u^k\,d^j\nu^l\, \nu^m \ell^n},~~ (\mathcal{O}_{\rm BNV}^{\text{L}}\text{ only})$};

  \draw[fermion] (7.45,1.30) -- (6.8,1.30)
    .. controls (5.60,1.30) and (5.4,0.98) .. (Ol);
  \draw[fermion] (7.45,0.65) -- (Ol);
  \draw[fermion] (7.45,0.00) -- (6.8,0.00)
    .. controls (5.60,0.00) and (5.4,0.35) .. (Ol);

  \node[right] at (7.45,1.30)
    {$e_L^+(k_3)\,/\,\mu_L^+(k_3)$};
  \node[right] at (7.45,0.65)
    {$\bar{\nu}_L^m(k_2)$};
  \node[right] at (7.45,0.00)
    {$\bar{\nu}_L^l(k_1)$};

  \node at (0.72,-1.38) {$P(P)$};

\end{tikzpicture}

%% file: tikz/proton_decay_l_fig1.tex
\begin{tikzpicture}[x=1cm,y=1cm,thick,>=Stealth]
	
	\draw[rounded corners=0.25cm] (0,1.7) rectangle (1.45,-0.27);
	\node at (0.58, 1.3) {$u_R$};
	\node at (0.58,0.65) {$d_{L}$};
	\node at (0.58,0.0) {$u_{L/R}$};
	
	\coordinate (O) at (4.45,0.65);
	\coordinate (Ol) at (4.62, 0.65);
	\coordinate (A) at (2.7,0.65);
	\coordinate (B) at (6.2,0.0);
	
	\draw[fermion] (1.02,1.3) -- (2.5,1.30)
	.. controls (3.5,1.30) and (4.0,1.0) ..  (O);
	
	\draw[fermion] (1.02,0.65) -- (A);
	\draw[fermion] (A) -- (O);
	
	\draw[fermion] (1.02,0.0) -- (2.5,0.0)
.. controls (3.5,0.0) and (4.0,0.3) ..  (O);
	
	\node[above] at (3.3,0.65) {$u^k$};
	\node[massinsert] at (3.8, 0.65) {$\boldsymbol{\times}$};
	
	
	\draw[wboson] (A).. controls (3.0, 0.2) and (5.9, 0.1).. (B);
	\node[below] at (4.38,0.1) {$W$};
	
	\node[
	draw,
	circle,
	fill=white,
	minimum size=0.34cm,
	inner sep=0pt
	] at (O) {$\times$};
	
	\node[above=9pt] at (4.8,0.8)
	{$\displaystyle\sum_{k,j}\mathcal{O}^{\text{L}/\text{R}}_{u u^k\,u \ell^j\,\nu^m \ell^n}$};
	
	\draw[fermion] (7.45,1.30) -- (5.75,1.30)
	.. controls (5.30,1.30) and (4.75,0.98) .. (Ol);
	\draw[fermion] (7.45,0.65) -- (Ol);
	\draw[fermion] (7.45,0.00) -- (B);
	\draw[fermion] (B) -- (Ol);
	
	\node[right] at (7.45,1.30)
	{$e_L^+(k_3)\,/\,\mu_L^+(k_3)$};
	\node[right] at (7.45,0.65)
	{$\bar{\nu}_L^m(k_2)$};
	\node[right] at (7.45,0.00)
	{$\bar{\nu}_L^l(k_1)$};
	\node[massinsert, rotate=60, text = red] at (5.147, 0.43) {$\boldsymbol{\times}$};
	
	\node at (0.72,-0.6) {$P(P)$};
	\node at (6.0, 0.3) {$\bar{\ell}^j$};
	
\end{tikzpicture}

%% file: tikz/proton-decay-q.tex
\begin{tikzpicture}
\node[font=\bfseries] at (2.2,1.1) {$LQ$};
    \begin{feynman}
        \vertex [label=left:$u_R(p_3)$](uR);
        \vertex [below = 1.cm of uR, label=left:$d_{L}(p_1)$] (dL);
        \vertex [below = 1.cm of dL, label=left:$u_L(p_2)$] (uL);
        \vertex [right = 1.2cm of uR] (uu1);
        \vertex [right =1.2cm of dL] (w1);
        \vertex [right=1.2cm of uL] (w2);
        \vertex [right = 1.2cm of uu1] (uu2);
        \vertex [right = 1.2cm of uu2] (LL);
        \vertex [above right = 0.5cm and 1.0cm of LL, label=right:$\bar{\nu}_{L}^m(k_2)$] (l2);
        \vertex [below right=0.5cm and 1.0cm of LL, label=right:$\bar{\ell}^{n}_{L}(k_3)$] (l3);
        \vertex [right = 1.2cm of w2] (LQ2);
        \vertex [right = 2.2cm of LQ2] (l1) {$\bar{\nu}^l_{L}(k_1)$};

        \diagram*{(uR) --[fermion] (uu1),
        (dL) --[fermion](w1) --[fermion, edge label=$u^k$](uu1),
        (uL) --[fermion](w2) --[fermion, edge label=$d_{L}^j$](LQ2),
        (w1) --[boson, edge label' = $W^{-}$](w2),
        (uu1) --[scalar, edge label=$\chi_{uu}$](uu2) --[scalar, edge label=$\chi_{LL}$](LL),
        (l2) --[fermion](LL),
        (l3) --[fermion] (LL),
        (uu2) --[scalar, edge label=$\chi_{LQ}^\dagger$](LQ2),
        (l1) --[fermion] (LQ2),
        };
    \end{feynman}
    \node[cross out, draw, minimum size=1.5pt, line width=0.6pt] at ($(w1)!0.8!(uu1)$) {};
\end{tikzpicture}

%% file: tikz/proton-decay-l.tex
\begin{tikzpicture}
\node[font=\bfseries] at (2.2,1.0) {$LQ/ul$};
    \begin{feynman}
        \vertex [label=left:$u_R(p_3)$](uR);
        \vertex [below = 1.cm of uR, label=left:$d_{L}(p_1)$] (dL);
        \vertex [below = 1.cm of dL, label=left:$u_{L/R}(p_2)$] (uL);
        \vertex [right = 1.2cm of uR] (uu1);
        \vertex [right =1.2cm of dL] (w1);
        \vertex [right=1.2cm of uL] (LQ2);
        \vertex [right = 1.2cm of uu1] (uu2);
        \vertex [right = 1.2cm of uu2] (LL);
        \vertex [above right = 0.5cm and 1.0cm of LL, label=right:$\bar{\nu}_{L}^m(k_2)$] (l2);
        \vertex [below right=0.5cm and 1.0cm of LL, label=right:$\bar{\ell}^n_{L}(k_3)$] (l3);
        \vertex [right = 1.2cm of LQ2] (w2);
        \vertex [right = 2.2cm of w2] (l1) {$\bar{\nu}_{L}^l(k_1)$};

        \diagram*{(uR) --[fermion] (uu1),
        (dL) --[fermion](w1) --[fermion, edge label=$u^k$](uu1),
        (uL) --[fermion](LQ2),
        (w2) --[fermion, edge label=$\bar{\ell}_{L/R}^j$](LQ2),
        (w1) --[boson, edge label = $W^{-}$, near end](w2),
        (uu1) --[scalar, edge label=$\chi_{uu}$](uu2) --[scalar, edge label=$\chi_{LL}$](LL),
        (l2) --[fermion](LL),
        (l3) --[fermion] (LL),
        (LQ2) --[scalar, edge label'=$\chi_{LQ}^\dagger$, near end](uu2),
        (l1) --[fermion] (w2),
        };
    \end{feynman}
    \node[cross out, draw, minimum size=1.2pt, line width=0.6pt] at ($(w1)!0.8!(uu1)$) {};
     \node[cross out, draw, minimum size=1pt, line width=0.6pt, color=red] at ($(w2)!0.25!(LQ2)$) {};
\end{tikzpicture}

%% file: tikz/collider_signature.tex
\begin{tikzpicture}
  \begin{feynman}
    \vertex (uu);
    \vertex [above left=0.85cm and \mylength of uu] (c) {$c$};
    \vertex [below left=0.85cm and \mylength of uu](u) {$u$};
    \vertex [right=\mylength of uu] (bnv);
    \vertex [above right=0.85cm and \mylength of bnv] (LQ);
    \vertex [below right=0.85cm and \mylength of bnv] (LL);
    \vertex [above right=0.5cm and \mylength of LQ] (t) {$\Bar{t}$};
    \vertex [below right=0.5cm and \mylength of LQ, label=right:$l^{+}_{1}$] (mu);
    \vertex [above right=0.5cm and \mylength of LL, label=right:$l^{+}_{2}$] (mu2);
    \vertex [below right=0.5cm and \mylength of LL] (nu) {$\Bar{\nu}_{l_{3}}$};

    \diagram* {
    (c) -- [fermion](uu) --[scalar, edge label=$\chi_{uu}^\dagger$](bnv) -- [scalar, edge label=$\chi_{LQ}^\dagger$](LQ) --[fermion] (t);
    (LQ)-- [fermion](mu);
    (bnv)-- [scalar, edge label'=$\chi_{LL}$](LL) -- [fermion](mu2);
    (LL) -- [fermion](nu);
    (u) -- [fermion](uu);
    };
  \end{feynman}
\end{tikzpicture}\label{fig:topology}

%% file: apssamp.bib
@article{Weinberg:1980bf,
    author = "Weinberg, Steven",
    title = "{Varieties of Baryon and Lepton Nonconservation}",
    reportNumber = "HUTP-80/A023",
    doi = "10.1103/PhysRevD.22.1694",
    journal = "Phys. Rev. D",
    volume = "22",
    pages = "1694",
    year = "1980"
}

@article{Babu:2006xc,
    author = "Babu, K. S. and Mohapatra, R. N. and Nasri, S.",
    title = "{Post-Sphaleron Baryogenesis}",
    eprint = "hep-ph/0606144",
    archivePrefix = "arXiv",
    reportNumber = "UMD-PP-06-004, OSU-HEP-06-02",
    doi = "10.1103/PhysRevLett.97.131301",
    journal = "Phys. Rev. Lett.",
    volume = "97",
    pages = "131301",
    year = "2006"
}

@article{Babu:2013yca,
    author = "Babu, K. S. and Bhupal Dev, P. S. and Fortes, Elaine C. F. S. and Mohapatra, R. N.",
    title = "{Post-Sphaleron Baryogenesis and an Upper Limit on the Neutron-Antineutron Oscillation Time}",
    eprint = "1303.6918",
    archivePrefix = "arXiv",
    primaryClass = "hep-ph",
    reportNumber = "OSU-HEP-13-02, MAN-HEP-2012-017, UMD-PP-013-003",
    doi = "10.1103/PhysRevD.87.115019",
    journal = "Phys. Rev. D",
    volume = "87",
    number = "11",
    pages = "115019",
    year = "2013"
}

@misc{pdg_ckm_matrix,
	author = {Ceccucci, A. and Ligeti, Z. and Sakai, Y.},
	title = {},
	howpublished = {\url{https://pdg.lbl.gov/2019/reviews/rpp2019-rev-ckm-matrix.pdf}},
	year = {},
	note = {[Accessed 20-08-2025]},
}

@misc{pdg_pmns_matrix,
	author = {Gonzalez-Garcia, M.C. and  Yokoyama, M.},
	title = {},
	howpublished = {\url{https://pdg.lbl.gov/2024/reviews/rpp2024-rev-neutrino-mixing.pdf}},
	year = {},
	note = {[Accessed 25-08-2025]},
}

@article{Sakharov:1967dj,
  author  = {Sakharov, A. D.},
  doi     = {10.1070/PU1991v034n05ABEH002497},
  journal = {Pisma Zh. Eksp. Teor. Fiz.},
  pages   = {32--35},
  title   = {{Violation of CP Invariance, C asymmetry, and baryon asymmetry of the universe}},
  volume  = {5},
  year    = {1967}
}

@article{McLaughlin:1999rr,
    author = "McLaughlin, G. C. and Ng, J. N.",
    title = "{A Study of the charged scalar in the Zee model}",
    eprint = "hep-ph/9903509",
    archivePrefix = "arXiv",
    reportNumber = "TRI-PP-99-03",
    doi = "10.1016/S0370-2693(99)00490-6",
    journal = "Phys. Lett. B",
    volume = "455",
    pages = "224--230",
    year = "1999"
}

@article{MEG:2016leq,
    author = "Baldini, A. M. and others",
    collaboration = "MEG",
    title = "{Search for the lepton flavour violating decay $\mu ^+ \rightarrow \mathrm {e}^+ \gamma $ with the full dataset of the MEG experiment}",
    eprint = "1605.05081",
    archivePrefix = "arXiv",
    primaryClass = "hep-ex",
    doi = "10.1140/epjc/s10052-016-4271-x",
    journal = "Eur. Phys. J. C",
    volume = "76",
    number = "8",
    pages = "434",
    year = "2016"
}

@article{MEGII:2023ltw,
    author = "Afanaciev, K. and others",
    collaboration = "MEG II",
    title = "{A search for $\mu ^+ \rightarrow \textrm{e}^+ \gamma $ with the first dataset of the MEG~II experiment}",
    eprint = "2310.12614",
    archivePrefix = "arXiv",
    primaryClass = "hep-ex",
    doi = "10.1140/epjc/s10052-024-12416-2",
    journal = "Eur. Phys. J. C",
    volume = "84",
    number = "3",
    pages = "216",
    year = "2024",
    note = "[Erratum: Eur.Phys.J.C 84, 1042 (2024)]"
}

@article{ATLAS:2020xov,
    author = "Aad, Georges and others",
    collaboration = "ATLAS",
    title = "{Search for pair production of scalar leptoquarks decaying into first- or second-generation leptons and top quarks in proton\textendash{}proton collisions at $\sqrt{s}$ = 13 TeV with the ATLAS detector}",
    eprint = "2010.02098",
    archivePrefix = "arXiv",
    primaryClass = "hep-ex",
    reportNumber = "CERN-EP-2020-160",
    doi = "10.1140/epjc/s10052-021-09009-8",
    journal = "Eur. Phys. J. C",
    volume = "81",
    number = "4",
    pages = "313",
    year = "2021"
}

@article{Wilczek:1979hc,
    author = "Wilczek, Frank and Zee, A.",
    title = "{Operator Analysis of Nucleon Decay}",
    reportNumber = "Print-79-0709 (PRINCETON)",
    doi = "10.1103/PhysRevLett.43.1571",
    journal = "Phys. Rev. Lett.",
    volume = "43",
    pages = "1571--1573",
    year = "1979"
}

@article{ATLAS:2021oiz,
    author = "Aad, Georges and others",
    collaboration = "ATLAS",
    title = "{Search for pair production of third-generation scalar leptoquarks decaying into a top quark and a $\tau$-lepton in $pp$ collisions at $ \sqrt{s} $ = 13 TeV with the ATLAS detector}",
    eprint = "2101.11582",
    archivePrefix = "arXiv",
    primaryClass = "hep-ex",
    reportNumber = "CERN-EP-2020-241",
    doi = "10.1007/JHEP06(2021)179",
    journal = "JHEP",
    volume = "06",
    pages = "179",
    year = "2021"
}

@article{Aoki:2017puj,
    author = "Aoki, Yasumichi and Izubuchi, Taku and Shintani, Eigo and Soni, Amarjit",
    title = "{Improved lattice computation of proton decay matrix elements}",
    eprint = "1705.01338",
    archivePrefix = "arXiv",
    primaryClass = "hep-lat",
    reportNumber = "RBRC-1235, KEK-CP-358",
    doi = "10.1103/PhysRevD.96.014506",
    journal = "Phys. Rev. D",
    volume = "96",
    number = "1",
    pages = "014506",
    year = "2017"
}

@article{Crivellin:2021ejk,
    author = "Crivellin, Andreas and Schnell, Luc",
    title = "{Complete Lagrangian and set of Feynman rules for scalar leptoquarks}",
    eprint = "2105.04844",
    archivePrefix = "arXiv",
    primaryClass = "hep-ph",
    reportNumber = "CERN-TH-2021-074, PSI-PR-21-08, ZU-TH 19/21",
    doi = "10.1016/j.cpc.2021.108188",
    journal = "Comput. Phys. Commun.",
    volume = "271",
    pages = "108188",
    year = "2022"
}

@article{Abbott:1989jw,
    author = "Abbott, L. F. and Wise, Mark B.",
    title = "{Wormholes and Global Symmetries}",
    reportNumber = "CALT-68-1523",
    doi = "10.1016/0550-3213(89)90503-8",
    journal = "Nucl. Phys. B",
    volume = "325",
    pages = "687--704",
    year = "1989"
}

@article{Kallosh:1995hi,
    author = "Kallosh, Renata and Linde, Andrei D. and Linde, Dmitri A. and Susskind, Leonard",
    title = "{Gravity and global symmetries}",
    eprint = "hep-th/9502069",
    archivePrefix = "arXiv",
    reportNumber = "SU-ITP-95-2",
    doi = "10.1103/PhysRevD.52.912",
    journal = "Phys. Rev. D",
    volume = "52",
    pages = "912--935",
    year = "1995"
}

@article{Lee:1988ge,
    author = "Lee, Ki-Myeong",
    title = "{Wormholes and Goldstone Bosons}",
    reportNumber = "FERMILAB-PUB-88-027-T",
    doi = "10.1103/PhysRevLett.61.263",
    journal = "Phys. Rev. Lett.",
    volume = "61",
    pages = "263--266",
    year = "1988"
}

@article{SNO:2001kpb,
    author = "Ahmad, Q. R. and others",
    collaboration = "SNO",
    title = "{Measurement of the rate of $\nu_e+d \to p+p+e^-$ interactions produced by $^8$B solar neutrinos at the Sudbury Neutrino Observatory}",
    eprint = "nucl-ex/0106015",
    archivePrefix = "arXiv",
    reportNumber = "UPR-0240E",
    doi = "10.1103/PhysRevLett.87.071301",
    journal = "Phys. Rev. Lett.",
    volume = "87",
    pages = "071301",
    year = "2001"
}

@article{Davis:1968cp,
    author = "Davis, Jr., Raymond and Harmer, Don S. and Hoffman, Kenneth C.",
    title = "{Search for neutrinos from the sun}",
    doi = "10.1103/PhysRevLett.20.1205",
    journal = "Phys. Rev. Lett.",
    volume = "20",
    pages = "1205--1209",
    year = "1968"
}

@article{Super-Kamiokande:1998kpq,
    author = "Fukuda, Y. and others",
    collaboration = "Super-Kamiokande",
    title = "{Evidence for oscillation of atmospheric neutrinos}",
    eprint = "hep-ex/9807003",
    archivePrefix = "arXiv",
    reportNumber = "BU-98-17, ICRR-REPORT-422-98-18, UCI-98-8, KEK-PREPRINT-98-95, LSU-HEPA-5-98, UMD-98-003, SBHEP-98-5, TKU-PAP-98-06, TIT-HPE-98-09",
    doi = "10.1103/PhysRevLett.81.1562",
    journal = "Phys. Rev. Lett.",
    volume = "81",
    pages = "1562--1567",
    year = "1998"
}

@article{Giudice:2011ak,
    author = "Giudice, Gian Francesco and Gripaios, Ben and Sundrum, Raman",
    title = "{Flavourful Production at Hadron Colliders}",
    eprint = "1105.3161",
    archivePrefix = "arXiv",
    primaryClass = "hep-ph",
    reportNumber = "CERN-PH-TH-2011-110, UMD-PP-11-006",
    doi = "10.1007/JHEP08(2011)055",
    journal = "JHEP",
    volume = "08",
    pages = "055",
    year = "2011"
}

@article{Hou:2005iu,
    author = "Hou, Wei-Shu and Nagashima, Makiko and Soddu, Andrea",
    title = "{Baryon number violation involving higher generations}",
    eprint = "hep-ph/0509006",
    archivePrefix = "arXiv",
    reportNumber = "SLAC-PUB-11461",
    doi = "10.1103/PhysRevD.72.095001",
    journal = "Phys. Rev. D",
    volume = "72",
    pages = "095001",
    year = "2005"
}

@article{Goodsell:2018tti,
    author = "Goodsell, Mark D. and Staub, Florian",
    title = "{Unitarity constraints on general scalar couplings with SARAH}",
    eprint = "1805.07306",
    archivePrefix = "arXiv",
    primaryClass = "hep-ph",
    reportNumber = "KA-TP-10-2018",
    doi = "10.1140/epjc/s10052-018-6127-z",
    journal = "Eur. Phys. J. C",
    volume = "78",
    number = "8",
    pages = "649",
    year = "2018"
}

@article{Hewett:1988xc,
    author = "Hewett, JoAnne L. and Rizzo, Thomas G.",
    title = "{Low-Energy Phenomenology of Superstring Inspired E(6) Models}",
    reportNumber = "MAD-PH-446, IS-J-3005",
    doi = "10.1016/0370-1573(89)90071-9",
    journal = "Phys. Rept.",
    volume = "183",
    pages = "193",
    year = "1989"
}

@article{Mohapatra:2007af,
    author = "Mohapatra, R. N. and Okada, Nobuchika and Yu, Hai-Bo",
    title = "{Diquark Higgs at LHC}",
    eprint = "0709.1486",
    archivePrefix = "arXiv",
    primaryClass = "hep-ph",
    reportNumber = "UMD-PP-07-010",
    doi = "10.1103/PhysRevD.77.011701",
    journal = "Phys. Rev. D",
    volume = "77",
    pages = "011701",
    year = "2008"
}

@article{Dorsner:2016wpm,
    author = "Dor{\v{s}}ner, I. and Fajfer, S. and Greljo, A. and Kamenik, J. F. and Ko{\v{s}}nik, N.",
    title = "{Physics of leptoquarks in precision experiments and at particle colliders}",
    eprint = "1603.04993",
    archivePrefix = "arXiv",
    primaryClass = "hep-ph",
    doi = "10.1016/j.physrep.2016.06.001",
    journal = "Phys. Rept.",
    volume = "641",
    pages = "1--68",
    year = "2016"
}

@article{Senjanovic:1982ex,
    author = "Senjanovic, Goran and Sokorac, Aleksandar",
    title = "{Light Leptoquarks in SO(10)}",
    reportNumber = "Print-82-0652 (BNL), BNL-31824",
    doi = "10.1007/BF01574858",
    journal = "Z. Phys. C",
    volume = "20",
    pages = "255",
    year = "1983"
}

@article{Buchmuller:1986iq,
    author = "Buchmuller, W. and Wyler, D.",
    title = "{Constraints on SU(5) Type Leptoquarks}",
    reportNumber = "CERN-TH-4461/86, ITP-UH 1/86",
    doi = "10.1016/0370-2693(86)90771-9",
    journal = "Phys. Lett. B",
    volume = "177",
    pages = "377--382",
    year = "1986"
}

@article{Zee:1980ai,
    author = "Zee, A.",
    title = "{A Theory of Lepton Number Violation, Neutrino Majorana Mass, and Oscillation}",
    reportNumber = "UPR-0150T",
    doi = "10.1016/0370-2693(80)90349-4",
    journal = "Phys. Lett. B",
    volume = "93",
    pages = "389",
    year = "1980",
    note = "[Erratum: Phys.Lett.B 95, 461 (1980)]"
}

@article{Babu:1988ki,
    author = "Babu, K. S.",
    title = "{Model of 'Calculable' Majorana Neutrino Masses}",
    reportNumber = "UR-1039, ER13065-517",
    doi = "10.1016/0370-2693(88)91584-5",
    journal = "Phys. Lett. B",
    volume = "203",
    pages = "132--136",
    year = "1988"
}

@article{Pascual-Dias:2020hxo,
    author = "Pascual-Dias, Bruna and Saha, Pratishruti and London, David",
    title = "{LHC Constraints on Scalar Diquarks}",
    eprint = "2006.13385",
    archivePrefix = "arXiv",
    primaryClass = "hep-ph",
    reportNumber = "UdeM-GPP-TH-20-279",
    doi = "10.1007/JHEP07(2020)144",
    journal = "JHEP",
    volume = "07",
    pages = "144",
    year = "2020"
}

@article{Super-Kamiokande:2020wjk,
    author = "Takenaka, A. and others",
    collaboration = "Super-Kamiokande",
    title = "{Search for proton decay via $p\to e^+\pi^0$ and $p\to \mu^+\pi^0$ with an enlarged fiducial volume in Super-Kamiokande I-IV}",
    eprint = "2010.16098",
    archivePrefix = "arXiv",
    primaryClass = "hep-ex",
    doi = "10.1103/PhysRevD.102.112011",
    journal = "Phys. Rev. D",
    volume = "102",
    number = "11",
    pages = "112011",
    year = "2020"
}

@article{McGrew:1999nd,
    author = "McGrew, C. and others",
    title = "{Search for nucleon decay using the IMB-3 detector}",
    doi = "10.1103/PhysRevD.59.052004",
    journal = "Phys. Rev. D",
    volume = "59",
    pages = "052004",
    year = "1999"
}

@article{Kamiokande-II:1989avz,
    author = "Hirata, K. S. and others",
    collaboration = "Kamiokande-II",
    title = "{Experimental Limits on Nucleon Lifetime for Lepton + Meson Decay Modes}",
    reportNumber = "ICR-184-89-1, KEK-Preprint-88-116, KOBE-AP-89-01, OULNS-88-11, UPR-0163E",
    doi = "10.1016/0370-2693(89)90058-0",
    journal = "Phys. Lett. B",
    volume = "220",
    pages = "308--316",
    year = "1989"
}

@article{Super-Kamiokande:2014otb,
    author = "Abe, K. and others",
    collaboration = "Super-Kamiokande",
    title = "{Search for proton decay via $p\to\nu K^+$ using 260  kiloton{\textperiodcentered}year data of Super-Kamiokande}",
    eprint = "1408.1195",
    archivePrefix = "arXiv",
    primaryClass = "hep-ex",
    doi = "10.1103/PhysRevD.90.072005",
    journal = "Phys. Rev. D",
    volume = "90",
    number = "7",
    pages = "072005",
    year = "2014"
}

@article{Super-Kamiokande:2005lev,
    author = "Kobayashi, K. and others",
    collaboration = "Super-Kamiokande",
    title = "{Search for nucleon decay via modes favored by supersymmetric grand unification models in Super-Kamiokande-I}",
    eprint = "hep-ex/0502026",
    archivePrefix = "arXiv",
    doi = "10.1103/PhysRevD.72.052007",
    journal = "Phys. Rev. D",
    volume = "72",
    pages = "052007",
    year = "2005"
}

@article{Super-Kamiokande:2014pqx,
    author = "Takhistov, V. and others",
    collaboration = "Super-Kamiokande",
    title = "{Search for Trilepton Nucleon Decay via $p \rightarrow e^+ \nu \nu$ and $p \rightarrow \mu^+ \nu \nu$ in the Super-Kamiokande Experiment}",
    eprint = "1409.1947",
    archivePrefix = "arXiv",
    primaryClass = "hep-ex",
    doi = "10.1103/PhysRevLett.113.101801",
    journal = "Phys. Rev. Lett.",
    volume = "113",
    number = "10",
    pages = "101801",
    year = "2014"
}

@article{Pati:1974yy,
    author = "Pati, Jogesh C. and Salam, Abdus",
    title = "{Lepton Number as the Fourth Color}",
    reportNumber = "IC-74-7",
    doi = "10.1103/PhysRevD.10.275",
    journal = "Phys. Rev. D",
    volume = "10",
    pages = "275--289",
    year = "1974",
    note = "[Erratum: Phys.Rev.D 11, 703--703 (1975)]"
}

@article{Georgi:1974sy,
    author = "Georgi, H. and Glashow, S. L.",
    title = "{Unity of All Elementary Particle Forces}",
    doi = "10.1103/PhysRevLett.32.438",
    journal = "Phys. Rev. Lett.",
    volume = "32",
    pages = "438--441",
    year = "1974"
}

@article{CMS:2019gwf,
    author = "Sirunyan, Albert M and others",
    collaboration = "CMS",
    title = "{Search for high mass dijet resonances with a new background prediction method in proton-proton collisions at $\sqrt{s} =$ 13 TeV}",
    eprint = "1911.03947",
    archivePrefix = "arXiv",
    primaryClass = "hep-ex",
    reportNumber = "CMS-EXO-19-012, CERN-EP-2019-222",
    doi = "10.1007/JHEP05(2020)033",
    journal = "JHEP",
    volume = "05",
    pages = "033",
    year = "2020"
}

@techreport{ATLAS-CONF-2019-007,
      collaboration = "ATLAS",
      title         = "{Search for New Phenomena in Dijet Events using 139
                       fb$^{−1}$ of $pp$ collisions at $\sqrt{s}$ = 13TeV
                       collected with the ATLAS Detector}",
      institution   = "CERN",
      reportNumber  = "ATLAS-CONF-2019-007",
      address       = "Geneva",
      year          = "2019",
      url           = "https://cds.cern.ch/record/2668385",
}

@article{Bigaran:2020jil,
    author = "Bigaran, Innes and Volkas, Raymond R.",
    title = "{Getting chirality right: Single scalar leptoquark solutions to the $(g-2)_{e,\mu}$ puzzle}",
    eprint = "2002.12544",
    archivePrefix = "arXiv",
    primaryClass = "hep-ph",
    doi = "10.1103/PhysRevD.102.075037",
    journal = "Phys. Rev. D",
    volume = "102",
    number = "7",
    pages = "075037",
    year = "2020"
}

@article{ATLAS:2011izm,
    author = "Aad, Georges and others",
    collaboration = "ATLAS",
    title = "{Inclusive search for same-sign dilepton signatures in $pp$ collisions at $\sqrt{s}=7$ TeV with the ATLAS detector}",
    eprint = "1108.0366",
    archivePrefix = "arXiv",
    primaryClass = "hep-ex",
    reportNumber = "CERN-PH-EP-2011-094",
    doi = "10.1007/JHEP10(2011)107",
    journal = "JHEP",
    volume = "10",
    pages = "107",
    year = "2011"
}

@article{CMS:2016mku,
    author = "Khachatryan, Vardan and others",
    collaboration = "CMS",
    title = "{Search for new physics in same-sign dilepton events in proton{\textendash}proton collisions at $\sqrt{s} = 13\,\text {TeV} $}",
    eprint = "1605.03171",
    archivePrefix = "arXiv",
    primaryClass = "hep-ex",
    reportNumber = "CMS-SUS-15-008, CERN-EP-2016-119",
    doi = "10.1140/epjc/s10052-016-4261-z",
    journal = "Eur. Phys. J. C",
    volume = "76",
    number = "8",
    pages = "439",
    year = "2016"
}

@article{ParticleDataGroupLeptoquarks:2024cfk,
    author = "Navas, S. and others",
    collaboration = "Particle Data Group, Volume I Chapter 94: Leptoquarks",
    title = "{Review of particle physics}",
    doi = "10.1103/PhysRevD.110.030001",
    journal = "Phys. Rev. D",
    volume = "110",
    number = "3",
    pages = "030001",
    year = "2024"
}

@article{CMS:2022nty,
    author = "Tumasyan, Armen and others",
    collaboration = "CMS",
    title = "{Inclusive nonresonant multilepton probes of new phenomena at $\sqrt s$=13{\,}{\,}TeV}",
    eprint = "2202.08676",
    archivePrefix = "arXiv",
    primaryClass = "hep-ex",
    reportNumber = "CMS-EXO-21-002, CERN-EP-2022-008",
    doi = "10.1103/PhysRevD.105.112007",
    journal = "Phys. Rev. D",
    volume = "105",
    number = "11",
    pages = "112007",
    year = "2022"
}

@article{Fritzsch:1974nn,
    author = "Fritzsch, Harald and Minkowski, Peter",
    title = "{Unified Interactions of Leptons and Hadrons}",
    reportNumber = "CALT-68-467",
    doi = "10.1016/0003-4916(75)90211-0",
    journal = "Annals Phys.",
    volume = "93",
    pages = "193--266",
    year = "1975"
}

@article{Heeck:2019kgr,
    author = "Heeck, Julian and Takhistov, Volodymyr",
    title = "{Inclusive Nucleon Decay Searches as a Frontier of Baryon Number Violation}",
    eprint = "1910.07647",
    archivePrefix = "arXiv",
    primaryClass = "hep-ph",
    reportNumber = "UCI-TR-2019-24",
    doi = "10.1103/PhysRevD.101.015005",
    journal = "Phys. Rev. D",
    volume = "101",
    number = "1",
    pages = "015005",
    year = "2020"
}

@article{Fonseca:2018ehk,
    author = "Fonseca, Renato M. and Hirsch, Martin and Srivastava, Rahul",
    title = "{$\Delta L = 3$ processes: Proton decay and the LHC}",
    eprint = "1802.04814",
    archivePrefix = "arXiv",
    primaryClass = "hep-ph",
    reportNumber = "IFIC-18-03",
    doi = "10.1103/PhysRevD.97.075026",
    journal = "Phys. Rev. D",
    volume = "97",
    number = "7",
    pages = "075026",
    year = "2018"
}

@article{Hyper-Kamiokande:2018ofw,
    author = "Abe, K. and others",
    collaboration = "Hyper-Kamiokande",
    title = "{Hyper-Kamiokande Design Report}",
    eprint = "1805.04163",
    archivePrefix = "arXiv",
    primaryClass = "physics.ins-det",
    month = "5",
    year = "2018"
}

@article{Cabibbo:1963yz,
    author = "Cabibbo, Nicola",
    title = "{Unitary Symmetry and Leptonic Decays}",
    doi = "10.1103/PhysRevLett.10.531",
    journal = "Phys. Rev. Lett.",
    volume = "10",
    pages = "531--533",
    year = "1963"
}

@article{Kobayashi:1973fv,
    author = "Kobayashi, Makoto and Maskawa, Toshihide",
    title = "{CP Violation in the Renormalizable Theory of Weak Interaction}",
    reportNumber = "KUNS-242",
    doi = "10.1143/PTP.49.652",
    journal = "Prog. Theor. Phys.",
    volume = "49",
    pages = "652--657",
    year = "1973"
}

@article{Maki:1962mu,
    author = "Maki, Ziro and Nakagawa, Masami and Sakata, Shoichi",
    title = "{Remarks on the unified model of elementary particles}",
    doi = "10.1143/PTP.28.870",
    journal = "Prog. Theor. Phys.",
    volume = "28",
    pages = "870--880",
    year = "1962"
}

@article{Pontecorvo:1957qd,
    author = "Pontecorvo, B.",
    title = "{Inverse Beta Processes and Nonconservation of Lepton Charge}",
    journal = "Sov. Phys. JETP",
    volume = "7",
    pages = "172--173",
    year = "1958"
}

@article{Herrero-Garcia:2014hfa,
    author = "Herrero-Garcia, Juan and Nebot, Miguel and Rius, Nuria and Santamaria, Arcadi",
    title = "{The Zee{\textendash}Babu model revisited in the light of new data}",
    eprint = "1402.4491",
    archivePrefix = "arXiv",
    primaryClass = "hep-ph",
    reportNumber = "FTUV-14-0218, IFIC-14-13",
    doi = "10.1016/j.nuclphysb.2014.06.001",
    journal = "Nucl. Phys. B",
    volume = "885",
    pages = "542--570",
    year = "2014"
}

@article{Pich:2013lsa,
    author = "Pich, Antonio",
    title = "{Precision Tau Physics}",
    eprint = "1310.7922",
    archivePrefix = "arXiv",
    primaryClass = "hep-ph",
    reportNumber = "IFIC-13-79, FTUV-13-1029, FTUV-13-1028",
    doi = "10.1016/j.ppnp.2013.11.002",
    journal = "Prog. Part. Nucl. Phys.",
    volume = "75",
    pages = "41--85",
    year = "2014"
}

@article{Dong:2011rh,
    author = "Dong, Zhe and Durieux, Gauthier and Gerard, Jean-Marc and Han, Tao and Maltoni, Fabio",
    title = "{Baryon number violation at the LHC: the top option}",
    eprint = "1107.3805",
    archivePrefix = "arXiv",
    primaryClass = "hep-ph",
    reportNumber = "CP3-11-24, MADPH-11-1573",
    doi = "10.1103/PhysRevD.85.016006",
    journal = "Phys. Rev. D",
    volume = "85",
    pages = "016006",
    year = "2012"
}

@article{CMS:2024dzv,
    author = "Hayrapetyan, Aram and others",
    collaboration = "CMS",
    title = "{Search for baryon number violation in top quark production and decay using proton-proton collisions at $\sqrt{s}$ = 13 TeV}",
    eprint = "2402.18461",
    archivePrefix = "arXiv",
    primaryClass = "hep-ex",
    reportNumber = "CMS-TOP-22-003, CERN-EP-2024-027",
    doi = "10.1103/PhysRevLett.132.241802",
    journal = "Phys. Rev. Lett.",
    volume = "132",
    pages = "241802",
    year = "2024"
}

@article{Alwall:2014hca,
    author = "Alwall, J. and Frederix, R. and Frixione, S. and Hirschi, V. and Maltoni, F. and Mattelaer, O. and Shao, H. -S. and Stelzer, T. and Torrielli, P. and Zaro, M.",
    title = "{The automated computation of tree-level and next-to-leading order differential cross sections, and their matching to parton shower simulations}",
    eprint = "1405.0301",
    archivePrefix = "arXiv",
    primaryClass = "hep-ph",
    reportNumber = "CERN-PH-TH-2014-064, CP3-14-18, LPN14-066, MCNET-14-09, ZU-TH-14-14",
    doi = "10.1007/JHEP07(2014)079",
    journal = "JHEP",
    volume = "07",
    pages = "079",
    year = "2014"
}

@article{Theroine:2016chp,
    author = "Theroine, Camille",
    editor = "Aguilar-Ben{\'\i}tez, M and Fuster, J and Mart{\'\i}-Garc{\'\i}a, S and Santamar{\'\i}a, A",
    collaboration = "$n\bar{n}$",
    title = "{A neutron-antineutron oscillation experiment at the European Spallation Source}",
    doi = "10.1016/j.nuclphysbps.2015.09.019",
    journal = "Nucl. Part. Phys. Proc.",
    volume = "273-275",
    pages = "156--161",
    year = "2016"
}

@article{Phillips:2014fgb,
    author = "Phillips, II, D. G. and others",
    title = "{Neutron-Antineutron Oscillations: Theoretical Status and Experimental Prospects}",
    eprint = "1410.1100",
    archivePrefix = "arXiv",
    primaryClass = "hep-ex",
    reportNumber = "FERMILAB-PUB-14-263-T",
    doi = "10.1016/j.physrep.2015.11.001",
    journal = "Phys. Rept.",
    volume = "612",
    pages = "1--45",
    year = "2016"
}

@article{Shima:2025aam,
    author = "Shima, Tatsushi",
    title = "{Experimental Search for Neutron{\textendash}Antineutron Oscillation with the Use of Ultra-Cold Neutrons Revisited}",
    eprint = "2508.07525",
    archivePrefix = "arXiv",
    primaryClass = "hep-ex",
    doi = "10.3390/sym17091524",
    journal = "Symmetry",
    volume = "17",
    number = "9",
    pages = "1524",
    year = "2025"
}

@article{Super-Kamiokande:2020bov,
    author = "Abe, K. and others",
    collaboration = "Super-Kamiokande",
    title = "{Neutron-antineutron oscillation search using a 0.37 megaton-years exposure of Super-Kamiokande}",
    eprint = "2012.02607",
    archivePrefix = "arXiv",
    primaryClass = "hep-ex",
    doi = "10.1103/PhysRevD.103.012008",
    journal = "Phys. Rev. D",
    volume = "103",
    number = "1",
    pages = "012008",
    year = "2021"
}

@article{QCDSF:2008qtn,
    author = "Braun, Vladimir M. and others",
    collaboration = "QCDSF",
    title = "{Nucleon distribution amplitudes and proton decay matrix elements on the lattice}",
    eprint = "0811.2712",
    archivePrefix = "arXiv",
    primaryClass = "hep-lat",
    reportNumber = "DESY-08-166, EDINBURGH-2008-45",
    doi = "10.1103/PhysRevD.79.034504",
    journal = "Phys. Rev. D",
    volume = "79",
    pages = "034504",
    year = "2009"
}

@article{Li:2025zfw,
    author = "Li, Lei-Yi and others",
    title = "{Lattice QCD calculation of charmed baryon decay constants in the continuum limit and at physical mass}",
    eprint = "2512.20372",
    archivePrefix = "arXiv",
    primaryClass = "hep-lat",
    doi = "10.1103/556x-5mgq",
    journal = "Phys. Rev. D",
    volume = "113",
    number = "5",
    pages = "054509",
    year = "2026"
}

@article{LHCb:2019ldj,
    author = "Aaij, Roel and others",
    collaboration = "LHCb",
    title = "{Precision measurement of the $\Lambda_c^+$, $\Xi_c^+$ and $\Xi_c^0$ baryon lifetimes}",
    eprint = "1906.08350",
    archivePrefix = "arXiv",
    primaryClass = "hep-ex",
    reportNumber = "LHCb-PAPER-2019-008, CERN-EP-2019-122",
    doi = "10.1103/PhysRevD.100.032001",
    journal = "Phys. Rev. D",
    volume = "100",
    number = "3",
    pages = "032001",
    year = "2019"
}

@article{Cowan:2010js,
    author = "Cowan, Glen and Cranmer, Kyle and Gross, Eilam and Vitells, Ofer",
    title = "{Asymptotic formulae for likelihood-based tests of new physics}",
    eprint = "1007.1727",
    archivePrefix = "arXiv",
    primaryClass = "physics.data-an",
    doi = "10.1140/epjc/s10052-011-1554-0",
    journal = "Eur. Phys. J. C",
    volume = "71",
    pages = "1554",
    year = "2011",
    note = "[Erratum: Eur.Phys.J.C 73, 2501 (2013)]"
}

@article{Durieux:2012gj,
    author = "Durieux, Gauthier and Gerard, Jean-Marc and Maltoni, Fabio and Smith, Christopher",
    title = "{Three-generation baryon and lepton number violation at the LHC}",
    eprint = "1210.6598",
    archivePrefix = "arXiv",
    primaryClass = "hep-ph",
    reportNumber = "CP3-12-44, LPSC12296",
    doi = "10.1016/j.physletb.2013.02.052",
    journal = "Phys. Lett. B",
    volume = "721",
    pages = "82--85",
    year = "2013"
}

@article{Heeck:2026dmh,
    author = "Heeck, Julian and Sokhashvili, Diana and Thapa, Anil",
    title = "{Opening up baryon-number-violating operators}",
    eprint = "2603.17050",
    archivePrefix = "arXiv",
    primaryClass = "hep-ph",
    month = "3",
    year = "2026"
}

@article{ATLAS:2025upm,
    author = "Aad, Georges and others",
    collaboration = "ATLAS",
    title = "{Search for resonant leptoquark production via lepton-jet signatures in pp collisions at $ \sqrt{s}=13 $ TeV and $ \sqrt{s}=13.6 $ TeV with the ATLAS detector}",
    eprint = "2507.03650",
    archivePrefix = "arXiv",
    primaryClass = "hep-ex",
    reportNumber = "CERN-EP-2025-142",
    doi = "10.1007/JHEP12(2025)180",
    journal = "JHEP",
    volume = "12",
    pages = "180",
    year = "2025"
}

@article{HeavyFlavorAveragingGroupHFLAV:2024ctg,
    author = "Banerjee, Sw. and others",
    collaboration = "Heavy Flavor Averaging Group (HFLAV)",
    title = "{Averages of b-hadron, c-hadron, and {\ensuremath{\tau}}-lepton properties as of 2023}",
    eprint = "2411.18639",
    archivePrefix = "arXiv",
    primaryClass = "hep-ex",
    doi = "10.1103/x87q-tld5",
    journal = "Phys. Rev. D",
    volume = "113",
    number = "1",
    pages = "012008",
    year = "2026"
}

@inproceedings{Pintucci:2024ude,
    author = "Pintucci, Laura",
    collaboration = "ATLAS, CMS",
    title = "{Measurements of single top quark production processes with the ATLAS and CMS experiments}",
    booktitle = "{16th International Workshop on Top Quark Physics}",
    eprint = "2401.07794",
    archivePrefix = "arXiv",
    primaryClass = "hep-ex",
    reportNumber = "ATL-PHYS-PROC-2024-005",
    month = "1",
    year = "2024"
}

@article{Bierlich:2022pfr,
    author = "Bierlich, Christian and others",
    title = "{A comprehensive guide to the physics and usage of PYTHIA 8.3}",
    eprint = "2203.11601",
    archivePrefix = "arXiv",
    primaryClass = "hep-ph",
    doi = "10.21468/SciPostPhysCodeb.8",
    journal = "SciPost Phys. Codeb.",
    volume = "2022",
    pages = "8",
    year = "2022"
}

@article{deFavereau:2013fsa,
    author = "de Favereau, J. and others",
    collaboration = "DELPHES 3",
    title = "{DELPHES 3: A modular framework for fast simulation of a generic collider experiment}",
    eprint = "1307.6346",
    archivePrefix = "arXiv",
    primaryClass = "hep-ex",
    doi = "10.1007/JHEP02(2014)057",
    journal = "JHEP",
    volume = "02",
    pages = "057",
    year = "2014"
}

@article{ParticleDataGroup:2022pth,
    author        = "Workman, R. L. and others",
    collaboration = "Particle Data Group",
    title         = "{Review of Particle Physics}",
    doi           = "10.1093/ptep/ptac097",
    journal       = "PTEP",
    volume        = "2022",
    pages         = "083C01",
    year          = "2022"
}

@article{Dev:2022jbf,
    author = "Dev, P. S. B. and others",
    title = "{Searches for baryon number violation in neutrino experiments: a white paper}",
    eprint = "2203.08771",
    archivePrefix = "arXiv",
    primaryClass = "hep-ex",
    doi = "10.1088/1361-6471/ad1658",
    journal = "J. Phys. G",
    volume = "51",
    number = "3",
    pages = "033001",
    year = "2024"
}

@article{ATLAS:2023afl,
    author = "Aad, Georges and others",
    collaboration = "ATLAS",
    title = "{Search for pair production of squarks or gluinos decaying via sleptons or weak bosons in final states with two same-sign or three leptons with the ATLAS detector}",
    eprint = "2307.01094",
    archivePrefix = "arXiv",
    primaryClass = "hep-ex",
    reportNumber = "CERN-EP-2023-123",
    doi = "10.1007/JHEP02(2024)107",
    journal = "JHEP",
    volume = "02",
    pages = "107",
    year = "2024"
}

@article{CMS:2020cpy,
    author = "Sirunyan, Albert M and others",
    collaboration = "CMS",
    title = "{Search for physics beyond the standard model in events with jets and two same-sign or at least three charged leptons in proton-proton collisions at $\sqrt{s}=$ 13 TeV}",
    eprint = "2001.10086",
    archivePrefix = "arXiv",
    primaryClass = "hep-ex",
    reportNumber = "CMS-SUS-19-008, CERN-EP-2020-001",
    doi = "10.1140/epjc/s10052-020-8168-3",
    journal = "Eur. Phys. J. C",
    volume = "80",
    number = "8",
    pages = "752",
    year = "2020"
}
